\documentclass[10pt,
aps,prd,
twocolumn,
superscriptaddress,nofootinbib,longbibliography]{revtex4-2}
\usepackage{cmap}
\usepackage[utf8]{inputenc}
\usepackage[T1]{fontenc}
\usepackage[english]{babel}
\usepackage{amsmath,amsfonts,amssymb,mathtools}
\usepackage{nccmath}
\usepackage[width=18cm,height=24cm]{geometry}
\usepackage[dvipsnames]{xcolor}
\usepackage{hyperref}
\hypersetup{
	colorlinks,%
	citecolor=RoyalPurple,%
	filecolor=blue,%
	linkcolor=ForestGreen,%
	urlcolor=WildStrawberry
}
\usepackage{indentfirst}
\usepackage{bbm}
\usepackage{enumitem}

\usepackage{setspace}
\makeatletter
\def\l@subsubsection#1#2{}
\makeatother

\numberwithin{equation}{section}
\renewcommand{\theequation}{\arabic{section}.\arabic{equation}}

\newcommand{\la}{\langle}
\newcommand{\ra}{\rangle}

\DeclareMathOperator{\Li}{Li}
\DeclareMathOperator{\Tr}{Tr}
\DeclareMathOperator{\tr}{tr}
\DeclareMathOperator{\Det}{Det}

\newcommand{\bk}{\boldsymbol{k}}
\newcommand{\bp}{\boldsymbol{p}}

\newcommand{\bx}{\boldsymbol{x}}

\begin{document}
	
	\title{One-loop QED effective action at finite temperature and density beyond HTL limit}
	\author{Farahmand Hasanov}
	\email{khasanov.mkh@phystech.edu}
	\affiliation{National Research Nuclear University MEPhI, 115409, Russia, 31 Kashirskoe shosse, Moscow, Russia}
	\affiliation{Institute for Theoretical and Mathematical Physics, Moscow State University, 119991, Leninskie Gory, GSP-1, Moscow, Russia}
	\author{Nikita Kolganov}
	\email{nikita.kolganov@phystech.edu}
	\affiliation{Theory Department, Lebedev Physics Institute, Leninsky Prospect 53, Moscow 119991, Russia}
	\affiliation{Moscow Institute of Physics and Technology, 141700, Institutskiy pereulok, 9, Dolgoprudny, Russia}
	
	\begin{abstract}
		We develop a framework for computing the nonlocal one-loop effective action in scalar and spinor QED at finite temperature and chemical potential in arbitrary even spacetime dimensions. Based on the heat kernel expansion to quadratic order in background fields, the framework reduces the effective action form factors to a common fundamental kernel. By resumming the external-momentum expansion, we obtain explicit closed-form expressions for the massless form factors in terms of polylogarithms and their derivatives with respect to the order. The same method determines the coefficients of the small-mass expansion to arbitrary order. The resulting expressions retain the full dependence on external Euclidean momentum, temperature, and chemical potential, without invoking the soft-momentum approximation. We also derive the corresponding photon polarization operators and evaluate their static limits, obtaining the Debye screening masses and electromagnetic susceptibilities.

	\end{abstract}
	
	\maketitle
		
	\vspace{-3ex}
	
\begingroup
	\setlength{\baselineskip}{10pt}      
	\setlength{\parskip}{0pt}
	\makeatletter
	\makeatother
	\tableofcontents
	\endgroup
%
%

\newpage
\section{Introduction and main results} \label{sec:main_res}

	The effective action provides a systematic description of the influence of quantum matter on background fields. In quantum electrodynamics (QED), integrating out charged particles produces corrections to Maxwell theory that encode vacuum polarization and photon interactions. The classic example is the Euler-Heisenberg effective action for spinor QED \cite{Euler-Heisenberg}, together with its scalar counterpart \cite{Weisskopf}. These one-loop results retain the full dependence on a constant electromagnetic field and describe nonlinear phenomena such as light-by-light scattering and vacuum pair production in an electric field. Their extensions and applications, including generalizations to finite temperature and density, are discussed in \cite{Dunne:2004nc,Hattori:2023StrongField}.
	
	For an inhomogeneous background, the effective action also depends on the spacetime variation of the field. Derivative corrections extend the constant field description to slowly varying backgrounds \cite{Karbstein:2021Derivative}. A complementary approach is to expand in the background amplitude while retaining the full dependence on its derivatives, as in covariant perturbation theory \cite{Barvinsky:1990CPT}. The resulting effective action is nonlocal, namely its kernels couple fields at distinct spacetime points and generally cannot be represented by a finite set of local derivative terms. At quadratic order, this information is encoded in the photon polarization tensor. Gauge invariance constrains its tensor structure, leaving scalar functions of momentum, called form factors, to be determined. Calculating the quadratic nonlocal effective action thus reduces to finding these form factors without expanding in the external momentum.
	
	At finite temperature and chemical potential, the same framework describes the electromagnetic response of a plasma. The medium selects a preferred rest frame, so the polarization tensor depends separately on frequency and spatial momentum and has distinct longitudinal and transverse components \cite{Weldon:1982Plasma,Haque:2024HTL}. The static longitudinal response determines Debye screening, while the momentum dependence of the static response also encodes electric and magnetic susceptibilities \cite{Ferreira:2023cqw}. After continuation to real frequencies, the polarization tensor governs photon propagation and collective excitations. Explicit form factors therefore connect equilibrium response coefficients to the behavior of spatially and temporally varying electromagnetic fields. They also allow different kinematic limits to be investigated within a common representation.
	
	Much of the analytic understanding of thermal form factors relies on a separation between soft external momenta and hard momenta carried by particles in the medium. The hard thermal loop (HTL) approximation captures the leading response in this regime \cite{Braaten:1990Soft}, with an analogous hard dense loop description at large chemical potential (see \cite{Haque:2024HTL} for a modern review). Recent developments include a complete next-to-leading-order calculation of the soft photon self energy in a hot and dense plasma \cite{Gorda:2023SoftPhoton}, leading fermion mass corrections to the HTL/HDL polarization tensor \cite{Comadran:2021Mass}, and mass corrections to its subleading momentum dependence \cite{Ferreira:2023cqw}. These results provide increasingly precise descriptions of the soft sector. However, a finite momentum expansion does not determine the full response outside that regime. More general integral representations are available, so the difficulty lies in obtaining explicit expressions retaining the external momentum dependence together with the dependence on temperature and chemical potential.
	
	An alternative approach is the Schwinger proper time representation, which expresses the one-loop functional determinant through the heat kernel of the matter wave operator \cite{Schwinger:1951Gauge}. This method organizes the calculation in powers of the background while allowing its derivative dependence to remain unexpanded \cite{Barvinsky:1990CPT}. Finite-temperature heat kernel techniques also provide a systematic route to effective theories expressed in terms of local operators, with gauge invariance and Polyakov loop dependence treated explicitly \cite{Megias:2004Thermal}. Recent work develops thermal effective actions obtained by integrating out heavy scalar and fermion fields \cite{Chakrabortty:2025Thermal}, while an application to hot scalar QED derives operators up to dimension six \cite{Bandyopadhyay:2026nrv}. These calculations characterize the response through an expansion in local operators. Determining nonlocal form factors instead requires retaining the external momentum dependence throughout the proper time calculation.
	
	We use the heat kernel approach within the imaginary time formalism, where temperature enters through compactified Euclidean time. The chemical potential is incorporated through twisted-periodic temporal boundary conditions. Working first with an imaginary chemical potential provides a convenient parametrization of this twist, whereas the physical chemical potential is subsequently reached by analytic continuation. This continuation requires attention to the singularities and branch structure of the resulting functions. The imaginary chemical potential used in the intermediate calculation must therefore be distinguished from the physical parameter specifying the density of the medium.
	
	Within this framework, we seek closed expressions combining full external momentum dependence, finite temperature and chemical potential, and a common treatment of scalar and spinor QED in arbitrary even spacetime dimension. Both theories can be formulated in terms of a second order wave operator, with their differences encoded in the normalization of the determinant, the matrix-valued potential term, and the temporal boundary conditions. Consequently, the same underlying form factor calculation provides the nonlocal kernels for both types of charged matter. Their physical interpretation after continuation to real chemical potential nevertheless differs. In the scalar theory, the limit $M\to0$ at fixed nonzero chemical potential leaves the stable normal phase domain and encounters the instability associated with Bose-Einstein condensation. The continued polarization operators and screening coefficients can then acquire imaginary contributions, signaling the instability of the uncondensed state rather than describing the equilibrium response of a condensate. We therefore preserve the general scalar polarization operators, which remain mathematically consistent at imaginary chemical potential, but restrict the explicit small-mass applications to the fermionic theory.
	
	We calculate the effective action to quadratic order in the background and organize the matter mass dependence as a small-mass expansion. This expansion is distinct from an HTL or high-temperature expansion. Specifically, expanding in the mass does not itself require the external momentum to be soft compared with the thermal scale. At each order, we keep the full dependence on external momentum, temperature, and the chemical potential parameter. 
	The massless form factors are obtained directly in closed form, and the same construction determines the coefficients of the small-mass expansion.
	
	All relevant form factors reduce to a common fundamental form factor. Evaluating it explicitly provides a starting point for separating vacuum and thermal contributions, extracting asymptotic regimes, and calculating electromagnetic response coefficients without repeating the proper time integration and Matsubara summation. The resulting representation complements soft momentum and local operator expansions by preserving the momentum dependence from which those limits can be investigated. 
	
	To formulate the problem and state our results, consider a charged matter field of mass $M$ coupled to a $U(1)$ gauge potential $A_\rho(x)$ at temperature $T=\beta^{-1}$ and finite chemical potential in an even spacetime dimension $D=d+1$. Its one-loop effective action is defined by
	\begin{equation} \label{eq:eff_act_intro}
		\varGamma[A] = \kappa \, \Tr \log \hat{F}[A],
		\quad
		\hat{F}[A] = -\hat{D}_\rho^2 + \hat{P} + M^2,
	\end{equation}
	where $\kappa$ is a normalization factor and $\hat{F}$ is the Euclidean wave operator of the matter field. It acts on the tensor product of the two-dimensional $U(1)$ charge space and a charge singlet factor such as spinor space, flavor space, etc. Here $\hat{D}_\rho = \partial_\rho + i \hat{\sigma} A_\rho$ is gauge covariant derivative, and $\hat{\sigma} = \operatorname{diag}(-1,1)$ is the $U(1)$ charge generator.
	We suppress the electromagnetic coupling $e$ throughout. The potential term $\hat{P}$ may represent the Hessian of a matter-field potential, non-minimal coupling to the gauge field, etc.
	The functional trace, denoted as $\Tr$ in the definition of $\varGamma[A]$, is taken over the space of functions satisfying the following twisted-periodic boundary conditions in Euclidean time $t$
	\begin{equation}
		\begin{aligned}
			&\varPhi(t, \bx)\bigr|_{t=\beta} = e^{i \beta\mu \hat{\sigma}} \varPhi(t, \bx)\bigr|_{t=0}, \\
			&\partial_t\varPhi(t, \bx)\bigr|_{t=\beta} = e^{i \beta\mu \hat{\sigma}} \partial_t\varPhi(t, \bx)\bigr|_{t=0},
		\end{aligned}
	\end{equation}
	where $\varPhi$ is a two-component column in the $U(1)$ charge space, with negative- and positive-charge components. The Euclidean time interval has length $\beta = T^{-1}$, whereas the twist parameter $\mu$ has the meaning of imaginary chemical potential and is discussed below.
	
	The effective action (\ref{eq:eff_act_intro}) covers both scalar QED and standard fermionic (spinor) QED. The two cases are distinguished by the choice of the normalization coefficient $\kappa$, the potential term $\hat{P}$ and the twist parameter $\mu$. Specifically, the relation between the imaginary chemical potential $\mu$ and the physical chemical potential $\mu_{\mathrm{phys}}$, differs for bosons and fermions. For the scalar and spinor QED, respectively, one has 
	\begin{subequations}
	\begin{align}
		&\kappa_{\mathrm{s}} = \frac12, &&
		\hat{P}_{\mathrm{s}} = 0, &&
		\mu_{\mathrm{s}} = i\mu_{\text{phys}}, \\
		&\kappa_{\mathrm{f}} = -\frac14, &&
		\hat{P}_{\mathrm{f}} = -\frac{i}{2} \hat{\sigma}\, \gamma^{\mu\nu} F_{\mu\nu}, &&
		\mu_{\mathrm{f}} = \frac{\pi}{\beta} + i\mu_{\text{phys}}, \hspace{-0.3em} \label{eq:intro_spinor_subs}
	\end{align}
	\end{subequations}
	where $\gamma^{\mu\nu} = \frac12[\gamma^\mu, \gamma^\nu]$.
	In what follows, we first keep the parameters general and then specialize to scalar and spinor QED.
	
	We calculate the effective action (\ref{eq:eff_act_intro}) using the heat kernel method within zeta-regularization technique as a perturbative expansion $\varGamma[A] = \varGamma_0[A] + \varGamma_1[A] + \varGamma_2[A] + \ldots$ in powers of $A_\nu$, $\varGamma_n \sim A^n$ up to quadratic order in $A_\nu$, assigning to the potential $\hat{P}$ the same order in expansion parameter as $A_\nu$. Decomposing the potential in powers of charge operator $\hat{P} = \hat{1} \, \mathbb{P}_{0} + \hat{\sigma} \, \mathbb{P}_{1}$ such that $\mathbb{P}_{0,1}$ act on charge singlets, the most relevant quadratic part of the effective action is obtained in the form
	\begin{widetext}
	\begin{align} \label{eq:G2_intro}
		&\varGamma_2[A] = -\frac12 \frac{\kappa}{(4\pi)^{D/2}} \int\limits_{(\omega_m, \bk)} \, \tr \Bigl[ k^2 \, \varphi_\|(k, M)\,  A^\|_{\rho}(k) A^\|_{\rho}(-k)  \mathbbm{1}+ 
		k^2 \, \varphi_\perp(k, M) \, A^\perp_{\rho}(k) A^\perp_{\rho}(-k) \mathbbm{1} \nonumber\\[-2ex]&\hspace{15em}+
		2 \varphi_{\times}(k, M) A^\|_t(k) \mathbb{P}_{0}(-k) + \varphi_{P}(k, M)\Bigl(\mathbb{P}_{0}(k) \mathbb{P}_{0}(-k) + \mathbb{P}_{1}(k) \mathbb{P}_{1}(-k) \Bigr)
		\Bigr].
	\end{align}
	\end{widetext}
	where we use the shorthand for the Fourier space sum and integral
	\begin{equation} \label{eq:intro_int_notation}
		\int\limits_{(\omega_m, \bk)} \equiv \frac1{\beta} \sum_{m=-\infty}^{\infty} \int\frac{d^d \bk}{(2\pi)^d},
	\end{equation}
	and the trace is taken over the charge-singlet space. All functions of $k$ are given in the Fourier space, parameterized by the momentum vector $k = (\omega_m, \bk)$, where $\omega_m = 2\pi m / \beta$ are discrete Matsubara frequencies. Here $A^\|_\rho$ and \(A^{\perp}_\rho\), appearing in (\ref{eq:G2_intro}), are the components of $A_\rho$ in the subspace orthogonal to $k=(\omega_m,\bk)$, split along and orthogonal to the Euclidean-time direction, defined by the unit vector $u^\rho = \delta^\rho_t$. With $\mathcal{P}_{\mu\nu}= \delta_{\mu\nu}-k_\mu k_\nu / k^2$ and $v_\mu=\mathcal{P}_{\mu\nu}u^\nu$ they read
	\begin{equation}
		A^\|_\rho(k)=\mfrac{v_\rho v_\nu}{v^2}A_\nu(k),
		\qquad
		A^\perp_\rho(k)=\Bigl(\mathcal{P}_{\rho\nu}
		-\mfrac{v_\rho v_\nu}{v^2}\Bigr)A_\nu(k).
	\end{equation}
	Both are orthogonal to $k_\rho$, and $A^{\|}_t(k)$ denotes the temporal component of $A^{\|}_\rho(k)$.
	
	The answer (\ref{eq:G2_intro}) shows that, at quadratic order, all information on the effective action is encoded in the four \emph{form factors}  $\varphi_I(k, M)$, $I=\|,\perp,\times, P$. As we will see, they can all be expressed through a single form factor, referred to as \emph{fundamental} form factor (``master integral''). It is labeled by an integer number $q$ and is defined as
	\begin{widetext}
	\begin{equation}
		\varphi_{0}^{q}(k, M) = \lim_{s\to0}\frac{\partial}{\partial s} \biggl(\frac{\Lambda^{2s}}{\Gamma(s)}\int_0^\infty d\tau \, \tau^{s-q-1} e^{-\tau M^2} \, \frac12 \sum_{\sigma=\pm1}\int_0^1 d\alpha\, e^{-\alpha(1-\alpha)\tau k^2}\,  \frac{\sqrt{4\pi\tau}}{\beta} \sum_{n=-\infty}^\infty e^{-\tau (\omega_n-\alpha\omega_m+\sigma\mu)^2}\biggr).
	\end{equation}
	\end{widetext}
	Here $s$ is a zeta-regularization parameter, $\Lambda$ is the ultraviolet regularization scale. The $\tau$-integral arises due to the Schwinger proper time parametrization, whereas  $\alpha$ has the meaning of the Feynman parameter. In turn, the sum over $\sigma$ comes from the trace over the $U(1)$ charge space. Throughout, we display only the explicit arguments $k$ and $M$ of the fundamental form factor $\varphi^q_0(k, M)$, whereas its dependence on the inverse temperature $\beta$, the imaginary chemical potential $\mu$ and the UV scale $\Lambda$ is kept implicit. 
	
	The fundamental form factor defines auxiliary quantities, referred to as the \emph{basic} form factors $\varphi_{1}^q$, $\psi_0^q$ and the basic function $\rho_0$, expressed in terms of the former as
	\begin{equation}
		\begin{aligned}
		&\varphi^q_1(k, M)= -\frac1{2} \frac{\partial}{\partial \mu} \varphi_0^{q+1}(k, M), \\
		&\psi_0^{q}(k, M) = \frac12 \frac{\partial}{\partial \log \Lambda} \varphi_0^{q}(k, M), \\
		&\rho_0^q(M) = \lim_{\bk\to0}\lim_{\omega_m\to 0} \varphi_0^q(k,M). \vphantom{\frac12}
		\end{aligned}
	\end{equation}
	The form factors $\varphi_I$,~$I=\|,\perp,\times, P$ defining the quadratic part of the effective action (\ref{eq:G2_intro}), are expressed as a linear combination of the fundamental and basic ones as
	\begin{align}
			&\begin{aligned}
				\varphi_\|(k,M) = -\mfrac{8}{k^2} \Bigl(\mfrac{D-2}{2}\varphi_0^{D/2-1}(k,M) - \psi_0^{D/2-1}(k,M)\Bigr) \\-2  \Bigl(1+\mfrac{4M^2}{k^2}\Bigr)\varphi_0^{D/2-2}(k,M),
			\end{aligned} \nonumber \\
			&\varphi_\perp(k,M) = \mfrac{4}{k^2} \bigl(\varphi^{D/2-1}_0(k,M) - \rho_0^{D/2-1}(M) \bigr), \vphantom{\mfrac{8}{k^2}}\nonumber \\
			&\varphi_{\times}(k,M) = 4 \, \varphi_1^{D/2-2}(k,M),\vphantom{\mfrac{8}{k^2}} \nonumber \\
			&\varphi_{P}(k,M) = 2 \, \varphi_0^{D/2-2}(k,M). \vphantom{\mfrac{8}{k^2}}
	\end{align}
	With the basic form factors defined in terms of the fundamental form factor, we now turn to the small-mass expansion of the fundamental form factor $\varphi_0^{q}(k, M)$, which will allow us to obtain their explicit closed forms.

	We express the fundamental form factor as a small-mass expansion
	\begin{equation} \label{eq:mass_exp_intro}
		\varphi_0^{q}(k, M) = \sum_{m=0}^\infty \frac{(-)^m}{m!} M^{2m} \, \varphi_0^{q-m}(k).
	\end{equation}
	In view of the obvious identity $\lim_{M\to 0}\varphi_0^{n}(k, M) = \varphi_0^{n}(k)$, we will refer to $\varphi_0^{n}(k)$ as \emph{massless} form factors. They are obtained in closed form, with expressions that differ significantly for $n\ge 0$ and $n<0$.
	
	In case of $n\ge0$ that gives the leading orders of small-mass expansion (\ref{eq:mass_exp_intro}), the fundamental form factor naturally splits into vacuum and pure thermal parts
	\begin{align}
		\varphi^n_0(k) = \varphi^n_{0v}(k) + \varphi^n_{0\beta}(k),
	\end{align}	
	where the vacuum part has the standard form
	\begin{equation}
		\varphi_{0v}^{n}(k) = -2 (-)^n \frac{n!}{(2n+1)!} k^{2n} \Bigl(\log\mfrac{k}{\Lambda}  - H_{2n+1} + \tfrac12 H_n \Bigr),
	\end{equation}
	where $H_n = \sum_{j=1}^n 1/j$ is the harmonic number, whereas the pure thermal part has the following explicit expression	
	\begin{equation} \label{eq:vphi_M0_pos_intro}
		\varphi^n_{0\beta}(k) =  (-)^n\Bigl(\frac2\beta\Bigr)^{\!2n}\, k^{2n} \,\partial_{k^2}^{n} \bigl( \varOmega \, \varPhi^n_0(|k|) \bigr).
	\end{equation}
	Here $\varOmega$ denotes shift-like operator, acting on test functions of the momentum norm $|k|$ as
	\begin{align}
		&\varOmega \, \varPhi(|k|) =
		\frac{1}{2|\bk|}\Bigl[\bigl(|\bk|+i\omega\bigr) \, \varPhi(|\bk|+i\omega) \nonumber\\&\hspace{9em} + \bigl(|\bk|-i\omega\bigr) \, \varPhi(|\bk|-i\omega)\Bigr].
	\end{align}
	The auxiliary function $\varPhi^n_0(|k|)$ has the form
	\begin{widetext}	
	\begin{align} \label{eq:vphi_aux_pos_intro}
		&\varPhi^n_0(|k|) = -  \sum_{r=0}^{n}   \frac{2}{(2r+1)!} \Bigl(\frac{\beta k}{2}\Bigr)^{\!2r} \bigl( \Li_{2n-2r}(e^{i\beta\mu}) + \Li_{2n-2r}(e^{-i\beta\mu}) \bigr) \bigl(\log (\tfrac{\beta k}{2}) + \gamma - H_{2r+1}\bigr) \nonumber \\ &\hspace{7em}+ \frac{2}{\beta k}\bigl( \Li^{(1)}_{2n+1}(e^{\beta(\frac{k}2+i\mu)}) +\Li^{(1)}_{2n+1}(e^{\beta(\frac{k}2-i\mu)}) -\Li^{(1)}_{2n+1}(e^{-\beta(\frac{k}2-i\mu)}) -\Li^{(1)}_{2n+1}(e^{-\beta(\frac{k}2+i\mu)}) \bigr),		 
	\end{align}
	\end{widetext}
	depending only on the absolute value of the momentum $|k|$. Here $\Li_s^{(1)}(x)$ denotes $\frac{\partial}{\partial s} \Li_s(x) $, and $\gamma$ is the Euler constant. We omit the subscript of $\omega_m$ for brevity reasons.
	
	Although the answers for the massless form factors have an explicit and rather compact form, it is useful to rewrite them in a more expanded form. For the case of $n \ge 0$, the massless form factor $\varphi^n_0$ naturally splits into three parts
	\begin{equation} \label{eq:intro_vphi_split}
		\varphi^{n}_{0}(k) = \varphi^{n,\mathrm{sing}}_{0}(k) + \varphi^{n,\mathrm{log}}_{0}(k) + \varphi^{n,\mathrm{reg}}_{0}(k)
	\end{equation}
	where the singular part $\varphi^{n,\mathrm{sing}}$ comes from the $r=1, \dots n$ terms in the sum in (\ref{eq:vphi_aux_pos_intro}), the logarithmic part corresponds to $r=0$ term of this sum, whereas the regular part $\varphi^{n,\mathrm{reg}}_{0}$ comes from
	the second line of~(\ref{eq:vphi_aux_pos_intro}). Substituting the corresponding parts into (\ref{eq:vphi_M0_pos_intro}) and summing with the vacuum part, we obtain the following expressions
	\begin{widetext}
	\begin{align}
		&\varphi^{n,\mathrm{sing}}_0(k) = 
		\sum_{r=1}^{n}  \frac{1}{(2r)!(2(n-r)+1)!} \, \Bigl(\frac{4\pi}{\beta}\Bigr)^{\!2r}  B_{2r}\bigl(\tfrac{\beta \mu}{2\pi}\bigr) \, k^{2n} \partial_{k^2}^{r-1} k^{2(n-r)} \partial_{k^2}^{n-r} \mfrac1{\bk^2} \Bigl[1 - \mfrac{i\omega}{2|\bk|} \log \mfrac{|\bk|+i\omega}{|\bk|-i\omega} \Bigr], \nonumber\\
		&\varphi^{n,\mathrm{log}}_0(k) = 2 (-)^n \frac{n!}{(2n+1)!}  \, k^{2n}  \Bigl[\log \mfrac{\beta \Lambda}{2} + \gamma\Bigr] +
		 \frac{2}{(2n+1)!} \, k^{2(2n+1)} \partial_{k^2}^n \Bigl[\mfrac1{k^2} \mfrac{i\omega}{2|\bk|} \log \mfrac{|\bk|+i\omega}{|\bk|-i\omega} \Bigr], \\
		 &\varphi^{n,\mathrm{reg}}_0(k) =  (-)^n\Bigl(\frac2\beta\Bigr)^{\!2n}\, k^{2n} \,\partial_{k^2}^{n} \Bigl[\mfrac{1}{\beta |\boldsymbol{k}|} \sum_{\{\sigma\}=\pm1} \sigma_k  \Li^{(1)}_{2n+1}\!\bigl( e^{\beta( \sigma_k\frac{|\boldsymbol{k}|}2 +i\sigma_\omega\frac{\omega}2 +i\sigma \mu)} \bigr)	\Bigr] \nonumber,
	\end{align}
	\end{widetext}
	where $B_n(x)$ denotes the Bernoulli polynomial, and the sum over $\{ \sigma \}$ means independent sums over $\sigma, \sigma_k, \sigma_\omega = \pm 1$. When differentiating w.r.t.\ $k^2$, one assumes that $\omega$ is fixed and $|\bk|=\sqrt{k^2-\omega^2}$.	
	This splitting has very intuitive interpretation in the limit $\beta k \to 0$, $\beta \mu = \mathrm{const}$. Indeed, in this limit $\varphi_0^{n,\mathrm{sing}}$ has a power-law divergence, $\varphi_0^{n,\mathrm{log}}$ is logarithmically divergent, whereas $\varphi_0^{n,\mathrm{reg}}$ is regular. This regularity can be read off from the expansion
	\begin{align}
		 &\varphi^{n,\mathrm{reg}}_0(k) = 2 (-)^n \, k^{2n}	\sum_{r=0}^{\infty} (-)^{r} \frac{n!}{\,(2r+2n+1)!} \Bigl(\frac{\beta k}{2}\Bigr)^{\!2r} \nonumber\\ &\hspace{2em} \times  \bigl( \Li^{(1)}_{-2r}(e^{i\beta\mu}) + \Li^{(1)}_{-2r}(e^{-i\beta\mu}) \bigr) 
		  \, C_{2r}^{n+1}(\tfrac{\omega}{k}),
	\end{align}
	where $C_{2r}^{n+1}(x)$ denotes the Gegenbauer polynomial.
	
	For $n < 0$, the massless form factors $\varphi^n_0(k)$ contribute an infinite series of subleading terms in the small-mass expansion (\ref{eq:mass_exp_intro}). Unlike the case of $n\ge0$, these form factors do not have UV divergences. It is parameterized in a form similar to that (\ref{eq:vphi_M0_pos_intro}) for nonnegative values of $n = -p-1$, again with the use of the operator $\varOmega$, as follows
	\begin{equation} \label{eq:vphi_M0_neg_intro}
		\varphi_0^{-p-1}(k) =   \frac{(-)^{p+1}}{k^{2p+2}} \, \beta |\bk| \, (2|\bk|)^{2p} \, \partial_{k^2}^p \bigl( \varOmega \, \varPhi_0^{-p-1}(k) \bigr) ,
	\end{equation}
	where $p \ge 0$ and the corresponding auxiliary function reads
	\begin{widetext}
	\begin{align} \label{eq:intro_vphi_neg_aux}
		&\varPhi^{-p-1}_0(k) = \frac{2}{\beta k} 
		\biggl[ \Li^{(1)}_{0}(e^{\beta(k/2+i\mu)}) + \Li^{(1)}_{0}(e^{\beta(k/2-i\mu)}) + \Li^{(1)}_{0}(e^{-\beta(k/2-i\mu)}) + \Li^{(1)}_{0}(e^{-\beta(k/2+i\mu)}) \nonumber\\ & \hspace{18.5em} - 2 \sum_{r=0}^{p} \frac{1}{(2r)!} \Bigl(\frac{\beta k}{2}\Bigr)^{\!2r} \bigl( \Li^{(1)}_{-2r}(e^{i\beta\mu}) + \Li^{(1)}_{-2r}(e^{-i\beta\mu})\bigr) \biggr].
	\end{align}
	\end{widetext}
	Similar to $\varphi^{n,\mathrm{reg}}_{0}(k)$ it is regular in the limit $\beta k \to 0$, $\beta\mu = \mathrm{const}$, that can be seen from the corresponding expansion
	\begin{align}
		&\varphi_0^{-p-1}(k) \nonumber\\ & \; = \frac{2}{k^{2(p+1)}} \,
		\sum_{r=0}^\infty \frac{(-)^{r}}{(2r+1)!} \, \Bigl(\frac{\beta k}{2}\Bigr)^{2(r+p+1)} \, \tilde{C}^{-p}_{2r+2p+2}(\tfrac{\omega}{k})  \nonumber \\  &\;\;\;\times \bigl( \Li^{(1)}_{-2(r+p+1)} (e^{i\beta\mu}) +  \Li^{(1)}_{-2(r+p+1)} (e^{-i\beta\mu})  \bigr) ,
	\end{align}
	where we have defined the regularized Gegenbauer polynomials as $\tilde{C}_{n}^{-p}(x) = \lim_{\lambda\to-p}\Gamma(\lambda) C^\lambda_n(x)$.
	The explicit expressions (\ref{eq:vphi_M0_pos_intro}) and (\ref{eq:vphi_M0_neg_intro}) for the form factors at each order in the small-mass expansion (\ref{eq:mass_exp_intro}), together constitute one of the main results of this work.

	Being specified to the most physically relevant case of the spinor QED, we substitute (\ref{eq:intro_spinor_subs}) into the general expression (\ref{eq:G2_intro}) for the quadratic part of the effective action and taking trace over spinor space, we obtain a simpler expression, having the form of the combination of longitudinal and transverse photon polarizations
	\begin{align}
		&\varGamma_{2}[A] = \frac12 \int\limits_{(\omega_m, \bk)} \Bigl[ \varPi_{\|}(k, M)\,  A^\|_{\rho}(k) A^\|_{\rho}(-k) \nonumber \\[-2ex]& \hspace{8em} + 
		\varPi_{\perp}(k, M) \, A^\perp_{\rho}(k) A^\perp_{\rho}(-k) \Bigr],
	\end{align}
	Here the \emph{polarization operators} $\varPi_{\|,\perp}(k, M)$, now absorb all the form factors and the normalization coefficients, and have the following expressions in terms of the basic form factors
	\begin{align}
		&\varPi_\|(k, M) = -\frac{2}{(2\pi)^{D/2}}  \Bigl[ \Bigl(\mfrac{D}2-1\Bigr) \varphi_0^{D/2-1}(k, M) \nonumber\\ & \hspace{4em} - \psi_0^{D/2-1}(k, M) +  M^2\varphi_0^{D/2-2}(k, M)\Bigr], \\
		&\varPi_\perp(k, M) = \frac{1}{(2\pi)^{D/2}}  \Bigl[\varphi^{D/2-1}_0(k, M) - \rho_0^{D/2-1}(M)  \nonumber\\ & \hspace{11.75em} + \mfrac{k^2}{2} \varphi_0^{D/2-2}(k, M)\Bigr],
	\end{align}
	 The small-mass expansion of the fundamental form factor (\ref{eq:mass_exp_intro}) along with the explicit form (\ref{eq:vphi_M0_pos_intro}),~(\ref{eq:vphi_M0_neg_intro}) of the expansion coefficients, allows us to obtain the explicit form of the corresponding small-mass expansion of the polarization operators	 
	 \begin{equation}
	 	\varPi_I(k, M) = \sum_{m=0}^\infty \varPi_I^{(m)}(k, M),\qquad
	 	 I = \|, \perp,
	 \end{equation}
	 where $\varPi_I^{(m)} \sim M^{2m}$.

		We illustrate these results in the physically most relevant case of spinor QED in $D=4$.
		The massless longitudinal polarization provides a simple example of the general structure. It separates into power-law, logarithmic, and regular contributions according to their behavior as $T\to\infty$ at fixed external momentum and $\mu/T$
		\begin{equation}
			\varPi_\|^{(0)}
			= \varPi_\|^{(0),\mathrm{sing}}
			+ \varPi_\|^{(0),\mathrm{log}}
			+ \varPi_\|^{(0),\mathrm{reg}}
		\end{equation}
		where the splitting is induced by the corresponding splitting of the fundamental form factor (\ref{eq:intro_vphi_split}).
	
	The explicit expressions are
			\begin{widetext}
			\begin{align}
				&\varPi_\|^{(0),\mathrm{sing}}(k) = 
				\Bigl(\mfrac{T^2}{3}+\mfrac{\mu^2}{\pi^2}\Bigr) \mfrac{k^2}{\bk^2} \Bigl(1-\mfrac{i \omega}{2|\bk|}\log \mfrac{|\bk|+i \omega}{|\bk|-i \omega}\Bigr),\\
				&\varPi_\|^{(0), \mathrm{log}}(k) = \frac{ k^{2}}{6\pi^{2}}  \Bigl[\log \mfrac{\Lambda}{2T} + \gamma + 1 - \Bigl(1+\mfrac{k^2}{2\bk^2}\Bigr) \Bigl(1-\mfrac{i \omega}{2|\bk|} \log \mfrac{|\bk|+i \omega}{|\bk|-i \omega}\Bigr) \Bigr], \\
				&\varPi_\|^{(0),\mathrm{reg}}(k) = - \frac{k^{2}}{\pi^2} \frac{T^3}{|\bk|^3} \sum_{\{\sigma\}=\pm1} \Bigl[ \sigma_k  \Li^{(1)}_{3}\!\bigl(-e^{\, \sigma_k\frac{|\boldsymbol{k}|}{2T} +\sigma_\omega\frac{i \omega}{2T} + \sigma \frac{\mu}{T}} \bigr) -\mfrac{|\bk|}{2T} \Li^{(1)}_{2}\!\bigl(-e^{\, \sigma_k\frac{|\boldsymbol{k}|}{2T} +\sigma_\omega\frac{i \omega}{2T} + \sigma \frac{\mu}{T}} \bigr) \Bigr] \nonumber,			
				\\&\hphantom{\varPi_\|^{(0),\mathrm{reg}}(k)} =  \frac{k^{2}}{\pi^{2}} 	\sum_{r=0}^{\infty} \frac{(-)^{r}}{\,(2r+3)!} \Bigl(\frac{ k}{2T}\Bigr)^{\!2r} \bigl( \Li^{(1)}_{-2r}(-e^{\frac{\mu}{T}}) + \Li^{(1)}_{-2r}(-e^{-\frac{\mu}{T}}) \bigr) 
				\, C_{2r}^{(2)}(\tfrac{\omega}{k}).
			\end{align}
		\end{widetext}
		The last equality gives the high-temperature series of the regular contribution. In particular, it makes explicit that the apparent $T^3$ growth of the closed expression cancels between the polylogarithm terms.
		The leading singular contribution has the familiar HTL/HDL structure, with Debye screening coefficient $T^2/3+\mu^2/\pi^2$ in our normalization.
		
		The organization into singular, logarithmic, and regular terms extends to higher orders in the small-mass expansion. At fixed even $D$, only finitely many orders can contain contributions growing as powers or logarithms of $T$. Increasing the mass-expansion order lowers the index of the fundamental massless form factors. Specifically, for $m\ge D/2$, only negative-index form factors and their derived quantities remain, and the corresponding contributions are regular in the high-temperature limit. 
		
		Every finite order can therefore be written explicitly using
		(\ref{eq:vphi_M0_pos_intro}) and
		(\ref{eq:vphi_M0_neg_intro}). The expressions become increasingly lengthy at high orders because both the number of subtraction terms in
		(\ref{eq:intro_vphi_neg_aux}) and the order of the momentum derivatives increase. The representation in terms of the fundamental form factor provides a compact all-order prescription, while the explicit massless example above displays the characteristic structure of the answer. Further longitudinal results are given in Section~\ref{sec:appl}, and the corresponding transverse expressions in $D=4$ are collected in Appendix~\ref{sec:app_Pi_explicit}.

	The rest of the paper is organized as follows. In Section~\ref{sec:ea_chemical}, we show how the effective action in scalar and fermionic QED at finite temperature and density is expressed in terms of a functional determinant of a second-order operator subject to twisted periodic boundary conditions. In Section~\ref{sec:heat_kernel}, we calculate the heat kernel trace needed for the zeta-regularization calculation of the effective action. The resulting heat trace is then used in Section~\ref{sec:eff_act} to obtain the effective action in terms of form factors, which we compute exactly at each order in the small-mass expansion. Section~\ref{sec:appl} is devoted to the particular cases of charged scalar and fermion fields and to the analytic continuation to real time and chemical potential. The appendices contain details of the calculations.

\section{Effective action at finite density} \label{sec:ea_chemical}

\subsection{Scalar QED}

Let us consider scalar QED, i.e.\ theory of complex scalar field $\phi$ coupled to a $U(1)$ gauge field $A_\mu$. The Minkowski-space action has the form
\begin{equation}
	S_{\mathrm{s}}[\phi, \phi^*\!, A] = \int d^D x \, \Bigl[
	-\frac14 F_{\rho\sigma}^2 - (D_\rho \phi)^* D^\rho\phi
	- M^2 \phi^* \phi \Bigr]
\end{equation}
where the covariant derivative is $D_\rho = \partial_\rho - i A_\rho$, the field strength is $F_{\rho\sigma} = \partial_\rho A_\sigma - \partial_\sigma A_\rho$ and we use mostly plus metric signature $\eta_{\rho\sigma} = \operatorname{diag}(-, +, \ldots +)$.  Equations of motion derived from this action read
\begin{align}
	&\partial_\rho F^{\rho\sigma} = J^\sigma, \label{eq:QED_eom}\\
	&J_\rho = i \bigl[\phi^* D_\rho \phi  - \phi \,(D_\rho \phi)^*\bigr]
\end{align}
where $J^\rho$ is the conserved Noether current associated with the global $U(1)$ symmetry $\phi \mapsto e^{-i\alpha} \phi$.

To study finite-density effects, we pass to the Hamiltonian formalism by decomposing spacetime into time $t$ and $d$ spatial dimensions, labeled by the coordinates $\bx$. The action becomes
\begin{align}
	&S_{\mathrm{s}}[\phi, \phi^*\!, A] = \int dt \, d^dx \biggl[
	\frac12(\dot A_i - \partial_i A_t)^2 - \frac14 F_{ij}^2 \nonumber \\& \hspace{4em} + (D_t \phi)^* D_t \phi - (D_i \phi)^* D_i \phi - M^2 \phi^* \phi
	\biggr]
\end{align}
where $i=1, \ldots d$ and $A_t \equiv A_0$, $\partial_t \equiv \partial_0$, etc. The canonical momenta conjugate to the fields are defined as
\begin{equation} \label{eq:legendre_mom}
	\begin{aligned}
	E_i &= \frac{\partial \mathcal{L}}{\partial\dot{A}_i} = \dot{A}_i - \partial_i A_t, \\ 
	\pi &= \frac{\partial \mathcal{L}}{\partial\dot{\phi}} = (D_t \phi)^*, \qquad \pi^* = \frac{\partial \mathcal{L}}{\partial\dot{\phi}^*} = D_t \phi
	\end{aligned}
\end{equation}
Using the Legendre transform
\begin{equation} \label{eq:legendre_ham}
	\mathcal{H} = \pi \dot{\phi} + \pi^* \dot{\phi}^* + E_i \dot{A}_i - \mathcal{L},
\end{equation}
we obtain the total Hamiltonian
\begin{align}
	&H = \int d^dx \biggl[ \pi^* \pi + (D_i \phi)^* D_i \phi + M^2 \phi^* \phi \nonumber \\&\hspace{2em} +\frac12 E_i^2 + \frac14 F_{ij}^2 - A_t \bigl(\partial_i E_i - i (\pi \phi - \pi^* \phi^*)\bigr)   \biggr]
\end{align}
where we neglected the boundary term $\partial_i (A_t E_i)$, which does not affect the bulk dynamics.
The conserved $U(1)$ charge associated with the current $J^\rho$ reads
\begin{equation}
	Q = \int d^dx \, J^t = \int d^dx \; i (\pi \phi - \pi^* \phi^*).
\end{equation}

At finite temperature $T = 1/\beta$ and chemical potential $\mu$ the partition function has the form
\begin{equation}
	Z = \Tr e^{-\beta (H - \mu Q)} .
\end{equation}
In the path integral formalism, this becomes
\begin{align} \label{eq:Z_scalar}
	&Z_{\mathrm{s}} = 
	\int \mathcal{D}[\phi, \phi^*\!, \pi, \pi^*\!, A_i, E_i, A_t] \nonumber\\&\;\times  \exp \biggl[ i \int d\tilde{t} \, d^dx \, (\pi \dot{\phi} + \pi^* \dot{\phi}^* + E_i \dot{A}_i)
	- \!\int\! d\tilde{t} \, \bigl(H - \mu Q\bigr)
	\biggr]
\end{align}
where all fields are assumed to be periodic in Euclidean time $\tilde{t} \in [0, \beta]$ and dot now denotes the $\tilde{t}$-derivative.
Note that the combination $H - \mu Q$ can be rewritten as
\begin{align} \label{eq:HQ_scalar}
	&H - \mu Q = \int d^dx \biggl[ \pi^* \pi + (D_i \phi)^* D_i \phi + M^2 \phi^* \phi  \nonumber\\ & \hspace{2em}+ \frac12 E_i^2 + \frac14 F_{ij}^2 - (A_t - \mu) \bigl(\partial_i E_i - i (\pi \phi - \pi^* \phi^*)\bigr)  \biggr]
\end{align}
up to the boundary term $\mu \,\partial_i E_i$, which was neglected. This shows that introducing chemical potential $\mu$ is equivalent to shifting the temporal gauge field, i.e.\  $A_t \mapsto A_t - \mu$. 

	Now, we explicitly integrate over momentum variables $\pi$, $\pi^*$, $E_i$. The integrals are Gaussian and the saddle point equations reproduce the inverse of  Legendre transformation (\ref{eq:legendre_mom}),~(\ref{eq:legendre_ham}) upon substitution $\partial_t \mapsto i \partial_{\tilde{t}}$ due to $i$ in the exponent (\ref{eq:Z_scalar}), and $A_t \mapsto A_t - \mu$. At this step, it is useful to introduce the Euclidean counterpart of $A_t$, namely $A_{\tilde{t}} \equiv A_D = - i A_t$, so that the substitution for $A_t$ takes the form $A_{\tilde{t}} \mapsto A_{\tilde{t}} + i \mu$. Thus, after integrating out the momenta, the partition function reads
\begin{equation} \label{eq:Z_shift}
	Z_{\mathrm{s}} = \int\limits_{\text{periodic}} \mathcal{D}[\phi, \phi^*\!, A_\rho] \; e^{-\tilde{S}_{\mathrm{sQED}}[\phi, \phi^*\!, A]}\Bigr|_{A_{\tilde{t}} \mapsto A_{\tilde{t}} + i \mu},
\end{equation}
where $\tilde{S}_{\mathrm{sQED}}$ is the Euclidean scalar QED action
\begin{equation} \label{eq:sqes_e}
	\tilde{S}_{\mathrm{sQED}}[\phi, \phi^*\!, A] = \int d^D \tilde{x} \, \biggl[
	\frac14 F_{\rho\sigma}^2 + (D_\rho \phi)^* D_\rho\phi + M^2 \phi^* \phi 
	\biggr]
\end{equation}
where $\rho, \sigma = 1, \ldots D$ and all the contractions are performed using Euclidean metric $\delta_{\rho\sigma}$. The integration measure $d^D \tilde{x}$ is defined as $d \tilde{t} \, d^d x$, with the range of integration $[0, \beta] \times R^d$.

The shift $A_{\tilde{t}} \mapsto A_{\tilde{t}} + i \mu$ in (\ref{eq:Z_shift}) can be effectively removed by a suitable transformation of functional integration measure. First, perform an analytic continuation of the chemical potential $\mu \mapsto -i \tilde{\mu}$.  Then, the shift for $A_{\tilde{t}}$ takes the form $A_{\tilde{t}} \mapsto A_{\tilde{t}} + \tilde{\mu}$. The analytically continued chemical potential $\tilde{\mu}$ resembles a constant background value for the Euclidean gauge field. Now, consider the following time-dependent gauge transformation in the path integral
\begin{equation} \label{eq:phi_subs}
	\phi(\tilde{t}, \bx) \mapsto e^{i\tilde{\mu} \tilde{t}} \phi(\tilde{t}, \bx), \qquad
	\phi^*(\tilde{t}, \bx) \mapsto e^{-i\tilde{\mu} \tilde{t}} \phi^*(\tilde{t}, \bx).
\end{equation}
Under this transformation, the covariant derivative shifts as
\begin{equation}
	D_{\tilde{t}} \phi \equiv (\partial_{\tilde{t}} - i A_{\tilde{t}}) \phi \;\mapsto\;  (\partial_{\tilde{t}} - i (A_{\tilde{t}}-\tilde{\mu})) \phi,
\end{equation}
where we omit phase factor, canceled by its complex conjugated counterpart in the Euclidean action. This transformation absorbs the chemical potential $\tilde{\mu}$ into the phase of the field. The price of this simplification is the modification of the boundary conditions for the matter fields. Originally, the fields $\phi$, $\phi^*$ are periodic in Euclidean time. After the transformation (\ref{eq:phi_subs}), the new fields satisfy twisted (or quasi-periodic) boundary conditions
\begin{equation} \label{eq:phi_twist}
	\begin{aligned}
	\phi(\tilde{t}{=}\beta, \bx) &= e^{-i\beta \tilde{\mu}} \, \phi(\tilde{t}{=}0, \bx), \\[0.25ex]
	\phi^*(\tilde{t}{=}\beta, \bx) &= e^{i\beta \tilde{\mu}} \, \phi^*(\tilde{t}{=}0, \bx).
	\end{aligned}
\end{equation}
After the transformation, the partition function simplifies to
\begin{equation}
	Z_{\mathrm{s}} = \int\limits_{\substack{\text{twisted}\\\text{-periodic}}} \mathcal{D}[\phi, \phi^*\!, A_\rho] \; e^{-\tilde{S}_{\mathrm{sQED}}[\phi, \phi^*\!, A]} ,
\end{equation}
where the functional integral is now over fields satisfying the twisted boundary conditions (\ref{eq:phi_twist}), and $\tilde{S}_{\mathrm{s}}$ is the standard Euclidean scalar QED action (\ref{eq:sqes_e}). Thus, the effect of the chemical potential $\mu$ is entirely encoded in the boundary conditions of the matter fields in Euclidean time.

Now, taking into account that the scalar QED action splits into the sum of the Maxwell action and the scalar on the gauge field background $\tilde{S}_{\mathrm{EM}}[A] + \tilde{S}_{\mathrm{s}}[\phi, \phi^*\!, A]$, split the integral into the integrals over $\phi$, $\phi^*$ and $A_\rho$ separately
\begin{equation}
	Z_{\mathrm{s}} = \int\limits_{\text{periodic}} \mathcal{D}[A_\rho] \, e^{-\tilde{S}_{\mathrm{EM}}[A] - \varGamma_{\mathrm{s}}[A]},
\end{equation}
where we introduce the effective action $\varGamma[A]$ as follows
\begin{equation} \label{eq:gamma_fun}
	e^{-\varGamma_{\mathrm{s}}[A]} = \int\limits_{\substack{\text{twisted}\\\text{-periodic}}}  \mathcal{D}[\phi,\phi^*] \, e^{-\tilde{S}_{\mathrm{s}}[\phi, \phi^*\!, A]}.
\end{equation}
Since the action $\tilde{S}_{\mathrm{s}}$ is quadratic in $\phi$, $\phi^*$ the integral is Gaussian and can be taken explicitly in terms of functional determinant of the corresponding wave operator $F$, namely 
\begin{equation} \label{eq:scalar_bi}
	\tilde{S}_{\mathrm{s}}[\phi, \phi^*\!, A] 
	= \frac12 \int d^D\tilde{x} \; \Phi^\dagger \, \hat{F}_{\mathrm{s}}[A] \, \Phi, \quad
	\Phi = \begin{bmatrix*}[l]
		\;\phi \\ \;\phi^*
	\end{bmatrix*}
\end{equation}
where $\Phi$ denotes a charged doublet of the fields $\phi$, $\phi^*$, whereas the explicit form of the wave operator reads
\begin{equation} 
	\hat{F}_{\mathrm{s}} = \begin{bmatrix}
		-D_\rho^2 + M^2 & 0 \\
		0 & -D_\rho^{*2}  + M^2
	\end{bmatrix} =  
	- \hat{D}_\rho^2 + \hat{1} \, M^2
\end{equation}
where $D_\mu^* = \partial_\mu + i A_\mu$, and the second form is obtained using matrices in $U(1)$ charge space
\begin{equation} \label{eq:D_hat_rho_def}
	\hat{D}_\rho = \hat{1} \, \partial_\rho + i \hat{\sigma} A_\rho, \quad
	\hat{1} = \begin{bmatrix}
		\;1 & 0\; \\ \;0 & 1\;
	\end{bmatrix}, \quad
	\hat{\sigma} = \begin{bmatrix*}[r]
		-1 & 0\; \\ 0 & 1\;
	\end{bmatrix*}.
\end{equation}
Here the $\hat{\sigma}$ distinguishes different charge sectors (in fact it is the $U(1)$ group generator). 

Thus, substituting the quadratic form (\ref{eq:scalar_bi}) into the functional integral (\ref{eq:gamma_fun}) we obtain the expression for the effective action
\begin{equation} \label{eq:eff_scalar}
	\varGamma_{\mathrm{s}}[A] = \frac12 \log \Det \hat{F}_{\mathrm{s}}[A], \\
\end{equation}
The functional determinant is subject to the twisted-periodic boundary conditions, i.e.\ its eigenfunctions obey 
\begin{equation}
	\Phi(\beta,\bx) = \begin{bmatrix}
		\;e^{-i\beta\tilde{\mu}} & & 0\; \\ 
		\;0 & & e^{+i\beta\tilde{\mu}}\;
	\end{bmatrix} \Phi(0,\bx)
\end{equation}
along with the same condition for its derivative $\dot{\Phi}(\tilde{t},\bx)$.

\subsection{Fermionic QED}
	Now, our aim is to obtain the expression for the effective action for the conventional QED with fermionic matter, in the form, similar to the scalar case (\ref{eq:eff_scalar}).
	The Dirac fermion action in mostly plus signature
	\begin{equation}
		S_{\mathrm{f}}[\psi, \bar{\psi}, A] = -\int d^D x
		\;\bar{\psi} \, (\gamma^\rho D_\rho + M) \, \psi
	\end{equation}
	where the algebra of $\gamma$-matrices is defined as $\{ \gamma^\mu, \gamma^\nu\} = 2 \eta^{\mu\nu} \mathbbm{1}$. In our conventions the spatial matrices $\gamma^i$ are Hermitian, whereas $\gamma^t \equiv \gamma^0$ is anti-Hermitian, and the Dirac conjugation is defined as $\bar{\psi} = \psi^\dagger i \gamma^{t}$. Equations of motion have the form (\ref{eq:QED_eom}) with the following expression for the current
	\begin{equation} \label{eq:current_D}
		J^\rho = -i \bar{\psi} \gamma^\rho \psi,
	\end{equation}
	The temporal component of the latter allows us to construct the conserved charge
	\begin{equation}
		Q_{\mathrm{f}} = \int d^d x \, \bigl[ -i \bar{\psi} \gamma^t \psi  \bigr].
	\end{equation}
	
	Performing manipulations analogous to the scalar case, we arrive at the following expression for the partition function	
	\begin{equation} \label{eq:Z_QED_pre}
		Z_{\mathrm{f}} = \int \mathcal{D}[\psi, \bar{\psi}, A_\rho] \; e^{-\tilde{S}_{\mathrm{QED}}[\psi, \bar{\psi}, A_\rho]} \biggr|_{A_{\tilde{t}} \mapsto A_{\tilde{t}} + i \mu} \hspace{-1em}
	\end{equation}
	that involves the Euclidean QED action
	\begin{equation}
		\tilde{S}_{\mathrm{QED}}[\psi, \bar{\psi}, A_\rho] = \int d^D \tilde{x} \, \biggl[ \frac14 F_{\rho\lambda}^2 + \bar{\psi} (\gamma^\rho D_\rho + M) \psi \biggr].
	\end{equation}
	where $\bar{\psi}$ is independent variable rather than Dirac conjugated spinor. Here the indices $\rho, \lambda = 1, \ldots D$ and the contraction is defined via Euclidean metric $\delta_{\rho\lambda}$. The analytic continuation convention has the form $\gamma^{\tilde{t}} \equiv \gamma^D = i \gamma^t$ and $A_{\tilde{t}} = -i A_t$ as before. The integration measure in (\ref{eq:Z_QED_pre}) implies periodic boundary conditions for the vector field $A_\rho$, and antiperiodic boundary conditions for $\psi$ and $\bar{\psi}$.
	
	As in the case of scalar QED, we can absorb the shift of $A_{\tilde{t}}$ in (\ref{eq:Z_QED_pre}) into the boundary conditions of the functional integration measure, making substitution
	\begin{equation} \label{eq:psi_subs}
		\psi(\tilde{t}, \bx) \mapsto e^{i\tilde{\mu} \tilde{t}} \, \psi(\tilde{t}, \bx), \qquad
		\bar{\psi}(\tilde{t}, \bx) \mapsto e^{-i\tilde{\mu} \tilde{t}} \, \bar{\psi}(\tilde{t}, \bx).
	\end{equation}
	After this substitution the boundary conditions take the form of so-called twisted antiperiodic boundary conditions
	\begin{equation} \label{eq:psi_twist}
		\psi(\beta, \bx) = -e^{-i\beta \tilde{\mu}} \, \psi(0, \bx), \quad
		\bar{\psi}(\beta, \bx) = -e^{i\beta \tilde{\mu}} \, \bar{\psi}(0, \bx).
	\end{equation}
	Now, splitting the integration measure over $\psi$, $\bar{\psi}$ and $A_\rho$, we can write the partition function $Z_{\mathrm{f}}$ in terms of effective action $\varGamma_{\mathrm{f}}[A]$, namely
	\begin{align} 
		&Z_{\mathrm{f}} = \int\limits_{\text{periodic}} \mathcal{D}[A] \; e^{-\tilde{S}_{\mathrm{EM}}[A] - \varGamma_{\mathrm{f}}[A]},\\
		&e^{-\varGamma_{\mathrm{f}}[A]} = \!\!\int\limits_{\substack{\text{twisted}\\\text{-antiperiodic}}} \!\!\mathcal{D}[\psi, \bar{\psi}] \; e^{-\tilde{S}_{\mathrm{f}}[\psi,\bar{\psi},A]}. \label{eq:Gam_QED_def}
	\end{align}
	
	Now, it is useful to rewrite the Dirac action in the symmetric Nambu-Gorkov-like form
	\begin{equation} \label{eq:dirac_bi}
		\tilde{S}_{\mathrm{f}}[\psi, \bar{\psi}, A]
		= \frac12 \int d^D \tilde{x} \; \bar{\Psi} \, \hat{D}[A] \, \Psi.
	\end{equation}
	Here we define the operator $\hat{D}[A]$
	\begin{equation} \label{eq:DA_def}
		\hat{D}[A] = \begin{bmatrix}
			\gamma^\rho D_\rho + M & 0 \\
			0 & \gamma^\rho D_\rho^* + M
		\end{bmatrix} \equiv \gamma^\rho\hat{D}_\rho + \hat{\mathbbm{1}} \, M,
	\end{equation}
	acting on the charge space doublet $\Psi$ and its conjugate $\bar{\Psi}$. They are constructed out of $\psi$ and $\bar{\psi}$ with the help of charge conjugation matrix $C$ satisfying $C(\gamma^\rho)^T C^{-1}=\eta\gamma^\rho$, with the $D$-dependent sign $\eta=\pm1$, and the matrix $R_\eta=\frac{1}{2}(\mathbbm{1}+\gamma_{D+1}) - \frac{\eta}{2}(\mathbbm{1}-\gamma_{D+1})$, where $\gamma_{D+1}$ is the chirality matrix. Thus, the explicit expressions for the doublets reads
	\begin{equation}
		\begin{aligned}
		&\Psi = \begin{bmatrix*}[l]
			\;\;\psi \\ R_\eta \psi^c
		\end{bmatrix*}, &
		\bar{\Psi} &= \begin{bmatrix*}[l]
			\;\bar{\psi} & \;\bar{\psi}^c R_\eta\,
		\end{bmatrix*}, \\[0.5ex]
		&\psi^c = C \bar{\psi}^T, &
		\bar{\psi}^c &= -\psi^T C^{-1}.
		\end{aligned}
	\end{equation}
	In its turn, the boundary conditions (\ref{eq:psi_twist}) can be rewritten in terms of Nambu-Gorkov spinor $\Psi$ as
	\begin{equation}
		\Psi(\beta,\bx) = -\begin{bmatrix}
			\;e^{-i\beta\tilde{\mu}} & & 0\; \\ 
			\;0 & & e^{i\beta\tilde{\mu}}\;
		\end{bmatrix} \Psi(0,\bx).
	\end{equation}
	
	Since the integral in (\ref{eq:Gam_QED_def}) is Gaussian, it can be evaluated explicitly, yielding the effective action in terms of the functional determinant
	\begin{equation}
		\varGamma_{\mathrm{f}}[A] = - \frac12 \log \Det \hat{D}[A]
	\end{equation}
	Using chirality matrix $\gamma_{D+1}$, giving $\gamma_{D+1} \,\gamma^\mu \, \gamma_{D+1} = -\gamma^\mu $, we can effectively flip the sign of $\gamma^\rho D_\rho$. Thus the determinant can be rewritten in various forms
	\begin{align}
		&\log \Det \hat{D}[A] = \log \Det ( \pm \gamma^\rho \hat{D}_\rho + \hat{\mathbbm{1}} M) \nonumber \\&\hspace{2em}= 
		\frac12  \log \Det (\gamma^\rho \hat{D}_\rho +  \hat{\mathbbm{1}}M)(-\gamma^\rho \hat{D}_\rho + \hat{\mathbbm{1}} M),
	\end{align}
	where in the second equality we use that $\log\Det \hat{D}[A] = \frac12 \log \Det \hat{D}^2[A]$. The latter form provides us the expression for the effective action in terms of the second order operator $\hat{F}_{\mathrm{f}}[A]$
	\begin{align}
		\varGamma_{\mathrm{f}}[A] &= - \frac14 \log \Det \hat{F}_{\mathrm{f}}[A], \\
		\hat{F}_{\mathrm{f}}[A] &= (\gamma^\rho \hat{D}_\rho + \hat{\mathbbm{1}} \, M)(-\gamma^\lambda \hat{D}_\lambda + \hat{\mathbbm{1}} \, M) \nonumber\\&= - \hat{D}_\rho^2 - \frac{i}{2}\hat{\sigma} \, \gamma^{\rho\lambda}F_{\rho\lambda} + \hat{\mathbbm{1}} \, M^2, \label{eq:F_fermi}
	\end{align}
	where $\gamma^{\rho\lambda} = \frac12[\gamma^\rho, \gamma^\lambda]$ and we took into account that $[D_\rho, D_\lambda] = -i F_{\rho\lambda}$. 
	
	\subsection{Unified treatment} 
	Thus, we conclude that both in scalar and fermionic case, the matter effective action $\varGamma[A]$ is written in terms of the functional determinant
	\begin{equation} \label{eq:ea_gen}
		\varGamma[A] = \kappa \Tr \log \hat{F}[A],
	\end{equation}
	though the constituents of r.h.s. strictly depend on the type of the QED.
	The value of the constant $\kappa$ depends on the type of matter field, namely $\kappa_{\mathrm{s}} = 1/2$ for the scalar QED and $\kappa_{\mathrm{f}} = -1/4$ for the fermionic QED, respectively.
	The operator $\hat{F}$ has the general form
	\begin{equation} \label{eq:F_gen}
		\hat{F}[A] = -\hat{D}_\rho^2 + \hat{P} + \hat{\mathbbm{1}} \, M^2
	\end{equation}
	where the potential $\hat{P}$ depends on the type of QED, namely we have
	\begin{equation} \label{eq:P_sf}
		\hat{P}_{\mathrm{s}} = 0, \qquad
		\hat{P}_{\mathrm{f}} = -\frac{i}2 \hat{\sigma} \, \gamma^{\rho\lambda} F_{\rho\lambda}
	\end{equation}
	in the scalar case and fermionic cases, respectively. In both cases the eigenfunctions obey twisted-periodic boundary conditions
	\begin{equation} \label{eq:twisted_bc_gen}
		\Phi(\beta, \bx) = \begin{bmatrix}
			\;e^{-i\theta} & 0 \\ 0 & \;e^{+i\theta}
		\end{bmatrix} \Phi(0, \bx)
	\end{equation}
	where $\theta =  \beta \tilde{\mu}$ in the scalar case and $\theta =  \beta \tilde{\mu} + \pi$ in the fermionic one. The additional term $\pi$ arises due to antiperiodic nature of fermionic functional integral. We also note that although the splitting into charge sectors may appear redundant from a physical point of view, it provides useful computational advantages. 
	
\section{Heat kernel at finite density} \label{sec:heat_kernel}
	Our aim is to calculate the effective action $\varGamma[A]$, defined via the functional trace in (\ref{eq:ea_gen}). The effective action can be expressed in terms of the heat kernel of the underlying operator $\hat{F}$ as
	\begin{equation}
		\hat{K}(\tau) = e^{-\tau \hat{F}}
	\end{equation}
	using zeta regularization formula
	\begin{equation} \label{eq:eff_act}
		\varGamma[A] = -\kappa \, \frac{\partial}{\partial s} \biggl[\frac{\Lambda^{2s}}{\Gamma(s)}\int_0^\infty d\tau \, \tau^{s-1} \Tr \hat{K}(\tau)\biggr]\biggr|_{s=0}. \
	\end{equation}
	Here the scale parameter $\Lambda$, having the dimension of mass, is introduced to ensure the right dimensionality of the result. 
	
	We will first calculate the heat kernel trace $\Tr \hat{K}(\tau)$, then employ it to compute the effective action $\varGamma$.
	In this calculation, the mass term of operator $\hat{F}$ factorizes, i.e. $\hat{K}(\tau) = e^{-\tau M^2}\hat{K}_{M=0}(\tau)$, where $\hat{K}_{M=0}(\tau)$ denotes the heat kernel for the massless case. This allows us to first compute $\Tr \hat{K}_{M=0}(\tau)$, and then multiply by $e^{-\tau M^2}$ to include the mass contribution if needed. 
	
	Treating $A_\mu$ as a small perturbation, we expand $\hat{F}$ in powers of $A_\mu$
	\begin{equation}
		\hat{F}_0 = -\hat{\mathbbm{1}} \, \partial_\mu^2, \quad
		\hat{F}_1 = -i\hat{\sigma}\{\partial_\mu, A_\mu\} + \hat{P}, \quad
		\hat{F}_2 = \hat{\mathbbm{1}} \, A_\mu A_\mu,
	\end{equation}
	where $\{ \cdot , \cdot\}$ denotes anticommutator, and we assume that the potential $\hat{P}$ is of first order in $A_\mu$. Here the indices $\mu$ run over the Euclidean range, i.e.\ $\mu=1,\ldots D$. 
	Hereinafter, we work only with Euclidean quantities and therefore omit the tilde for them. Thus, $t$ now denotes the Euclidean time, $A_t \equiv A_D$, etc.
	This expansion induces a corresponding expansion for the heat kernel trace
	\begin{equation} \label{eq:K_expansion}
		\Tr \hat{K}(\tau) = \sum_{n=0}^\infty \Tr \hat{K}_n(\tau), \qquad
		\hat{K}_n(\tau) \sim A^n.
	\end{equation}
	
	The first few orders of this expansion are derived using the exponential map variation formula
	\begin{equation}
		\delta e^{-\tau \hat{F}} = -\tau \int_0^1 d\alpha \; e^{-\alpha \tau \hat{F}} \, \delta \hat{F} \, e^{-(1-\alpha) \tau \hat{F}}.
	\end{equation}
	Applying this iteratively to $\hat{F} = \hat{F}_0 + \hat{F}_1 + \hat{F}_2$, we obtain 
	\begin{subequations} \label{eq:K_orders}
		\begin{align}
			&\Tr \hat{K}_0(\tau) = \Tr e^{-\tau \hat{F}_0}, \label{eq:K0}\\[0em]
			&\Tr \hat{K}_1(\tau) = - \tau \Tr \bigl(e^{-\tau \hat{F}_0} \hat{F}_1 \bigr), \label{eq:K1} \\
			&\Tr \hat{K}_2(\tau) = - \tau \Tr \bigl(e^{-\tau \hat{F}_0} \hat{F}_2 \bigr)\nonumber \\ &\hspace{2em}+ \frac12 \tau^2 \int_0^1 d\alpha \, \Tr \bigl(e^{-\alpha \tau \hat{F}_0} \, \hat{F}_1 \, e^{-(1-\alpha)\tau \hat{F}_0} \, \hat{F}_1\bigr) \label{eq:K2}.
		\end{align}
	\end{subequations}
	Substituting the expansion $\Tr \hat{K}(\tau) = \sum_n \Tr \hat{K}_n(\tau$) into (\ref{eq:eff_act}), we obtain a perturbative expansion for the effective action
	\begin{equation}
		\varGamma[A] = \sum_{n=0}^\infty \varGamma_n[A], \qquad
		\varGamma_n \sim A^n,
	\end{equation}
	where each $\varGamma_n$ is computed by inserting $\Tr \hat{K}_n(\tau)$ into the zeta regularization formula (\ref{eq:eff_act}).
	
\subsection{Perturbation theory}

	To calculate the orders of the perturbative expansion of the heat kernel trace (\ref{eq:K0})--(\ref{eq:K2}), we first need to define a suitable basis on the domain of the operator $\hat{F}$. Specifically, we need to establish a basis in the space of twisted‑periodic functions (\ref{eq:twisted_bc_gen}). We construct this basis starting from the space of periodic functions.
	
	Let $|p\ra$ be an eigenstate of the momentum operator $\hat{p}_\rho \equiv -i\partial_\rho$, satisfying
	\begin{equation}
	\partial_\rho |p\ra = i p_\rho |p\ra.
	\end{equation}
	Here, $p_\mu$ is a $D$-dimensional momentum vector composed of a spatial $d$-dimensional component $\bp$ and a temporal component $\omega_n = 2\pi n / \beta$, known as the Matsubara frequency, so that $p_\mu = (\omega_n, \bp)$.	The discreteness of $\omega_n$ follows from the requirement of $\beta$-periodicity in Euclidean time.
	Projecting onto the coordinate representation basis vector $|x\ra$, with $x^\mu = (t, \bx)$, gives
	\begin{equation}
		\la x | p \ra = e^{i \bp \bx + i \omega_n t},
	\end{equation}
	which is manifestly $\beta$-periodic, i.e.\ $\la t + \beta, \bx | p \ra = \la t, \bx | p \ra$ (recall that $t$ now denotes the Euclidean time).
	
	To obtain twisted‑periodic basis functions, we apply the transformation $e^{i \theta \hat{t} / \beta}$ to the periodic eigenstate $|p\ra$. This yields
	\begin{equation}
		\la x | e^{i \theta \hat{t} / \beta} |p\ra = e^{i \bp  \bx + i (\omega_n + \theta/\beta) t}.
	\end{equation}
	Under a shift by the period $\beta$ in time, this function acquires the desired phase shift
	\begin{equation}
		\la t + \beta, \bx | e^{i \theta \hat{t} / \beta} |p\ra = e^{i\theta} \la t, \bx | e^{i \theta \hat{t} / \beta} |p\ra,
	\end{equation}
	as required by the twisted boundary conditions (\ref{eq:twisted_bc_gen}).

	To account for the boundary conditions (\ref{eq:twisted_bc_gen}) in both charge sectors, we define the basis states $|\sigma\ra$ as eigenvectors of the matrix $\hat{\sigma}$
	\begin{equation}
		\hat{\sigma} |\sigma\ra = \sigma |\sigma\ra,
	\end{equation}
	where $\sigma = \pm 1$ labels the two charge sectors.
	Using the periodic basis vectors in the charge sectors, $|p, \sigma\ra \equiv |p\ra |\sigma\ra$, we define the twisted‑periodic basis vectors as
	\begin{equation} \label{eq:p_sigma_theta}
		|p, \sigma\ra_\theta = e^{i \theta \hat{\sigma}  \hat{t}/\beta} |p, \sigma\ra \equiv e^{i \mu \hat{\sigma} \hat{t}} |p, \sigma\ra.
	\end{equation}
	where $\mu = \theta / \beta$ is the analytically continued chemical potential. We will use this twisted‑periodic basis in the partition of unity
	\begin{equation}
	\hat{I} = \sum_{\sigma=\pm1}\int\limits_{(\omega_n, \bp)}  |p, \sigma\ra_{\theta} \la p, \sigma|_{\theta} \, \mathbbm{1},
	\label{eq:un_part}
	\end{equation}
	which will be used evaluating functional traces. Recall the notation (\ref{eq:intro_int_notation}), used here.
	
	From the technical point of view, it will be convenient to use the periodic basis $|p, \sigma\ra$ rather than the twisted-periodic one $|p, \sigma\ra_\theta$. For this purpose we evaluate the similarity‑transformed operator $\hat{F}_\theta$ in the basis $|p, \sigma\ra$. Since the two bases are related by (\ref{eq:p_sigma_theta}), transforming to the periodic basis yields
	\begin{equation}
		\hat{F}_\theta = e^{-i\mu\hat{\sigma} \hat{t}} \, \hat{F} \, e^{i\mu\hat{\sigma} \hat{t}}.
	\end{equation}
	The transformation operator commutes with $A_\mu(x)$ and $\partial_i$, $i = 1, \ldots d$, whereas the transformation of $\partial_t \equiv \partial_D$ is
	\begin{equation}
	e^{-i\mu\hat{\sigma} \hat{t}} \, \partial_t \, e^{i\mu\hat{\sigma} \hat{t}} = \hat{1} \, \partial_t + i \hat{\sigma} \mu
 	\end{equation}
 	As a result, the operator $\hat{F}_\theta$ is expanded in powers of $A_\nu$ as $\hat{F}_\theta = (\hat{F}_{0})_\theta + (\hat{F}_{1})_\theta + (\hat{F}_{2})_\theta$ with the components
 	\begin{subequations} \label{eq:F_th_orders}
		\begin{align}
			&(\hat{F}_{0})_\theta = -  \,  \bigl(\partial_i^2 + (\partial_t + i \hat{\sigma} \mu )^2\bigr) \hat{\mathbbm{1}}, \\
			&(\hat{F}_{1})_\theta = -i \hat{\sigma} \bigl( \{\partial_i, A_i\} + \{\partial_t + i \hat{\sigma} \mu, A_t\} \bigr) \hat{\mathbbm{1}} + \hat{P} , \\
			&(\hat{F}_{2})_\theta =  A_\rho A_\rho \, \hat{\mathbbm{1}},
		\end{align}
	\end{subequations}
	where we denoted $\hat{\mathbbm{1}} \equiv \hat{1} \otimes \mathbbm{1}$.
	At this step, it is convenient to decompose the potential $\hat{P}$ as $\hat{P}(x) = \hat{1} \, \mathbb{P}_{0}(x) + \hat{\sigma} \, \mathbb{P}_{1}(x)$. Here, the term $\hat{\sigma} \, \mathbb{P}_{1}$ corresponds to the potential term in the original fermionic operator (\ref{eq:F_fermi}), whereas $\hat{1} \, \mathbb{P}_{0}$ can be employed to absorb the mass term of the operator $\hat{F}$ if one aims to treat the mass perturbatively. Note that the operator $\hat{P}$ acts on the tensor product of the charge space and the Lorentz group (scalar or spinor) representation space, whereas the operators $\mathbb{P}_{0}$ and $\mathbb{P}_{1}$ act only on the Lorentz group representation space (scalar or spinor).
	Now, we express these components in the basis $|p, \sigma\ra \equiv |p\ra |\sigma\ra$. The action of $(\hat{F}_0)_\theta$ in this basis reads
	\begin{equation} \label{eq:F0_th_p}
		(\hat{F}_0)_\theta |p, \sigma\ra =  \bigl(\bp^2 + (\omega_n +{\sigma} \mu)^2\bigr) |p, \sigma\ra \, \mathbbm{1}.
	\end{equation}
	For the purposes of perturbation theory, we decompose the operator $(\hat{F}_1)_\theta$ in the momentum-space basis as
	\begin{align}
		&(\hat{F}_{1})_\theta = \int\limits_{(\omega_n, \bp)} \int\limits_{(\omega_m, \bk)}  |p\ra\la p-k|   \nonumber  \\&\hspace{3em} \times  \bigl( \hat{v}_\rho(p,k) \, A_\rho(k) \mathbbm{1} + \mathbb{P}_{0}(k) \hat{1} + \mathbb{P}_{1}(k)  \hat{\sigma} \bigr)  \label{eq:F1_dec}\\[1ex]
		&\la p -k | (\hat{F}_{1})_\theta | p \ra = \hat{v}_\rho(p-k,-k) A_\rho(-k) \mathbbm{1} \nonumber\\ &\hspace{7.5em}+  \mathbb{P}_{0}(-k) \,\hat{1} + \mathbb{P}_{1}(-k)  \,\hat{\sigma}, \label{eq:F1_me}
	\end{align}
	where the operator function on the charge space $\hat{v}_\rho$ is defined as
	\begin{equation}
		\hat{v}_\rho(p, k) =  \hat{\sigma} \Bigl[2 \bigl(p_\rho + \hat{\sigma} \mu \, \delta^t_\rho\bigr) - k_\rho \Bigr].
	\end{equation}
	For brevity we omit the explicit charge‑space identity $\hat{1} = \sum_{\sigma=\pm1} |\sigma\ra\la \sigma|$.

	For the second-order part of the operator $(\hat{F}_2)_\theta$ only the diagonal matrix element is required
	\begin{equation} \label{eq:F2_me}
		\la p | (\hat{F}_2)_\theta | p \ra = \int\limits_{(\omega_m, \bk)}  A_\rho(k) A_\rho(-k) \, \hat{\mathbbm{1}}.
	\end{equation} 
	
	The expressions (\ref{eq:F1_dec}),~(\ref{eq:F1_me}) and (\ref{eq:F2_me}) establish the necessary ingredients for the perturbation theory for the heat kernel, i.e.\ substitution to (\ref{eq:K_orders}).
	
	\hspace{1ex}
	
\subsubsection*{Zeroth-order part}
	Let us begin the calculation of the heat kernel trace with its zeroth order (\ref{eq:K0}). Expanding the trace in momentum space representation and employing (\ref{eq:F1_me}) for $k=0$ and (\ref{eq:F0_th_p}), we get
	\begin{equation}
		\Tr \hat{K}_0(\tau) = \int\limits_{(\omega_n, \bp)} \tr_{\sigma,L} \, \la p | e^{-\tau (\hat{F}_0)_\theta}| p \ra
	\end{equation}
	where trace is taken both over charge space and Lorentz group representation space. Using (\ref{eq:F0_th_p}) we get
	\begin{align}
		\Tr {}\hat{K}_0(\tau) &= \int\limits_{(\omega_n, \bp)} e^{-\tau\bp^2} \!\tr_\sigma  \bigl[e^{-\tau (\omega_n + \hat{\sigma} \mu)^2} \bigr] \int d^Dx \tr_L \mathbbm{1} \nonumber\\ &=
		\frac1{(4\pi \tau)^{D/2}} \, 2r_0(\tau) \int d^D x \, \tr_L \mathbbm{1} \label{eq:K0_ans}
	\end{align}
	where going from the first to the second line we take Gaussian integral over $\bp$ and use that $\la p | p \ra = \int d^D x\, 1$. Here we introduce a function
	\begin{equation} \label{eq:r0_def}
		r_0(\tau) = \frac12\sum_{\sigma=\pm1}\frac{\sqrt{4\pi\tau}}{\beta} \sum_{n}  e^{-\tau (\omega_n + \sigma \mu)^2} 
	\end{equation}
	which is can be expressed in terms of Jacobi theta function. Note that the $1/2$ introduced such that the form factors correspond to a single charge sector in the $\mu \to 0$ limit.

\subsubsection*{First-order part}
	Next, let us calculate the first-order term $\Tr \hat{K}_1(\tau)$. Inserting the partition of unity (\ref{eq:un_part}) to (\ref{eq:K1}) and using (\ref{eq:F1_me}) with $k=0$, we obtain
	\begin{align} \label{eq:K1_pre}
		\Tr \hat{K}_1(\tau) = - \tau \int\limits_{(\omega_n, \bp)} \tr_{\sigma,L} \, \la p |\,  e^{-\tau (\hat{F}_0)_\theta}  (\hat{F}_1)_\theta \,| p \ra.
	\end{align}
	Employing (\ref{eq:F0_th_p}) and (\ref{eq:F1_me}) for $k=0$, we get
	\begin{widetext}
	\begin{align}
		\Tr \hat{K}_1(\tau) &= - \tau \int\limits_{(\omega_n, \bp)} e^{-\tau\bp^2} \tr_{\sigma,L}\Bigl[  e^{-\tau (\omega_n + \hat{\sigma} \mu)^2} \Bigl(2\hat{\sigma} (\omega_n + \hat{\sigma} \mu) A_t(0) \mathbbm{1} + \hat{1} \, \mathbb{P}_{0}(0)\Bigr)\Bigr] \nonumber\\[-1ex]&=
		-\frac{\tau}{(4\pi\tau)^{D/2}} \int d^D x \, \tr_L \bigl[4r_1(\tau) A_t(x) \mathbbm{1} + 2r_0(\tau) \mathbb{P}_{0}(x)  \bigr], \label{eq:K1_ans}
	\end{align}
	\end{widetext}
	where $A_t(0) \equiv A_t(k{=}0)$, $\mathbb{P}(0) \equiv \mathbb{P}(k{=}0)$.
	Here integral over $\bp$ and sum over $n$ decouple from the functional dependence on $A_\nu$ and $\hat{P}$, hence the integral can be taken explicitly, whereas the sums are expressed in terms of the function (\ref{eq:r0_def}) and the related function	
	\begin{align} \label{eq:r1_def}
		r_1(\tau) 
		= -\frac1{2\tau} \frac{\partial}{\partial\mu} r_0(\tau).
	\end{align}
	Note that the $A_i$ and $\mathbb{P}_1$ do not appear in the expression for $\Tr \hat{K}_1(\tau)$ for parity reasons.
	
\subsubsection*{Second-order part} \label{sec:K2_YM}
	Now let us proceed to the calculation of the second-order part $\Tr \hat{K}_2(\tau)$ of the heat kernel trace. The general form of the second-order part can be written as
	\begin{widetext}
	\begin{align}
		&\Tr \hat{K}_2 = \frac12 \frac1{(4\pi\tau)^{D/2}} \int\limits_{(\omega_m, \bk)} \tr_L \biggl[ \Pi_{\rho\lambda}(\tau, k) \, A_\rho(k) A_\lambda(-k) \mathbbm{1} + 2 (\Pi_{\times})_\rho(\tau, k) \, A_\rho(k) \mathbb{P}_0(-k)	\nonumber\\[-2ex] &\hspace{18em}
			+\Pi_{P_0}(\tau, k) \, \mathbb{P}_0(k) \mathbb{P}_0(-k) +
			\Pi_{P_1}(\tau, k) \, \mathbb{P}_1(k) \mathbb{P}_1(-k)
		\biggr]
	\end{align}
	for some functions $\Pi(\tau, k)$ to be calculated. Note that we do not include the cross terms proportional to $A$ times $\mathbb{P}_1$ and $\mathbb{P}_0$ times $\mathbb{P}_1$, since these vanish in our case for parity reasons.
	
	According to (\ref{eq:K2}), the second-order part of the heat trace consists of two terms, i.e.\ $\Tr \hat{K}_2(\tau) = \Tr \hat{K}_{2,1}(\tau) + \Tr \hat{K}_{2,2}(\tau)$. At the diagrammatic level, these terms correspond to tadpole and self-energy diagrams, respectively. Let us begin with the first term. Inserting the partition of unity to $\Tr K_{2,1}(\tau)$ and using (\ref{eq:F2_me}), we obtain	
	\begin{align}
		\Tr \hat{K}_{2,1}(\tau) & = - \tau\int\limits_{(\omega_n, \bp)} \tr \, \la p | \, e^{-\tau (\hat{F}_0)_\theta} (\hat{F}_2)_\theta \, | p \ra =
		\frac12\frac1{(4\pi\tau)^{D/2}} \int\limits_{(\omega_m, \bk)} \tr_L \Bigl[ (\Pi_1)_{\rho\lambda}(\tau) A_{\rho}(k) A_{\lambda}(-k) \, \mathbbm{1} \Bigr],
	\end{align}
	where the Gaussian integral over $\bp$ is taken explicitly and the sum over $n$ is expressed in terms of $r_0(\tau)$, defined in (\ref{eq:r0_def}). Here we also introduce the quantity 
	\begin{equation}
		(\Pi_1)_{\rho\lambda}(\tau) = 2 r_0(\tau) \delta_{\rho\lambda}.
	\end{equation}
	To calculate the second term in (\ref{eq:K2}) we employ (\ref{eq:F1_dec}) for the rightmost $(\hat{F}_\theta)_1$ in $\Tr \hat{K}_{2,2}(\tau)$ and use (\ref{eq:F1_me}) afterwards. This yields
	\begin{align}
		&\Tr \hat{K}_{2,2} = \frac12 \frac1{(4\pi\tau)^{D/2}} \int\limits_{(\omega_m, \bk)} \tr_L \Bigl[ (\Pi_2)_{\rho\lambda}(\tau, k) \, A_\rho(k) A_\lambda(-k) \mathbbm{1} + 2 (\Pi_{\times})_\rho(\tau, k) \, A_\rho(k) \mathbb{P}_{0}(-k) \nonumber \\[-2ex] &\hspace{20.5em}
		+\Pi_{P_0}(\tau, k) \, \mathbb{P}_{0}(k) \mathbb{P}_{0}(-k) +
		\Pi_{P_1}(\tau, k) \, \mathbb{P}_{1}(k) \mathbb{P}_{1}(-k)
		\Bigr],
	\end{align}
	where the corresponding ``polarization operators'' $\Pi$ are defined as
	\begin{subequations}
	\begin{align}
		(&\Pi_{2})_{\rho\lambda}(\tau, k) = \tau^2 \int_0^1 d\alpha\,  \int\limits_{(\omega_n, \bp)} e^{- \alpha \tau (\bp-\bk)^2 - (1-\alpha)\tau \bp^2} \tr_{\sigma} \Bigl[e^{- \alpha \tau (\omega_n-\omega_m + \hat{\sigma}\mu)^2 - (1-\alpha)\tau (\omega_n + \hat{\sigma}\mu)^2}
			\hat{v}_\rho(p,k) \, \hat{v}_\lambda(p-k,-k)\Bigr],
		 \label{eq:Pi_2_YM_def}\\
		(&\Pi_{\times})_\rho(\tau, k) = \tau^2 \int_0^1 d\alpha\,   \int\limits_{(\omega_n, \bp)} e^{- \alpha \tau (\bp-\bk)^2 - (1-\alpha)\tau \bp^2}\tr_\sigma \Bigl[e^{- \alpha \tau (\omega_n-\omega_m + \hat{\sigma}\mu)^2 - (1-\alpha)\tau (\omega_n + \hat{\sigma}\mu)^2}
		\hat{v}_\rho(p,k)\Bigr]\\
		&\Pi_{P_0}(\tau, k) = \Pi_{P_1}(\tau, k) = \tau^2 \!\int_0^1 \! d\alpha\,  \int\limits_{(\omega_n, \bp)} e^{- \alpha \tau (\bp-\bk)^2 - (1-\alpha)\tau \bp^2}\tr_\sigma \Bigl[e^{- \alpha \tau (\omega_n-\omega_m + \hat{\sigma}\mu)^2 - (1-\alpha)\tau (\omega_n + \hat{\sigma}\mu)^2} \Bigr].
	\end{align}
	\end{subequations}
	\end{widetext}
	Thus, the part of $\Tr \hat{K}_2(\tau)$ quadratic in $A_\rho$ is given by 
	\begin{equation} \label{eq:Pi_sum_YM}
		\Pi_{\mu\nu} = (\Pi_{1})_{\mu\nu} + (\Pi_{2})_{\mu\nu}.
	\end{equation}
	
	To determine the explicit form of these ``polarization operators'' $\Pi$, we shift the momentum $\bp \mapsto \bp + \alpha \bk$ which completes the square in the exponent. The momentum integral then reduces to a Gaussian integral with weight $e^{-\tau \boldsymbol{p}^2}$ and can be performed analytically. After integrating over $\bp$, each polarization operator becomes a polynomial in $\alpha$ and $\omega_n$ and $\hat{\sigma}$. This polynomial can be reorganized in powers of $1-2\alpha$ (appearing at most linearly), $\alpha(1-\alpha)$, $\omega_n-\alpha \omega_m + \hat{\sigma}\mu$ and $\hat{\sigma}$. The remaining $\alpha$-integral and sum over $n$ act on these monomials with the common weight $e^{-\tau \alpha(1-\alpha)k^2-\tau(\omega_n - \alpha\omega_m + \hat{\sigma}\mu)^2}$.
	
	Consequently, the final result for the polarization operators can be expressed as a linear combination of auxiliary quantities defined as
	\begin{align} \label{eq:theta_Chem}
		&\Theta_{p, p'}^{(l,l')}(\tau, k) = \!\sum_{\sigma=\pm1} \int_0^1 \! \! d\alpha \, (1-2\alpha)^p  \alpha^{l'}(1-\alpha)^{l'} \! e^{-\tau \alpha(1-\alpha)k^2} \nonumber \\ &\hspace{0.25em} 
		\times \frac1{\beta}\sum_{n} \sigma^{p'} (\omega_n - \alpha\omega_m + \sigma \mu)^{l} \, e^{-\tau (\omega_n-\alpha\omega_m + \sigma \mu)^2},\hspace{-0.25em} 
	\end{align}
	defining the result of integration and summation of each monomial.
	In terms of the components, the answer for $(\Pi_{2})_{\mu\nu}$ reads
	\begin{equation} \label{eq:Pi_2_YM}
		\begin{aligned}
			&(\Pi_{2})_{tt} =  \sqrt{4\pi\tau} \, \tau^2 \Bigl(\omega_m^2 \Theta^{(0,0)}_{0,0} - 4\omega_m^2 \Theta^{(0,1)}_{0,0} \\& \hspace{12.25em} - 4 \omega_m \Theta^{(1,0)}_{1,0} + 4\Theta^{(2,0)}_{0,0}\Bigr), \\
			&(\Pi_{2})_{ti} = \sqrt{4\pi\tau}\, \tau^2 \Bigl(\omega_m \Theta^{(0,0)}_{0,0}-4\omega_m\Theta^{(0,1)}_{0,0} - 2 \Theta^{(1,0)}_{1,0} \Bigr) \bk_i, \\
			&(\Pi_{2})_{ij} = \sqrt{4\pi\tau} \Bigl[ 2\tau \Theta^{(0,0)}_{0,0} \delta_{ij} + \tau^2 \bigl(\Theta^{(0,0)}_{0,0}- 4 \Theta^{(0,1)}_{0,0}\bigr) \bk_i \bk_j \Bigr].
		\end{aligned}
	\end{equation}
	Similarly, the remaining components read
	\begin{equation} \label{eq:pi_p_def}
		\begin{aligned}
		(&\Pi_{\times})_{t} = \sqrt{4\pi\tau} \, \tau^2 \Bigl(-\omega_m \Theta^{(0,0)}_{1,1} + 2 \Theta^{(1,0)}_{0,1}\Bigr),\\
		(&\Pi_{\times})_{i} = \sqrt{4\pi\tau} \, \tau^2 \bigl(-\Theta^{(0,0)}_{1,1}\bigr)  \,\bk_i,\\[0.25ex]
		&\Pi_{P_0} = \Pi_{P_1} =
		\sqrt{4\pi\tau} \, \tau^2 \Theta^{(0,0)}_{0,0}.
		\end{aligned}
	\end{equation}
	
	In the Appendix~\ref{sec:app_Theta_rec} we derive a recurrence relation for $\Theta^{(l,l')}_{p,p'}(\tau, k)$ relating different values of the parameters $l$,~$l'$ and the number of $\tau$-derivatives. Specifically, we show that $\Theta^{(l,l')}_{p,p'}(\tau, k)$ can be expressed as linear combinations of the two basic quantities referred to as the basic form factors
	\begin{align} 
		&f_0(\tau, k) \equiv \frac12 \sqrt{4\pi\tau} \, \Theta^{(0,0)}_{0,0}(\tau, k) \nonumber\\ &= \frac12 \sum_{\sigma=\pm1}\int_0^1\!\! d\alpha\, e^{-\alpha(1-\alpha)\tau k^2}  \frac{\sqrt{4\pi\tau}}{\beta} \sum_{n} e^{-\tau (\omega_n-\alpha\omega_m+\sigma\mu)^2},\label{eq:f0_def}\\
		&f_1(\tau, k) \equiv \frac12\sqrt{4\pi\tau} \,\Theta^{(1, 0)}_{0,1}(\tau, k) = -\frac1{2\tau}\frac{\partial}{\partial\mu} f_0(\tau, k) \label{eq:f1_def}
	\end{align}
and the quantities $r_0(\tau)$ and $r_1(\tau)$, defined in (\ref{eq:r0_def}) and (\ref{eq:r1_def}) along with a finite number of their $\tau$-derivatives. The factor $\sqrt{4\pi\tau}$ in the definition of $f_0$,~$f_1$ absorbs those from the expressions for $\Pi$. Note that the second form factor $f_1$ can be expressed in terms of the first one $f_0$, which will prove very useful. Moreover, the functions $r_0(\tau)$,~$r_1(\tau)$ equal to $f_0(\tau, k)$ and $f_0(\tau, k)$, respectively, at vanishing external momentum $k$
\begin{equation} \label{eq:f_r_lim}
	r_{p'}(\tau) = \lim_{k\to 0} f_{p'}(\tau, k), \qquad
	p' = 0,1.
\end{equation}
Applying the aforementioned recurrence relations to (\ref{eq:Pi_2_YM}) we find the following answer for the sum~(\ref{eq:Pi_sum_YM})
	\begin{subequations}
		\begin{align}
			&\Pi_{tt} = - 2\tau \Bigl[ 4\tau\partial_\tau f_0 + \tau k^2 f_0\Bigr] \frac{\bk^2}{k^2} \\
			&\Pi_{ti} = -2\tau \Bigl[ 4\tau\partial_\tau f_0 + \tau k^2 f_0\Bigr] \Bigl(-\frac{\omega_m}{k^2}\bk_i\Bigr) \\
			&\Pi_{ij} = 4\tau (f_0 - r_0) \Bigl(\delta_{ij} - \frac{\bk_i \bk_j}{\bk^2}\Bigr)\nonumber \\&\hspace{3em}- 2\tau \Bigl[4\tau\partial_\tau f_0 + \tau k^2 f_0\Bigr]\frac{\omega_m^2}{\bk^2 k^2}\bk_i \bk_j.
		\end{align}
	\end{subequations}
	Similarly, for the remaining terms of the second-order part, we have
	\begin{align}
		&(\Pi_{\times})_t = 4\tau^2 \frac{\bk^2}{k^2} f_1, \qquad (\Pi_{\times})_i = -4\tau^2 \frac{\omega_m\bk_i}{k^2} f_1, \nonumber \\
		& \Pi_{P_0} = \Pi_{P_1} = 2\tau^2 f_0.
	\end{align}

\subsection{Invariant form}
	The full heat kernel trace $\Tr \hat{K}$ must be gauge invariant under infinitesimal local gauge transformations $\delta_\alpha A_\nu(x) = \partial_\nu \alpha(x)$. The zero-order part $\Tr \hat{K}_0$ is trivially invariant because of it is independent of $A_\nu$. However, invariance of the first- and second-order parts $\Tr \hat{K}_1$ and $\Tr \hat{K}_2$ require explicit verification.

	Assuming that the potential term $\hat{P}$ is gauge-invariant (as happens for the fermionic case $\hat{P}_{\mathrm{f}} = - \frac{i}2\hat{\sigma} \gamma^{\nu\lambda} F_{\nu\lambda}$), the invariance of the first-order part (\ref{eq:K1_ans}) follows form the periodic boundary conditions for $A_\nu(x)$. A direct calculation gives $\delta_\alpha \Tr \hat{K}_1$ is proportional to the boundary term $\int d^D x \, \partial_t \alpha(x) = \int d^d\bx \, (\alpha(\beta, \bx) - \alpha(0, \bx))$. The latter difference vanishes, because $\alpha(x)$ is periodic in Euclidean time, as required to preserve the periodic boundary conditions for $A_\nu$.
	
	The invariance of the second-order part $\Tr \hat{K}_2$ is ensured by the Ward-like identities
	\begin{equation}
		k^\rho \Pi_{\rho\lambda}(\tau, k) = 0, \qquad
		k^\rho (\Pi_{\times})_{\rho} (\tau, k) = 0,
	\end{equation} 
	which are precisely the transversality conditions. 
	It is convenient to express the polarization operators $\Pi_{\rho\lambda}$, $(\Pi_{\times})_{\rho}$ in the basis of independent $SO(d)$-invariant transverse tensors. As a result, the heat kernel will take a manifestly gauge invariant form. The natural basis for decomposing $\Pi_{\rho\lambda}$ is the two-dimensional basis of longitudinal/transverse projectors, constructed as follows. Let $\mathcal{P}_{\mu\nu}$ be a projector on the subspace orthogonal to $k^\mu$
	\begin{equation}
		\mathcal{P}_{\mu\nu} = \delta_{\mu\nu} - \frac{k_\mu k_\nu}{k^2}.
	\end{equation}
	Let $u^\rho = \delta^\rho_t$ be the unit vector in the Euclidean-time direction. Denoting its projection on the space orthogonal to $k^\mu$ as $v^\mu = \mathcal{P}^{\mu}_{\nu} u^\nu$, we can split the space orthogonal to $k^\mu$ into two subspaces, via the corresponding projectors, defined as
	\begin{equation}
		\mathcal{P}^\|_{\mu\nu} = \frac{v_\mu v_\nu}{v^2}, \qquad
		\mathcal{P}^\perp_{\mu\nu} = 	\mathcal{P}_{\mu\nu} - 	\mathcal{P}^\|_{\mu\nu}.
	\end{equation}
	with the following explicit form of the components
	\begin{align}
		\mathcal{P}^\|_{tt}(k) &= \frac{\bk^2}{k^2}, 
		&\mathcal{P}^\|_{ti}(k) &= -\frac{\omega_m \bk_i}{k^2},
		&\mathcal{P}^\|_{ij}(k) &= \frac{\omega_m^2}{\bk^2 k^2}\bk_i \bk_j,\nonumber \\
		\mathcal{P}^\perp_{tt}(k) &= 0, 
		&\mathcal{P}^\perp_{ti}(k) &= 0,
		&\mathcal{P}^\perp_{ij}(k) &=  \delta_{ij} - \frac{\bk_i \bk_j}{\bk^2}.
	\end{align}
	In particular, in this basis the photon propagator is diagonal. In turn, $(\Pi_{\times})_{\rho}$ should be proportional to $\mathcal{P}^\|_{0\rho}$, which is the only $SO(d)$ invariant vector, transverse to~$k_\rho$.

	Thus, we can express the ``polarization operators'' $\Pi_{\rho\lambda}$ and $(\Pi_{\times})_{\rho}$, which appear in $\Tr \hat{K}_2$, in terms of $\mathcal{P}^\|_{\rho\lambda}$,~$\mathcal{P}^\perp_{\rho\lambda}$ and $\mathcal{P}^\|_{t\rho}$, respectively
	\begin{align} 
		&\Pi_{\rho\lambda}(\tau, k) = 
		\tau^2 \Bigl[ f_\|(\tau, k^2) \, k^2 \, \mathcal{P}^\|_{\rho\lambda}(k) + f_\perp(\tau, k) \, k^2 \, \mathcal{P}^\perp_{\rho\lambda}(k)\Bigr], \nonumber\\
		(&\Pi_{\times})_\rho(\tau, k) = \tau^2 f_{\times}(\tau, k) \,  \mathcal{P}^\|_{t\rho}(k), \nonumber\\[0.7ex]
		 &\Pi_{P_0}(\tau, k) = \Pi_{P_1}(\tau, k) = f_{P}(\tau, k), 
	\end{align}
	where we introduce the potential form factor $f_{P}$, which gives the common value of the two equal coefficients $\Pi_{P_0}$ and $\Pi_{P_1}$ for the completeness of notation.
	The explicit form of the longitudinal, transverse, mixed and potential form factors reads
	\begin{subequations}
	\begin{align}
		&f_\|(\tau, k) = - \frac2{\tau k^2} \Bigl[ 4\tau\partial_\tau f_0(\tau, k) + \tau k^2 f_0(\tau, k)\Bigr], \label{eq:f_L} \\
		&f_\perp(\tau, k) = \frac{4}{\tau k^2} \Bigl[f_0(\tau, k) - r_0(\tau) \Bigr], \label{eq:f_T}\\
		&f_{\times}(\tau, k) = 4 f_1(\tau, k), \phantom{\frac2{\tau k^2}}\\
		&f_{P}(\tau, k) = 2 f_0(\tau, k). \phantom{\frac2{\tau k^2}} \label{eq:f_P}
	\end{align}
	\end{subequations}
	The factor $k^2$ in the expression for $\Pi_{\rho\lambda}(\tau, k)$ is artificially introduced so that these form factors become dimensionless. Note that the form factor $f_{\times}$ is dimensionful because $f_1$ has the mass dimension one.
	
	With the polarization operators expressed in terms of the projectors and form factors, we can now write the second‑order part of the heat kernel trace
	\begin{widetext}	
	\begin{multline} \label{eq:K2_ans_TL}
		\Tr \hat{K}_2(\tau) = \frac12 \frac{\tau^2}{(4\pi\tau)^{D/2}}\int\limits_{(\omega_m, \bk)} \tr_L \Bigl[ k^2 \, f_\|(\tau, k)\,  A^\|_{\rho}(k) A^\|_{\rho}(-k) \mathbbm{1} + 
		k^2 \, f_\perp(\tau, k) \, A^\perp_{\rho}(k) A^\perp_{\rho}(-k) \mathbbm{1} \\[-2ex]+
		2 f_{\times}(\tau, k) A^\|_t(k) \mathbb{P}_0(-k) + f_{P}(\tau, k)\Bigl(\mathbb{P}_0(k) \mathbb{P}_0(-k) + \mathbb{P}_1(k) \mathbb{P}_1(-k) \Bigr)
		 \Bigr],
	\end{multline}
	\end{widetext}
	where we introduce the longitudinal and transverse parts of the gauge potential 
	\begin{equation}
		A^\|_\rho(k) = \mathcal{P}^\|_{\rho\nu}(k) 	A_\nu(k), \quad
		A^\perp_\rho(k) = \mathcal{P}^\perp_{\rho\nu}(k) A_\nu(k).
	\end{equation}
	
	An alternative basis, which directly separates electric and magnetic components of the electromagnetic stress tensor $F_{\rho\lambda}$, is the electromagnetic basis defined by
	\begin{equation}
		\begin{aligned}
			&\mathcal{P}^E_{\rho\lambda}(k) \, A_\rho(k) A_{\lambda}(-k) = F_{ti}(k) F_{ti}(-k), \\
			&\mathcal{P}^M_{\rho\lambda}(k) \, A_\rho(k) A_{\lambda}(-k) = \frac12 F_{ij}(k) F_{ij}(-k).
		\end{aligned}
	\end{equation}
	The equivalence between the longitudinal/transverse and electromagnetic bases implies the linear transformation
	\begin{equation}
		\mathcal{P}^\|_{\rho\lambda}(k) = \frac1{k^2} \mathcal{P}^E_{\rho\lambda}(k) - \frac{\omega_m^2}{\bk^2 k^2} \mathcal{P}^M_{\rho\lambda}(k), \quad 
		\mathcal{P}^\perp_{\rho\lambda}(k) = \frac1{\bk^2} \mathcal{P}^M_{\rho\lambda}(k).
	\end{equation}
	Consequently, the second‑order part of the heat kernel trace can be rewritten as
	\begin{widetext}
	\begin{multline}
		\Tr \hat{K}_2(\tau) = \frac12 \frac{\tau^2}{(4\pi\tau)^{D/2}} \int\limits_{(\omega_m, \bk)} \tr_L \Bigl[ f_E(\tau, k)\,  F_{ti}(k) F_{ti}(-k) \, \mathbbm{1} + 
	 f_M(\tau, k) \,\tfrac12 F_{ij}(k) F_{ij}(-k) \, \mathbbm{1} \\[-2ex]+
		2 i \, f_{E P_0}(\tau, k) \, \bk_i F_{ti}(k) \, \mathbb{P}_0(-k) + f_{P}(\tau, k)\Bigl(\mathbb{P}_0(k) \mathbb{P}_0(-k) + \mathbb{P}_1(k) \mathbb{P}_1(-k) \Bigr)
		\Bigr].
	\end{multline}
	\end{widetext}
	The form factors $f_{E}$,~$f_{M}$, and~$f_{EP_0}$ are obtained from $f_\|$,~$f_\perp$, and~$f_{\times}$ by the linear transformation given above.

\subsection{Fundamental heat kernel form factor}

	The fundamental heat kernel form factor $f_0(\tau, k)$, defined in (\ref{eq:f0_def}), is expressed in terms of the $\alpha$-integral over the finite interval $[0, 1]$, where $\alpha$ is the standard Feynman parameter.
	In Appendix~\ref{sec:f_0_eta_par}, we derive an alternative form of the basic form factor, where the $\alpha$-integral is replaced by an integral over a new variable $\eta$ over from $0$ to $\infty$, namely
	\begin{multline} \label{eq:f_0_eta}
		f_0(\tau, k) = \frac12\sum_{\sigma=\pm1}\int_0^\infty d\eta\, \frac{2\sin(\sqrt{\tau} k \eta)}{\sqrt{\tau}k}\, e^{-\eta^2} \\\times \frac{\sqrt{4\pi\tau}}{\beta}
		\sum_{n} e^{-\tau (\omega_n+i\frac{\eta}{\sqrt{\tau}}\frac{\omega}{k}+\sigma \mu)^2}.
	\end{multline}
	Hereinafter we omit the discrete label subscript of the external Matsubara frequency, assuming $\omega \equiv \omega_m$.
	This form is much more useful in effective action considerations, allowing both a simpler calculation scheme for the form factors defining the effective action and a proper high-temperature expansion procedure.
	A similar form for the second form factor $f_1$ is obtained using the identity $f_1 = -\frac1{2\tau} \frac{\partial}{\partial \mu} f_0$. We note that the transition from $\alpha$ to $\eta$ is not a simple change of variables, because it relies crucially on the properties of the integrand.
	
	Now, let us split the basic form factor $f_0$ into vacuum and pure thermal parts. This is done by applying the Poisson resummation to the sum over $n$ in (\ref{eq:f_0_eta}), that gives
	\begin{multline}
		\frac12\sum_{\sigma=\pm1}\frac{\sqrt{4\pi\tau}}{\beta}
		\sum_{n} e^{-\tau (\omega_n + \sigma \mu+i\frac{\eta}{\sqrt{\tau}}\frac{\omega}{k})^2} \\ = 1 + 2\sum_{n=1}^\infty e^{-\frac{\beta^2 n^2}{4\tau}} \cosh\bigl(n \tfrac{\beta}{\sqrt{\tau}} \tfrac{\omega}{k} \eta \bigr) \cos( n\theta).
	\end{multline}
	The two-term structure on r.h.s.\ leads to desired splitting of the basic form factor
	\begin{equation} \label{eq:f0_split}
		f_0(\tau, k) = f_{0v}(\tau k^2) + f_{0\beta}(\tau, k)
	\end{equation}
	where we define the vacuum and pure thermal basic form factors
	\begin{align}
		&f_{0v}(\tau k^2) = 2\int_0^\infty d\eta\, \frac{\sin(\sqrt{\tau} k \eta)}{\sqrt{\tau}k}\, e^{-\eta^2}, \label{eq:f_v} \\ 
		&f_{0\beta}(\tau, k) = 4\int_0^\infty d\eta\, \frac{\sin(\sqrt{\tau} k \eta)}{\sqrt{\tau}k}\, e^{-\eta^2} \nonumber\\&\hspace{4em}\times \sum_{n=1}^\infty e^{-\frac{\beta^2 n^2}{4\tau}} \cosh\bigl(n \tfrac{\beta}{\sqrt{\tau}} \tfrac{\omega}{k} \eta \bigr) \cos(n\theta). \label{eq:f_beta}
	\end{align}
	The distinctive feature of the vacuum form factor is that it is expressible in terms of so-called Dawson function $F_D(x)$ (related to the error function) as follows $f_{0v}(\tau k^2) = \frac2{\sqrt{\tau}k} F_{\mathrm{D}}(\frac{\sqrt{\tau}k}2)$. As a consequence, the vacuum form factor satisfies the first-order differential equation	
	\begin{equation}
		\tau \partial_\tau f_{0v}(\tau k^2) + \frac14 (2 + \tau k^2) f_{0v}(\tau k^2) = \frac12.
	\end{equation}
	Applying this identity to the vacuum part of the form factors (\ref{eq:f_L})--(\ref{eq:f_P}), we get the answer for the 
	\begin{equation}
		\begin{aligned}
			&f_{\|v}(\tau k^2) = f_{\perp v}(\tau k^2) = \frac{4}{\tau k^2} \Bigl[f_{0v}(\tau k^2) - 1\Bigr],  \\
			&f_{\times}(\tau k^2) = 0,\qquad
			f_{Pv}(\tau k^2) = 2 f_{0v}(\tau k^2).
		\end{aligned}
	\end{equation}
	The equality $f_{\|v} = f_{\perp v}$ manifests the Euclidean $SO(D)$ invariance of the vacuum part of the heat kernel trace.

\section{Effective action} \label{sec:eff_act}
	Now, let us proceed to the calculation of the effective action. As it was mentioned at the beginning of Section~\ref{sec:heat_kernel}, the effective action will be calculated using the heat kernel via zeta-regularization procedure (\ref{eq:eff_act}). The answer for the heat kernel is given for the massless operator $\hat{F} = - \hat{D}_\mu^2 + \hat{P}$. Inclusion of nonzero mass via shift $\hat{F} \mapsto \hat{F} + M^2 \, \hat{\mathbbm{1}}$ yields multiplication of the heat kernel $\hat{K}(\tau) \mapsto e^{-\tau M^2} \hat{K}(\tau)$. Thus, in the massive case we will use the answers (\ref{eq:K0_ans}),~(\ref{eq:K1_ans}) and~(\ref{eq:K2_ans_TL}) multiplied by $e^{-\tau M^2}$.
	
\subsection{Effective action in terms of form factors}	
	Substitution of the zeroth- and first-order part of the heat kernel (\ref{eq:K0_ans}),~(\ref{eq:K1_ans}) to the expression for zeta-regularized effective action (\ref{eq:eff_act}) (with the massive factor $e^{-\tau M^2}$ inserted) gives the answer for the corresponding parts of the effective action 	
	\begin{align}
		&\varGamma_0[A] = - \frac{\kappa}{(4\pi)^{D/2}} \int d^D x \, \tr_L \bigl[ 2\rho^{D/2}_0 \, \mathbbm{1}\bigr], \label{eq:eff_act_0}\\
		&\varGamma_1[A] = \phantom{-} \frac{\kappa}{(4\pi)^{D/2}} \int d^D x \, \tr_L \bigl[4\rho^{D/2-1}_1 \, A_t(x) \, \mathbbm{1} \nonumber\\&\hspace{13.25em} + 2\rho^{D/2-1}_0 \, \mathbb{P}_{0}(x)  \bigr].  \label{eq:eff_act_1}
	\end{align}
	Here we introduce the basic function
	\begin{equation} \label{eq:rho_def}
		\rho^q_{p'}(M) = 
		\lim_{s\to0}\frac{\partial}{\partial s} \biggl[\frac{\Lambda^{2s}}{\Gamma(s)}\int_0^\infty d\tau \, \tau^{s-q-1} \, e^{-\tau M^2} \,  r_{p'}(\tau)\biggr],
	\end{equation}
	where $p'=0,1$. It implicitly depends on the inverse temperature $\beta$ and the (imaginary) chemical potential $\mu \equiv \theta / \beta$.

	Substitution of the second-order part (\ref{eq:K2_ans_TL}) of the heat kernel into the zeta-regularized effective action (\ref{eq:eff_act}) gives
	\begin{widetext}
	\begin{multline} \label{eq:eff_act_2}
		\varGamma_2[A] = -\frac12 \frac{\kappa}{(4\pi)^{D/2}} \int\limits_{(\omega, \bk)} \tr_L \Bigl[ k^2 \, \varphi_\|(k, M)\,  A^\|_{\rho}(k) A^\|_{\rho}(-k)  \mathbbm{1}+ 
		k^2 \, \varphi_\perp(k, M) \, A^\perp_{\rho}(k) A^\perp_{\rho}(-k) \mathbbm{1} \\[-2ex]+ 
		2 \varphi_{\times}(k, M) A^\|_t(k) \mathbb{P}_{0}(-k) + \varphi_{P}(k, M)\Bigl(\mathbb{P}_{0}(k) \mathbb{P}_{0}(-k) + \mathbb{P}_{1}(k) \mathbb{P}_{1}(-k) \Bigr)
		\Bigr],
	\end{multline}
	\end{widetext}
	where we introduce the form factors, associated with the quadratic gauge invariant combinations of $A$ and $P$
	\begin{equation} \label{eq:vphi_def}
		\varphi_{I}(k, M) = 
		\lim_{s\to0}\frac{\partial}{\partial s}
		\biggl[\frac{\Lambda^{2s}}{\Gamma(s)}\int_0^\infty \!\! d\tau \, \tau^{s-D/2+1} e^{-\tau M^2} f_I(\tau, k)\biggr].
	\end{equation}

	To express the effective‑action form factors $\varphi_I(k)$ through a single set of basic form factors to be introduced below, we note that the heat‑kernel form factors $f_I(\tau, k)$, defined in (\ref{eq:f_L})--(\ref{eq:f_P}), can all be written in the general form
	\begin{align} \label{eq:f_gen}
		&\tau^2 f_I(\tau, k) = \sum_{p,p'} \Bigl[\tau^p  c^{(p,p')}_{I}(k,\tau\partial_\tau) f_{p'}(\tau, k) \nonumber \\[-1ex] &\hspace{10em} + \tau^p d^{(p,p')}_{I}(k) \,r_{p'}(\tau) \Bigr]
	\end{align}
	where $I$ runs over four values $I=\|,\perp,\times, P$, and the ranges of summation are $p=1,2$, $p'=0,1$. Here $c^{(p,p')}_{I}$ and $d^{(p,p')}_{I}$ are the coefficient functions, whose explicit form can be straightforwardly extracted from the definition (\ref{eq:f_L})--(\ref{eq:f_P}). The only nonzero coefficients read
	\begin{equation} \label{eq:cd_expl}
		\begin{aligned}
			&c_\|^{(1,0)} = -\mfrac{8}{k^2} \tau\partial_\tau,  &&c_\|^{(2,0)} = -2,\\
			&c_\perp^{(1,0)} = \mfrac{4}{k^2}, & & d_\perp^{(1,0)} = -\mfrac{4}{k^2}, \\
			&c_\times^{(2,1)} = 4, &
			&c_P^{(2,0)} = 2.
		\end{aligned}
	\end{equation}
	The coefficient functions $c^{(p,p')}_{I}(k, \tau\partial_\tau)$ are polynomial in $\tau\partial_\tau$ and Laurent polynomial in $k^2$. In turn, $d^{(p,p')}_{I}(k)$ is also Laurent polynomial in $k^2$.
	
	Substituting (\ref{eq:f_gen}) into the definition of the form factors of the effective action (\ref{eq:vphi_def}), we arrive at the following expression for the form factors $\varphi_I$
	\begin{widetext}
	\begin{multline} \label{eq:vphi_pre_1}
			\varphi_{I}(k, M) = \lim_{s\to0}\frac{\partial}{\partial s} \biggl[\frac{\Lambda^{2s}}{\Gamma(s)}\sum_{p,p'} \biggl(
		 	\int_0^\infty d\tau \, \tau^{s-D/2+p-1} \, e^{-\tau M^2}  c^{(p,p')}_{I}(k,\tau\partial_\tau) f_{p'}(\tau, k)	\\+
		 	d^{(p,p')}_{I}(k) \, \int_0^\infty d\tau \, \tau^{s-D/2+p-1} \, e^{-\tau M^2} \, r_{p'}(\tau)		 	
		 	\biggr)\biggr].
	\end{multline}
	\end{widetext}
	The terms containing local function $r_{p'}(\tau)$ can be expressed via basic function defined in (\ref{eq:rho_def}). In turn, the terms containing the non-local form factors $f_{p'}$, involve new integrals of the form
	\begin{align}
		&\int_0^\infty d\tau \, \tau^{s'-1} \,e^{-\tau M^2} \, c(\tau\partial_\tau)\, f(\tau) \nonumber\\[-0.5ex]& \hspace{1em} = c(-s' - M^2 \partial_{M^2}) \int_0^\infty d\tau \, \tau^{s'-1} \,e^{-\tau M^2} \, f(\tau), 
	\end{align}
	where $s' = s-D/2+p$, and the r.h.s.\ is obtained using integration by parts along with an obvious identity
	\begin{equation}
		\tau\partial_\tau (\tau^{s'} e^{-\tau M^2}) =  (s' + M^2 \partial_{M^2}) (\tau^{s'} e^{-\tau M^2}).
	\end{equation}
	No boundary terms arise due to suitable analytic continuation in $s$.
	Employing it in (\ref{eq:vphi_pre_1}), we find the following expression for $\varphi_I$
	\begin{align} \label{eq:vphi_pre_2}
		\varphi_{I}(k, M) = \sum_{p,p'}  \Bigl[&c_{I}^{(p,p')}(k,\tfrac{D}2{-}p{-}M^2\partial_{M^2}) \, \varphi_{p'}^{D/2-p}(k, M) 
		\nonumber\\-\, &{c}_{I}{}^{\!\!\!\prime(p,p')}(k,\tfrac{D}2{-}p{-}M^2\partial_{M^2}) \,  \psi_{p'}^{D/2-p}(k, M) 
		 \nonumber\\ + \, &d^{(p,p')}_{I}(k) \, \rho^{D/2-p}_{p'}(M) \Bigr]
	\end{align}
	where $c'(k,  \mathcal{E}) = \frac{\partial}{\partial  \mathcal{E}} c(k, \mathcal{E})$ and we introduced the following effective action form factors basic integrals
	\begin{align}
		&\varphi_{p'}^{q}(k, M) = \lim_{s\to0}\frac{\partial}{\partial s} \biggl(\frac{\Lambda^{2s}}{\Gamma(s)}\int_0^\infty \! \!d\tau \,\tau^{s-q-1} e^{-\tau M^2} f_{p'}(\tau, k)\biggr), \label{eq:vphi_q} \\
		&\psi_{p'}^{q}(k, M) = \lim_{s\to0} \frac{\Lambda^{2s}}{\Gamma(s)}\int_0^\infty d\tau \, \tau^{s-q-1} \, e^{-\tau M^2} \, f_{p'}(\tau, k), \label{eq:psi_def}
	\end{align}
	with $p'=0,1$. Such defined basic form factors are not independent, and can be derived from the single fundamental one $\varphi^q_0(k, M)$. Indeed, due to relation (\ref{eq:f1_def}) between $f_1$ and $f_0$ the basic form factor $\varphi^q_1(k)$ can be expressed in terms of $\varphi^q_0(k)$ as
	\begin{equation} \label{eq:vphi1_vphi0}
		\varphi^q_1(k, M)= -\frac1{2} \frac{\partial}{\partial \mu} \varphi_0^{q+1}(k, M).
	\end{equation}
	Moreover, from the definition (\ref{eq:vphi_q}) we deduce that $\psi_0^q$ itself can be expressed as a logarithmic derivative of $\varphi^q_0$ w.r.t.\ scale $\Lambda$, namely
	\begin{equation} \label{eq:psi_vphi0} 
		\psi_0^{q}(k, M) = \frac12 \frac{\partial}{\partial \log \Lambda} \varphi_0^{q}(k, M).
	\end{equation}
	Thus, we treat $\psi_0^{q}(k, M)$ as a finite renormalization.
	In analogy with (\ref{eq:vphi1_vphi0}), the identity (\ref{eq:f1_def}) yields $\psi^q_1= -\frac1{2} \frac{\partial}{\partial \mu} \psi_0^{q+1}$. However, the quantity $\psi^q_1$ vanishes identically, because $\psi^q_0$ is independent of the chemical potential~$\mu$. Indeed, the renormalization scale $\Lambda$ enters $\varphi_0^{q}$ only through its vacuum part, hence the $\Lambda$-derivative 
	in (\ref{eq:psi_vphi0}) removes all state-dependent contributions, including the dependence on inverse temperature $\beta$ and the chemical potential $\mu$.
	
	Note also that the derivatives $\partial_{M^2}$, acting on the basic form factors in (\ref{eq:vphi_pre_2}), do not lead to additional basic form factors. Indeed, from the definitions (\ref{eq:vphi_q}),~(\ref{eq:psi_def}) it follows that
	\begin{equation} \label{eq:par_M2}
		\partial_{M^2} \,  \varphi^{q}_{p'}(k, M) = - \varphi^{q-1}_{p'}(k, M).
	\end{equation}
	The same identity holds for $\psi_0^q(k, M)$.
	
	For the completeness, we also provide the identity of the basic function $\rho_{p'}^q(M)$, expressing it in terms of the basic form factor 
	\begin{equation} \label{eq:rho_vphi_lim}
		\rho_{p'}^q(M) = \lim_{\bk\to 0} \lim_{\omega\to 0} \varphi_{p'}^q(k, M).
	\end{equation}
	It directly follows from the definition of the basic function (\ref{eq:f_r_lim}), basic form factor (\ref{eq:rho_def}) and the limit (\ref{eq:vphi_q}).

	Finally, substituting the explicit form (\ref{eq:cd_expl}) of the coefficients $c$, $d$ into (\ref{eq:vphi_pre_2}) and employing the identity (\ref{eq:par_M2}) for the action of $\partial_{M^2}$, we get the desired expressions for the form factors, defining the second-order part (\ref{eq:eff_act_2}) of the effective action
	\begin{subequations}
	\begin{align}
		&\varphi_\| = -\mfrac{8}{k^2} \Bigl[\Bigl(\mfrac{D}2-1\Bigl)\varphi_0^{D/2-1} - \psi_0^{D/2-1}\Bigr] \nonumber \\ &\hspace{3em}- 2 \Bigl(1+\mfrac{4M^2}{k^2}\Bigr)\varphi_0^{D/2-2}, \label{eq:vphi_L} \\
		&\varphi_\perp = \mfrac{4}{k^2} \Bigl[\varphi^{D/2-1}_0 - \rho_0^{D/2-1} \Bigr], \label{eq:vphi_T}\\
		&\varphi_{\times} = 4 \, \varphi_1^{D/2-2}, \vphantom{\Big[}\\
		&\varphi_{P} = 2 \, \varphi_0^{D/2-2}. \vphantom{\Big[} \label{eq:vphi_P}
	\end{align}
	\end{subequations}	
	Here we omit the arguments $k$ and $M^2$ of the form factors for brevity reasons. In the following subsections, we calculate the explicit form of the fundamental form factor, along with that of the dependent the basic form factors, entering the r.h.s.\ of (\ref{eq:vphi_L})--(\ref{eq:vphi_P}).

\subsection{Fundamental form factor}
	We now turn to the fundamental form factor $\varphi_{0}^{q}(k, M)$, defined in (\ref{eq:vphi_q}) with $p'=0$. Using the relations (\ref{eq:vphi1_vphi0}), (\ref{eq:psi_vphi0}) and \eqref{eq:rho_vphi_lim} we can express the form factors $\varphi_1^q$ and $\psi_0^q$, as well as the basic function $\rho_0^q$ in terms of the fundamental one~$\varphi_0^q$. The remaining dependent basic form factors are determined in the following subsections.
	
	Let us reproduce the definition (\ref{eq:vphi_def}) of the fundamental form factor in the following form
	\begin{align}
		&\varphi_0^{q}(k, M) = \lim_{s\to0}\frac{\partial}{\partial s} \frac{\Lambda^{2s}}{\Gamma(s)} \tilde{\varphi}(s-q, k, M), \label{eq:vphi_0_spectral}\\
		&\tilde{\varphi}(s, k, M) = \int_0^\infty d\tau \, \tau^{s-1} e^{-\tau M^2} f_{0}(\tau, k), \label{eq:spectral_fun}
	\end{align}
	where we have introduced an auxiliary function $\tilde{\varphi}(s, k, M)$. It has the meaning of the spectral zeta function, and will be referred to as spectral function.
	
	We will calculate $\varphi_0^q(k, M)$ by using the Mellin transform of the massless heat kernel form factor $f_{0}(\tau, k)$. Specifically, substituting the Mellin representation of the exponential 
	\begin{equation}
		e^{-\tau M^2} = \frac1{2\pi i} \int_{c-i\infty}^{c+i\infty} dw \, \Gamma(w) M^{-2w} \tau^{-w}, 
	\end{equation}
	into (\ref{eq:spectral_fun}) and interchanging orders of integration, we get the following form of the spectral function
	\begin{equation} \label{eq:spectral_mellin}
		\tilde{\varphi}(s, k, M) = \frac1{2\pi i} \int_{c-i\infty}^{c+i\infty} dw \, M^{-2w} \, \Gamma(w) \, \tilde{f}_0(s-w, k),
	\end{equation}
	where $\tilde{f}_0(s, k)$ is the Mellin transform of $f_0(\tau, k)$, i.e.\
	\begin{equation} \label{eq:ff_0_Mellin}
		\tilde{f}_0(s, k) = \int_0^\infty d\tau \, \tau^{s-1} \, f_0(\tau, k).
	\end{equation}
	For the Mellin form (\ref{eq:spectral_mellin}) of the spectral function to be well-defined, the constant $c$ in its definition should be chosen in the fundamental strip. This is defined by inequalities $c > 0$, $s - 1 < c < s$, which come from the convergence conditions for the Mellin transform of the exponential and $f_0(\tau, k)$, respectively.
	
	In Appendix~\ref{sec:app_f0_Mellin} we obtain the following series representation for the Mellin transform of the fundamental heat kernel form factor
	\begin{align} \label{eq:ff_Mellin_Li}
		&\tilde{f}_0(s, k) =   \Bigl(\frac{\beta}{2}\Bigr)^{\!2s} 
		\sum_{r=0}^\infty (-)^r \frac{r!}{(2r+1)!} \, 	z^{2r} \nonumber \\ &\hspace{1em}\times \Bigl[ \Li_{-2(s+r)} (e^{i\theta}) +  \Li_{-2(s+r)} (e^{-i \theta})  \Bigr] \nonumber \\[1ex] &\hspace{2em}\times 
		  \Gamma(-r-s)\, {}_2F_1(-s-r, r+1, \tfrac12, \sin^2\vartheta).
	\end{align}
	Examination of the summand shows that the only source of poles in $r$-th term is the simple poles of gamma function $\Gamma(-r-s)$. For fixed $r$, these poles occur at $s+r=l$ for \emph{nonnegative} integers $l$. The poles at \emph{positive} $l$, however, turn out to be spurious, because these are canceled by vanishing combination $\Li_{-2l}(e^{i\theta})+\Li_{-2l}(e^{-i\theta})=0$ for positive integers $l$, which is the property of polylogarithm function. This cancellation does not occur for $l=0$, since $\Li_{0}(e^{i\theta})+\Li_{0}(e^{-i\theta})=-1$. Thus, the poles of $\tilde{f}_0(s, k)$ are simple poles at $s = -r$ for all nonnegative integers $r$.
	
	These poles come exactly from the vacuum part of $\tilde{f}_0(s, k)$. Indeed, employing splitting (\ref{eq:f0_split}) of the fundamental form factor $f_0(\tau, k)$ into the vacuum and pure thermal parts, we calculate the Mellin transform of the vacuum part $f_{0v}(\tau, k)$, given in (\ref{eq:f_v}), as
	\begin{equation} \label{eq:ff_Mellin_vac}
		\tilde{f}_{0v}(s, k) = k^{-2s} \frac{\Gamma(s) \Gamma(1-s)^2}{\Gamma(2-2s)}.
	\end{equation}
	Direct inspection shows that $\tilde{f}_{0v}(s, k)$ has exactly the same residues at poles $s=-r$ as $\tilde{f}_0(s, k)$. Note however that $\tilde{f}_{0v}(s, k)$ also has poles at positive integer values of $s$, which cancel with those of the pure thermal part $\tilde{f}_{0\beta}(s, k)$.
	
	As we observed, the function $\tilde{f}_0(s, k)$ is meromorphic in $s$ with the poles at $s = -r$ for nonnegative integers $r$. Thus, let us first take $s > 0$ to have a nonempty range of $c$ in the expression (\ref{eq:spectral_mellin}) for the spectral function $\tilde{\varphi}(s, k, M)$. For this choice, the poles of the integrand coming from the gamma function at $w = -n$ for all nonnegative integers $n$, lie to the left of the integration contour. In turn, the poles coming from $\tilde{f}_0(s-w)$ at $w = s + r$ lie to the right of the contour of integration. Thus, we can shift the contour to the left, picking up the poles at $w = -n$. This yields
	\begin{align}
		\tilde{\varphi}(s, k, M) &= \sum_{n=0}^\infty \operatorname{res}_{w = -n} M^{-2w} \, \Gamma(w) \, \tilde{f}_0(s-w, k) \nonumber\\ &=
		 \sum_{m=0}^\infty \frac{(-)^m}{m!} M^{2m}\, \tilde{f}_0(s+m, k).
	\end{align}
	where, in passing from the first to the second line, we used that all the picked-up poles are the simple poles of the gamma function. Now, we can safely analytically continue $\tilde{\varphi}(s, k, M)$ expression to the values of $s$ near $-q$, when substituted into (\ref{eq:vphi_0_spectral}). This yields the following small-mass series representation of effective action fundamental form factor
	\begin{align}
		&\varphi_0^{q}(k, M) = \sum_{m=0}^\infty \frac{(-)^m}{m!} M^{2m}\, \varphi_0^{q-m}(k), \label{eq:vphi_mass_exp}
	\end{align}
	where the coefficients are defined as
	\begin{equation}
		\varphi_0^n(k) = \lim_{s\to0}\frac{\partial}{\partial s} \frac{\Lambda^{2s}}{\Gamma(s)} \tilde{f}_0(s-n, k), \label{eq:vphi_M0_def}
	\end{equation}
	and will be referred to as massless effective action form factors, because series representation (\ref{eq:vphi_mass_exp}) implies $\lim_{M\to 0} \varphi_0^{n}(k, M) = \varphi_0^n(k)$.
	We also note that for negative integers $n$ in (\ref{eq:vphi_M0_def}) (equivalently, for the terms $m > q$ in (\ref{eq:vphi_mass_exp})), zeta regularization can be lifted, because $\tilde{f}_0(s, k)$ has poles only at $s=-r$. This gives the simpler form $\varphi_0^{-p-1}(k) = \tilde{f}_0(p+1)$ for nonnegative integers~$p$.
	
\subsection{Massless form factor calculation}
	
	Having expressed the massive fundamental form factor through its massless counterparts, we now evaluate these massless coefficients explicitly. Using the expression (\ref{eq:ff_Mellin_Li}) for $\tilde{f}_0(s, k)$, we can now compute the massless effective action fundamental form factors $\varphi_0^n(k)$, defined (\ref{eq:vphi_M0_def}), which are the coefficients of expansion (\ref{eq:vphi_mass_exp}) of the massive ones $\varphi_0^q(k, M)$. The main challenge is to perform the $r$-sum in (\ref{eq:ff_Mellin_Li}) explicitly after substituting it into (\ref{eq:vphi_M0_def}), in order to obtain a closed form for $\varphi_0^n(k)$.

	We begin with negative integer $n = -p-1$. As stated above, $\varphi_0^{-p-1}(k)$ does not require regularization and simply equals $\tilde{f}_0(p+1)$. Taking the limit $s \to p+1$ in (\ref{eq:ff_Mellin_Li}), we get
	\begin{align} \label{eq:vphi_neg_series}
		&\varphi_0^{-p-1}(k) = 2 (-)^{p+1} \Bigl(\frac{\beta}{2}\Bigr)^{\!2(p+1)} 
		\sum_{r=0}^\infty \frac{r!}{(2r+1)!} \, z^{2r} \nonumber\\ & \hspace{0.5em}\times \bigl( \Li^{(1)}_{-2(r+p+1)} (e^{i\theta}) +  \Li^{(1)}_{-2(r+p+1)} (e^{-i \theta})  \bigr) \\ &\hspace{1em}\times 
		 \frac1{(r+p+1)!} \, {}_2F_1(-r-p-1, r+1, \tfrac12, \sin^2\vartheta),  \nonumber
	\end{align}
	where we introduce notation $\Li^{(1)}_s(x) = \frac{\partial}{\partial s} \Li_s(x)$. The main object that allows us to calculate the $r$-sum explicitly is the shift-like operator $\varOmega$, defined by its action on the test function
	\begin{equation}
		\varOmega \, \varPhi(k) = \frac{e^{i\vartheta}\varPhi(k e^{i\vartheta}) + e^{-i\vartheta} \varPhi(k e^{-i\vartheta})}{e^{i\vartheta} + e^{-i\vartheta}}.
	\end{equation}
	This operator absorbs all $\vartheta$-dependence into itself. So by factoring this operator outside the $r$-summation sign, we are able to compute the sum explicitly. Specifically, employing the identity (\ref{eq:app_Omega_z_d_Euler}), derived in Appendix~\ref{sec:app_Omega}
	\begin{align} \label{eq:Omega_d_z}
		&\partial_{z^2}^p\bigl(\varOmega z^{2r+2p+1}\bigr)	= \frac1{2^{2p+1}}\frac{r!}{(2r+1)!} \frac{(2r+2p+2)!}{(r+p+1)!} \\&\hspace{0.5em}\times z^{2r+1} (\cos\vartheta)^{-2p-1} {}_2F_1(-p-r-1,r+1;\tfrac12;\sin^2\vartheta), \nonumber
	\end{align}
	where $\partial_{z^2} \equiv \partial / \partial z^2$, we rewrite $\varphi_0^{-p-1}(k)$ as
	\begin{align}
		&\varphi_0^{-p-1}(k) = \frac{(-)^{p+1}}{z} \beta^{2(p+1)} (\cos\vartheta)^{2p+1} \nonumber \\ &\hspace{1.5em}\times
		\partial_{z^2}^p\biggl(\varOmega \, \frac1{z} \sum_{r=p+1}^\infty  \frac{z^{2r}}{(2r)!} \bigl( \Li^{(1)}_{-2r}(e^{i\theta}) + \Li^{(1)}_{-2r}(e^{-i\theta}) \bigr)
		 \biggr) .
	\end{align}
	Note that when taking the $z^2$-derivative, we treat $\omega$, entering the expressions via $\vartheta$, as constant. Then $\sin \vartheta = y / z$, where $y = \beta \omega / 2$ is fixed during the differentiation. 
	
	Now, the sum over $r$ can be taken explicitly in terms of polylogarithm, using the series expansion of the latter
	\begin{equation} \label{eq:Li_Taylor}
		\Li_s(e^{z+i\theta}) = \sum_{n=0}^\infty \frac1{n!} \Li_{s-n}(e^{i\theta}) \, z^n.
	\end{equation}
	Employing it, we obtain the explicit expression for $\varphi_0^{-p-1}(k)$, now having no infinite sums, $\varPhi_0^{-p-1}(k)$ as
	\begin{equation}
		\varphi_0^{-p-1}(k) =  \frac{(-)^{p+1}}{2z} \beta^{2(p+1)} (\cos\vartheta)^{2p+1} \partial_{z^2}^p\Bigl(\varOmega \, \varPhi_0^{-p-1}(k) 
		\Bigr), \label{eq:vphi_neg_ans} \\
	\end{equation}
	which we write introducing an auxiliary function, defined as
	\begin{widetext}
	\begin{equation}
			\varPhi_0^{-p-1}(k) = \frac1{z} 
			\biggl( \Li^{(1)}_{0}(e^{z+i\theta}) + \Li^{(1)}_{0}(e^{z-i\theta}) + \Li^{(1)}_{0}(e^{-z+i\theta}) + \Li^{(1)}_{0}(e^{-z-i\theta}) - 2 \sum_{r=0}^{p}\bigl( \Li^{(1)}_{-2r}(e^{i\theta}) + \Li^{(1)}_{-2r}(e^{-i\theta})\bigr)\frac{z^{2r}}{(2r)!}	\biggr).
	\end{equation}
	\end{widetext}
	Note that the sum on the second line of the definition of $\varPhi_0^{-p-1}(k)$ represents exactly the first $p+1$ nontrivial terms of the Taylor expansion of the four-term polylogarithm combination appearing on the first line.
	
	
	Now, let us turn to the calculation of $\varphi_0^{n}(k)$ for nonnegative integer $n$. Unlike the previously handled case of negative integer $n$, in this case one cannot lift the zeta regularization in its definition~(\ref{eq:vphi_M0_def}). Nevertheless, we can isolate the singular part of $\varphi^n_0(k)$ in (\ref{eq:vphi_M0_def}), by subtracting and adding back a singular part of $\tilde{f}_0(s, k)$, defined by its vacuum part (\ref{eq:ff_Mellin_vac}). This procedure effectively splits $\varphi_0^{n}(k)$ into the vacuum and the pure thermal parts as
	\begin{equation} \label{eq:vphi_0_vac_beta_split}
		\varphi_0^{n}(k) = \varphi_{0v}^{n}(k) + \varphi_{0\beta}^{n}(k),
	\end{equation}
	where the individual terms are defined as
	\begin{align}
		\varphi_{0v}^{n}(k) &= \lim_{s\to 0} \frac{\partial}{\partial s} \frac{\Lambda^{2s}}{\Gamma(s)} \tilde{f}_{0v}(s-n, k), \label{eq:vphi_vac_def}\\
		\varphi_{0\beta}^{n}(k) &= \lim_{s\to 0} \Bigl[\tilde{f}_{0}(s-n, k) - \tilde{f}_{0v}(s-n, k)\Bigl], \label{eq:vphi_beta_def}
	\end{align}
	where the factor $\Lambda^{2s} / \Gamma(s)$ is absent in the expression for $\varphi_{0\beta}^{n}(k)$ due to the finiteness of $\tilde{f}_{0\beta}(s-n, k)$ in $s\to 0$ limit.
	The vacuum part has a simple expression
	\begin{equation} \label{eq:vphi_0_v}
		\varphi_{0v}^{n}(k) = -2 (-)^n \frac{n!}{(2n+1)!} k^{2n} \Bigl(\log\mfrac{k}{\Lambda}  - H_{2n+1} + \tfrac12 H_n \Bigr),
	\end{equation}
	obtained by a direct substitution of (\ref{eq:ff_Mellin_vac}) into zeta regularized expression (\ref{eq:vphi_vac_def}). In turn, the pure thermal part is defined by the finite limit (\ref{eq:vphi_beta_def}), where the zeta regularization is lifted since the singular (pole) part is isolated to $\varphi_{0v}^{n}(k)$.

	To calculate the pure thermal part $\varphi_{0\beta}^n(k)$, we substitute the series representation (\ref{eq:ff_Mellin_Li}) for $\tilde{f}_0$ and the expression (\ref{eq:ff_Mellin_vac}) for its vacuum part $\tilde{f}_{0v}$ into the definition (\ref{eq:vphi_beta_def}) of the pure thermal massless form factor $\varphi^n_{0\beta}$. We observe that the only singularity in $\tilde{f}_0(s-n, k)$ occurs in the $r=n$ term of its series representation (\ref{eq:ff_Mellin_Li}) and is canceled exactly by the vacuum part. 
	In Appendix~\ref{sec:app_vaphi_massless} we perform the calculation of the remaining finite part in terms of series expansion, which is analogous to (\ref{eq:vphi_neg_series}).	
	Then, following the same algorithm used above for $\varphi^{-p-1}_{0}$, we absorb all $\vartheta$-dependence contained in the hypergeometric function into the action of $\varOmega$.
	This result in the explicit closed form expression for the pure thermal part $\varphi^n_{0\beta}(k)$, namely
	\begin{equation} 
		\varphi^n_{0\beta}(k) =  (-)^n k^{2n} \, \partial_{z^2}^{n} \Bigl( \varOmega \, \varPhi_{0}^{n}(k) \Bigr), \label{eq:vphi_n_ans}
	\end{equation}
	where we introduced the auxiliary function
	\begin{widetext}
		\begin{align}
			&\varPhi_{0}^{n}(k) = - 2 \sum_{r=0}^{n}   \frac{z^{2r}}{(2r+1)!}  \bigl( \Li_{2n-2r}(e^{i\theta}) + \Li_{2n-2r}(e^{-i\theta}) \bigr) \bigl(\log z + \gamma - H_{2r+1}\bigr) \nonumber\\ &\hspace{4em} + \frac{1}{z}\bigl( \Li^{(1)}_{2n+1}(e^{z+i\theta}) +\Li^{(1)}_{2n+1}(e^{z-i\theta}) -\Li^{(1)}_{2n+1}(e^{-z+i\theta}) -\Li^{(1)}_{2n+1}(e^{-z-i\theta}) \bigr).
		\end{align}
	\end{widetext}
	The $z^2$-differentiation rules in (\ref{eq:vphi_n_ans}) are the same as for (\ref{eq:vphi_neg_ans}).
	
	The total fundamental form factor $\varphi_0^n(k)$ is obtained by summing the expressions (\ref{eq:vphi_0_v}) and (\ref{eq:vphi_n_ans}) for the vacuum and the pure thermal parts, respectively. It naturally splits into three parts
	\begin{equation} \label{eq:vphi_0_sing_log_reg_split}
		\varphi_0^n(k) = \varphi^{n,\mathrm{sing}}_{0}(k) + \varphi^{n,\mathrm{log}}_{0}(k) + \varphi^{n,\mathrm{reg}}_{0}(k),
	\end{equation}
	according to its behavior in the high-temperature limit $\beta \to 0$, namely singular, logarithmic and regular (recall that $z = \beta k / 2$). The expressions for the individual terms read
	\begin{widetext}
	\begin{subequations}
		\begin{align}
			&\varphi^{n,\mathrm{sing}}_{0}(k) = \, (-)^n k^{2n}
			\sum_{r=1}^{n} (-)^{r}  \frac{(r-1)!(n-r)!}{(2n-2r+1)!} \, z^{-2r}   \bigl( \Li_{2r}(e^{i\theta}) + \Li_{2r}(e^{-i\theta}) \bigr) \, {}_2F_1(r, n-r+1, \tfrac12, \sin^2\vartheta), \label{eq:vphi_sing_series} \\
			&\varphi^{n,\mathrm{log}}_{0}(k) =
			2 (-)^n k^{2n} \frac{n!}{(2n+1)!} \, \Bigl(\log \tfrac{\beta \Lambda}{2} + \gamma -
			(n+1) \sin^2\vartheta \,\, {}_3F_2(1,1,n+2;2,\tfrac32;\sin^2\vartheta) \Bigr), \label{eq:vphi_log_series} \\
			&\varphi^{n,\mathrm{reg}}_{0}(k) =
			(-)^n k^{2n} \partial_{z^2}^{n} \Bigl( \varOmega \frac{1}{z}\bigl( \Li^{(1)}_{2n+1}\!(e^{z+i\theta}) {+}\Li^{(1)}_{2n+1}\!(e^{z-i\theta}) - \Li^{(1)}_{2n+1}(e^{-z+i\theta}) - \Li^{(1)}_{2n+1}(e^{-z-i\theta}) \bigr) \Bigr) \label{eq:vphi_reg_resum}\\ & \hphantom{\varphi^{n,\mathrm{reg}}_{0}(k)} =  2 (-)^n k^{2n} 
			\sum_{r=0}^{\infty} \frac{(r+n)!}{r!\,(2r+2n+1)!} \, z^{2r} \bigl( \Li^{(1)}_{-2r}(e^{i\theta}) + \Li^{(1)}_{-2r}(e^{-i\theta}) \bigr)\, {}_2F_1(-r, r+n+1, \tfrac12, \sin^2\vartheta). \label{eq:vphi_reg_series}
		\end{align}
	\end{subequations}	
	\end{widetext}
	The decomposition (\ref{eq:vphi_0_sing_log_reg_split}) for $\varphi_0^n(k)$ corresponds to the closed form (\ref{eq:vphi_n_ans}) (along with (\ref{eq:vphi_0_v})) in the following way.
	After applying $\varOmega$ and $\partial_{z^2}^n$ in (\ref{eq:vphi_n_ans}), the $r=0, \ldots n-1$ terms in the first line of $\varPhi_{0}^{n}(k)$ yield $\varphi_0^{n,\mathrm{sing}}$. The $r=n$ term, combined with the vacuum part $\varphi_{0v}^n$ gives $\varphi_0^{n,\mathrm{log}}$. The second line of $\varPhi_{0}^{n}(k)$ leads to the closed form expression (\ref{eq:vphi_reg_resum}) for $\varphi_0^{n,\mathrm{reg}}$. Equivalently, (\ref{eq:vphi_reg_resum}) is obtained by explicitly summing the $r$-sum in (\ref{eq:vphi_reg_series}).
	
	Before collecting the final results, we note that hypergeometric functions appearing in the series representations of $\varphi_{0}^n(k)$ both for negative and nonnegative values of $n$, namely (\ref{eq:vphi_neg_series}) and (\ref{eq:vphi_sing_series})--(\ref{eq:vphi_reg_series}) respectively, can be reduced to elementary functions. Specifically, the singular and logarithmic parts (\ref{eq:vphi_sing_series}) and (\ref{eq:vphi_log_series}) involve the following hypergeometric functions, which can be written as iterated $k^2$-derivatives of a simple combination of elementary functions
	\begin{align}
		&{}_2 F_1(r, n-r+1, \tfrac12, \sin^2\vartheta) =
		\frac{(-)^{n-1}}{(n-r)! (r-1)!} \nonumber \\& \hspace{0.25em}\times k^{2 r} \, \partial_{k^2}^{r-1} k^{2(n-r)} \partial_{k^2}^{n-r} \, \mfrac1{\bk^2} \Bigl[1 - \mfrac{i\omega}{2|\bk|} \log \mfrac{|\bk|+i\omega}{|\bk|-i\omega} \Bigr], \\[1ex]
		&\sin^2\vartheta \, {}_3F_2(1,1,n+2; 2,\tfrac32;\sin^2\vartheta) 
		 \nonumber \\& \hspace{1em}=-\frac{(-)^{n}}{(n+1)!}  k^{2(n+1)} \partial_{k^2}^n \Bigl(\mfrac1{k^2} \mfrac{i\omega}{2|\bk|} \log \mfrac{|\bk|+i\omega}{|\bk|-i\omega} \Bigr),
	\end{align}
	where we explicitly substitute $\sin\vartheta = \frac{\omega}{k}$.
	Here, in all $k^2$-derivatives $\omega$ is held fixed, whereas $|\bk|=\sqrt{k^2-\omega^2}$ is treated as function of $k^2$. Similarly, the regular part (\ref{eq:vphi_reg_series}) of $\varphi_0^n$ for nonnegative $n$ and series representation (\ref{eq:vphi_neg_series}) for $\varphi_0^n$ itself for negative $n$, involve hypergeometric functions, that reduce to known polynomial families
	\begin{align}
		&{}_2F_1(-r, r+n+1, \tfrac12, \sin^2\vartheta) = (-)^r \frac{n!\,r!}{(n+r)!} \, C_{2r}^{n+1}(\sin\vartheta), \nonumber\\
		&{}_2F_1(-r-p-1, r+1, \tfrac12, \sin^2\vartheta) \\ & \hspace{3em}= (-)^{r+p+1} \frac{(r+p+1)!}{r!} \,  \tilde{C}^{-p}_{2r+2p+2}(\sin\vartheta).\nonumber
	\end{align}
	Here $C_n^\lambda(x)$ denotes the Gegenbauer polynomials, and we also denote by $\tilde{C}_n^{-p}(x)$ the regularized Gegenbauer polynomials for negative parameter $\tilde{C}_n^{-p}(x) = \lim_{\lambda\to -p} \Gamma(\lambda) \, C_n^{\lambda}(x)$ (these are proportional to the Jacobi polynomials $P_n^{(-p-\frac12,-p-\frac12)}(x)$, which are more standard, though the proportionality coefficient is cumbersome). An additional simplification of the singular part (\ref{eq:vphi_sing_series}) follows from the identity
	\begin{equation}
		\Li_{2r}(e^{i\theta}) + \Li_{2r}(e^{-i\theta}) = (-)^{r+1}\frac{(2\pi)^{2r}}{(2r)!} B_{2r}(\tfrac{\theta}{2\pi}).
	\end{equation}
	which expresses a combination of polylogarithms in terms of Bernoulli polynomials.
	Collecting all these simplifications, the most explicit expressions for the fundamental form factor $\varphi^n_{0}(k)$ are given in Section~\ref{sec:main_res} among the main results of this work.

\subsection{Basic (dependent) form factors}

	With the fundamental form factor $\varphi_0^q(k, M)$ now known, we can determine the remaining basic form factors. Specifically, $\varphi_1^q(k, M)$ is determined through (\ref{eq:vphi1_vphi0}), $\psi_0^q(k, M)$ through (\ref{eq:psi_vphi0}), $\psi_1^q(k, M)=0$, and $\rho_{0,1}^q(M)$ through the $k \to 0$ limit of $\varphi_{0,1}^q(k, M)$. Consequently, all ingredients entering the effective-action form factors (\ref{eq:vphi_L})--(\ref{eq:vphi_P}) are now known in closed form. Since identities (\ref{eq:vphi1_vphi0}), (\ref{eq:psi_vphi0}), and (\ref{eq:rho_vphi_lim}) express basic quantities linearly in terms of the fundamental one, one can easily obtain the analogues of the small-mass expansion (\ref{eq:vphi_mass_exp}) for these quantities. Specifically, applying $\frac12\frac{\partial}{\partial\log\Lambda}$ to both sides of (\ref{eq:vphi_mass_exp}), we obtain	
	\begin{equation}
		\begin{aligned}
			&\psi_0^q(k, M) =  \sum_{m=0}^{q} \frac{(-)^m}{m!} M^{2m} \psi_0^{q-m}(k), \\
			&\psi_0^n(k) = (-)^n \frac{n!}{(2n+1)!} k^{2n},
		\end{aligned}
	\end{equation}
	where we use the expression (\ref{eq:vphi_0_v}) for the vacuum part of $\varphi_0^n(k)$, which is the only $\Lambda$-dependent part of the latter, and take into account that $\varphi_0^n(k)$ is $\Lambda$-dependent for $n < 0$.

	Similarly, taking the $k \to 0$ limit of both sides of (\ref{eq:vphi_mass_exp}), we get
	\begin{equation}
		\rho_0^q(M) =  \sum_{m=0}^\infty \frac{(-)^m}{m!} M^{2m} \, \rho^{q-m}_{0}
	\end{equation}
	where we define the coefficients $\rho_0^n = \lim_{k\to 0} \varphi_0^n(k)$, which have the following explicit form
		\begin{equation}
			\begin{aligned}
			&\rho_0^n = (n-1)! \, \Bigl(\frac{\beta}{2}\Bigr)^{\!\!-2n} \, \bigl( \Li_{2n}(e^{i\theta}) + \Li_{2n}(e^{-i\theta}) \bigr), \\[0.5ex]
			&\rho^0_0 = 
			2 \Bigl(\log \tfrac{\beta\Lambda}2 + \gamma  +
			\Li_{0}^{(1)}(e^{i\theta})+\Li_{0}^{(1)}(e^{-i\theta}) \Bigr),\\
			&\rho^{-p-1}_0 = (-)^{p+1} \frac{2}{(p+1)!} \Bigl(\frac{\beta}{2}\Bigr)^{\!2(p+1)}  \\ &\hspace{4em}\times
			\bigl( \Li^{(1)}_{-2(p+1)} (e^{i\theta}) +  \Li^{(1)}_{-2(p+1)} (e^{-i \theta})  \bigr),
		\end{aligned}
	\end{equation}
	where $n > 0$ and $p \ge 0$.
	Together with (\ref{eq:vphi_mass_exp}) and known massless form factors $\varphi_0^n(k)$, derived in the previous subsection, this allows us to obtain the small-mass expansion of the form factors (\ref{eq:vphi_L})--(\ref{eq:vphi_P}).
	
	To summarize, the second-order part of the effective action is expressed through the four form factors (\ref{eq:vphi_L})--(\ref{eq:vphi_P}). These form factors are determined by the fundamental form factor $\varphi_0^q(k,M)$ and the basic quantities $\varphi_1^q(k, M)$, $\psi_0^q(k, M)$, and $\rho_0^q(M)$, where the latter are expressed through $\varphi_0^q(k, M)$ by the identities (\ref{eq:vphi1_vphi0}), (\ref{eq:psi_vphi0}), and (\ref{eq:rho_vphi_lim}). 
	The massive fundamental form factor $\varphi_0^q(k,M)$ is given by the small-mass expansion (\ref{eq:vphi_mass_exp}), whose coefficients $\varphi_0^n(k)$ (the massless form factors) are obtained in explicit form. For $n\ge0$, $\varphi_0^n(k)$ splits into vacuum and pure thermal parts according to (\ref{eq:vphi_0_vac_beta_split}). The vacuum contribution is given by (\ref{eq:vphi_0_v}), while the pure thermal contribution is given by (\ref{eq:vphi_n_ans}). Alternatively, for $n\ge0$, the massless fundamental form factor $\varphi_0^n(k)$ admits decomposition (\ref{eq:vphi_0_sing_log_reg_split}) into singular, logarithmic and regular parts, according to its behavior in the high-temperature limit. These terms constitute the high-temperature expansion of the massless fundamental form factors, and the infinite series representing the regular part is resummed in closed form expression (\ref{eq:vphi_reg_resum}). For negative integer $n=-p-1$, $p \ge 0$, the massless fundamental form factor $\varphi_0^{-p-1}(k)$ is similarly represented by the high-temperature expansion (\ref{eq:vphi_neg_series}), which can be resummed into the closed form expression (\ref{eq:vphi_neg_ans}). This small-mass expansion induces the corresponding expansion of four effective action form factors (\ref{eq:vphi_L})--(\ref{eq:vphi_P}), which will be developed in the following section.

\section{Applications} \label{sec:appl}

	Having derived the general effective action form factors in Section~\ref{sec:eff_act}, we now specialize to scalar and spinor QED. For this purpose, we substitute the corresponding expressions for the potentials (\ref{eq:P_sf}) into the results obtained for the first few orders of the effective action expansion (\ref{eq:eff_act_0}),~(\ref{eq:eff_act_1}),~(\ref{eq:eff_act_2}) and evaluate the remaining traces explicitly.

	In both cases, the second-order part is a linear combination of longitudinal and transverse contributions. It is therefore convenient to parameterize it in terms of two polarization operators $\varPi_\|$ and $\varPi_\perp$ as
	\begin{align} \label{eq:ea2_Pi_par}
		&\varGamma_{2}[A] = \frac12 \int\limits_{(\omega, \bk)} \Bigl[ \varPi_{\|}(k, M)\,  A^\|_{\rho}(k) A^\|_{\rho}(-k)\nonumber \\[-2.5ex]&\hspace{8.5em}+ 
		\varPi_{\perp}(k, M) \, A^\perp_{\rho}(k) A^\perp_{\rho}(-k) \Bigr],
	\end{align}
	where $\varPi_\|$ and $\varPi_\perp$ are linear combinations of the form factors $\varphi_I$,~$I=\|,\perp,\times,P$, which also absorb the prefactor $- \kappa (4\pi)^{-D/2} k^2$ along with any factors arising from taking the trace in (\ref{eq:eff_act_2}).
	
	The aim of this section is to obtain explicit expressions for $\varPi_\|$ and $\varPi_\perp$ in terms of the small-mass expansions, using the corresponding expansion (\ref{eq:vphi_mass_exp}) of the fundamental form factor $\varphi^q_0(k, M)$. We write this expansion as
	\begin{equation} \label{eq:Pi_mass_exp}
		\varPi_I(k, M) = \sum_{m=0}^\infty \varPi_I^{(m)}(k, M), \qquad I = \|, \perp,
	\end{equation}
	where $\varPi_I^{(m)}(k, M)$ denotes the contributions of order $M^{2m}$ (including the prefactor $M^{2m}$). The polarization operators $\varPi_I(k,M)$ are linear combinations of fundamental form factor $\varphi^{q}_0(k, M)$ (along with derived basic form factors) for certain $D$-dependent values of $q$. Consequently, the expansion (\ref{eq:vphi_mass_exp}) implies that the term $\varPi^{(m)}$ contains massless form factors $\varphi_0^{q-m}(k)$. As the expansion order $m$ of $\varPi_I(k, M)$ increases, the parameter $n = q - m$ of the fundamental massless form factors $\varphi_0^{n}(k)$ decreases and become negative. This makes the explicit expressions for $\varPi_I^{(m)}$ more involved at higher orders, since the number of terms in $\varphi^n_0(k)$ grows for large negative $n$ (cf.~(\ref{eq:vphi_neg_ans})). Thus, although $\varPi_I^{(m)}(k,M)$ can be written explicitly at any finite order $m$, writing out the full all-order small-mass expansion is not particularly useful.
	
	However, for quantities defined in the zero-frequency, long-wavelength limit ($\omega \to 0$ and $\bk \to 0$), such as the Debye masses
	\begin{equation}
		M^2_{\mathrm{D},I} = \lim_{\bk \to 0} \lim_{\omega\to 0}\varPi_I(k, M), \qquad I = \|, \perp
	\end{equation}
	and the susceptibilities
	\begin{equation}
		\begin{aligned}
			&\xi = \phantom{-} \frac12\lim_{\bk \to 0} \lim_{\omega\to 0}\partial_{k^2} \varPi_\|(k, M), \\
			&\chi = -\frac12\lim_{\bk \to 0} \lim_{\omega\to 0} \partial_{k^2} \varPi_\perp(k, M).
		\end{aligned}
	\end{equation}
	we can obtain the small-mass expansion to all orders. Indeed, taking the limit $\omega \to 0$, $\bk \to 0$ in the expansions (\ref{eq:vphi_sing_series}),~(\ref{eq:vphi_log_series}),~(\ref{eq:vphi_reg_series}) and (\ref{eq:vphi_neg_series}) leaves a finite (and order-independent) number of terms with compact expressions, which can then be substituted into the small-mass expansion (\ref{eq:vphi_mass_exp}).
	This yields simple expressions for the above zero-frequency, long-wavelength limit quantities. Although not strictly necessary for such quantities, we include this limit here as a cross-check against known results. The main objects of interest in this work are the polarization operators $\varPi_I(k, M)$ and, in particular, the terms $\varPi^{(m)}_I(k, M)$ of its small-mass expansions, whose explicit expressions are derived in the remainder of this section.

	\subsection{Scalar QED form factors}
	
	The scalar case is obtained by substituting $\mathbb{P}_{0} = \mathbb{P}_{1} = 0$ in (\ref{eq:eff_act_0}),~(\ref{eq:eff_act_1}),~(\ref{eq:eff_act_2}) and omitting the matrix trace, which is trivial in this case. The zeroth- and first-order contributions to the effective action are
	\begin{align}
		&\varGamma_{0\mathrm{s}}[A] = - \frac{\rho^{D/2}_0}{(4\pi)^{D/2}} \int d^D x \, 1, \\
		&\varGamma_{1\mathrm{s}}[A] = \frac{2\rho^{D/2-1}_1}{(4\pi)^{D/2}}  \int d^D x \, A_t(x).
	\end{align}
	For the quadratic contribution in the parametrization (\ref{eq:ea2_Pi_par}), collecting the overall factors gives
	$\varPi_I(k, M) = - \frac12 \frac1{(4\pi)^{D/2}} k^2 \varphi_I(k, M)$, $I=\|,\perp$, where $\varphi_I$ are given in (\ref{eq:vphi_L}),~(\ref{eq:vphi_T}). The scalar polarization operators therefore read
	\begin{align}
		&\varPi_\|(k,M) =  \frac{4}{(4\pi)^{D/2}}\Bigl[ \Bigl(\mfrac{D}{2}-1\Bigr)\varphi_0^{D/2-1}(k,M) \nonumber \\ & - \psi_0^{D/2-1}(k,M) + \Bigl(\mfrac{k^2}{4}+M^2\Bigr)\varphi_0^{D/2-2}(k,M)\Bigr], \\
		&\varPi_\perp(k,M) = - \frac{2}{(4\pi)^{D/2}}  \Bigl[ \varphi^{D/2-1}_0(k,M) - \rho_0^{D/2-1}(M)\Bigr].
	\end{align}
	Using the small-mass expansion (\ref{eq:vphi_mass_exp}), we obtain the coefficients in (\ref{eq:Pi_mass_exp})
	\begin{align}
		&\varPi_\|^{(m)}(k, M) = \frac{4}{(4\pi)^{D/2}} \frac{(-)^m}{m!} M^{2m} \nonumber \\ & \times\Bigl[ \Bigl(\mfrac{D}{2}-1-m\Bigr)\varphi_0^{D/2-1-m}(k)  - \psi_0^{D/2-1 - m}(k) \nonumber \\ & \hspace{12.5em}+ \mfrac{k^2}{4}\varphi_0^{D/2-2-m}(k)\Bigr], \\ 
		&\varPi_\perp^{(m)}(k,M) = - \frac{2}{(4\pi)^{D/2}} \frac{(-)^m}{m!} M^{2m} \nonumber \\ &\hspace{6em}\times\Bigl[ \varphi^{D/2-1-m}_0(k) - \rho_0^{D/2-1-m}\Bigr].
	\end{align}
	Although the scalar determinant with twisted-periodic boundary condition and the corresponding form factors are mathematically well defined, continuing them to real physical $\mu$ is of limited physical relevance within the small-mass expansion. For real $\mu$, the underlying free scalar ensemble has a stable normal phase only for $|\mu|<M$. At $|\mu|=M$ Bose-Einstein condensation begins, so for $|\mu| > M$ the uncondensed determinant contains unstable modes. Hence for fixed $\mu$ an expansion around $M=0$ is not an expansion about a stable normal state. Formal continuation past threshold can produce complex polarization operators, particularly a complex longitudinal Debye mass squared. Its imaginary part signals the instability of the uncondensed background and its effect on the quadratic electromagnetic response. For imaginary $\mu$, the temporal twist remains a consistent mathematical parametrization, and the scalar form factors can be evaluated. We therefore retain the general scalar formulas but omit their explicit forms, and focus on the fermionic results, which have no analogous Bose condensation obstruction.

\subsection{Fermion QED form factors}
	The case of the Dirac fermion is slightly more subtle, since it requires explicit calculation of the trace in (\ref{eq:eff_act_0}),~(\ref{eq:eff_act_1}), and (\ref{eq:eff_act_2}). Recalling the expansion $\hat{P}_{\mathrm{f}} = \hat{1} \, \mathbb{P}_{0\mathrm{f}} + \hat{\sigma} \, \mathbb{P}_{1\mathrm{f}}$ in powers of charge operator and concluding that $\mathbb{P}_{0\mathrm{f}} = 0$, $\mathbb{P}_{1\mathrm{f}} = -\frac{i}2 \gamma^{\rho\lambda} F_{\rho\lambda}$ and recalling the dimension of spinor space in even dimension $\tr_L \mathbbm{1} = 2^{D/2}$, we obtain the following expression for the zeroth- and first-order parts of the effective action
	\begin{equation}
		\begin{aligned}
			&\varGamma_{0\mathrm{f}}[A] = \frac12 \frac{1}{(2\pi)^{D/2}} \rho^{D/2}_0(M) \int d^D x \, 1 , \\
			&\varGamma_{1\mathrm{f}}[A] = - \frac{1}{(2\pi)^{D/2}}  \rho^{D/2-1}_1(M) \int d^D x  \, A_t(x)
		\end{aligned}
	\end{equation}
	To calculate the second-order part, we need to compute the trace
	\begin{align}
		\tr_L \bigl(\mathbb{P}_{1\mathrm{f}}(k) \mathbb{P}_{1\mathrm{f}}(-k)\bigr) &= - \frac14 F_{\mu\nu}(k) F_{\rho\sigma}(-k) \, \tr_L (\gamma^{\mu\nu} \gamma^{\rho\sigma}) \nonumber\\ &=  2^{D/2-1} F_{\mu\nu}(k) F_{\mu\nu}(-k),
	\end{align}
	where we used the well-known identity for the Dirac matrices: $\tr_L (\gamma^{\mu\nu} \gamma^{\rho\sigma}) = 2^{D/2}\,(\delta^{\mu\sigma}\delta^{\nu\rho} -\delta^{\mu\rho}\delta^{\nu\sigma})$.
	Finally, decomposing the square of the electromagnetic stress tensor in the transverse/longitudinal basis
	\begin{align}
		&F_{\mu\nu}(k) F_{\mu\nu}(-k)= 2 k^2  A_{\mu}(k)  \mathcal{P}_{\mu\nu}(k) A_{\nu}(-k) \nonumber \\ & \hspace{3em} = 2 k^2 \bigl( A^\|_{\mu}(k) A^\|_{\mu}(-k) + A^\perp_{\mu}(k) A^\perp_{\mu}(-k) \bigr)
	\end{align}
	we deduce that $\varphi_{\|} \mapsto \varphi_{\|} + \varphi_P$ and $\varphi_{\perp} \mapsto \varphi_{\perp} + \varphi_P$, so in terms of the parametrization (\ref{eq:ea2_Pi_par}) the quadratic part of the effective action reads	
	\begin{align}
		&\varPi_\|(k, M) = -\frac{2}{(2\pi)^{D/2}}  \Bigl[ \Bigl(\mfrac{D}2-1\Bigr) \varphi_0^{D/2-1}(k, M)\nonumber \\ &\hspace{4em}- \psi_0^{D/2-1}(k, M) +  M^2\varphi_0^{D/2-2}(k, M)\Bigr], \\
		&\varPi_\perp(k, M) = \frac{1}{(2\pi)^{D/2}}  \Bigl[\varphi^{D/2-1}_0(k, M) - \rho_0^{D/2-1}(M) \nonumber \\ &\hspace{11.75em}+ \mfrac{k^2}{2} \varphi_0^{D/2-2}(k, M)\Bigr],
	\end{align}
	Mass expansion (\ref{eq:Pi_mass_exp}) gives the coefficients
	\begin{align}
		&\varPi_\|^{(m)}(k, M) = -\frac{2}{(2\pi)^{D/2}} \frac{(-)^m}{m!} M^{2m} \\ &\hspace{1.5em}\times\Bigl[ \Bigl(\mfrac{D}{2}-1-m\Bigr)\varphi_0^{D/2-1-m}(k) - \psi_0^{D/2-1 - m}(k)\Bigr], \nonumber \\ 
		&\varPi_\perp^{(m)}(k,M) = \frac{1}{(2\pi)^{D/2}} \frac{(-)^m}{m!} M^{2m} \\ &\hspace{1.5em}\times \Bigl[ \varphi^{D/2-1-m}_0(k) - \rho_0^{D/2-1-m}  + \mfrac{k^2}{2}\varphi_0^{D/2-2-m}(k)\Bigr]. \nonumber 
	\end{align}
	For the fermion we provide the explicit expressions for the first few orders of the small-mass expansion in $D=4$, obtained by direct substitution of the expressions for the massless form factors, obtained in the previous section.

\subsubsection*{Explicit form in $D=4$}
	
	Here we give the expressions for the longitudinal part $\varPi_\|^{(0)}$. The expressions for the transverse part resemble the same structure and given in Appendix~\ref{sec:app_Pi_explicit}.
	The leading in the small-mass expansion coefficient $\varPi^{(0)}_\|$ representing the massless longitudinal form factor naturally splits into the sum $\varPi_\|^{(0),\mathrm{sing}}+\varPi_\|^{(0), \mathrm{log}}+\varPi_\|^{(0),\mathrm{reg}}$ (cf.\ the corresponding splitting (\ref{eq:vphi_0_sing_log_reg_split}) of $\varphi_0^n(k)$), according to its behavior (singular/logarithmic/regular) in high-temperature $T\to \infty$ limit
	\begin{widetext}
	\begin{align}
		&\varPi_\|^{(0),\mathrm{sing}}(k, M) = 
		 \Bigl(\mfrac{T^2}{3}+\mfrac{\mu^2}{\pi^2}\Bigr) \mfrac{k^2}{\bk^2} \Bigl(1-\mfrac{i \omega}{2|\bk|}\log \mfrac{|\bk|+i \omega}{|\bk|-i \omega}\Bigr),\\
		&\varPi_\|^{(0), \mathrm{log}}(k, M) = \frac{ k^{2}}{6\pi^{2}}  \Bigl[\log \mfrac{\Lambda}{2T} + \gamma + 1 - \Bigl(1+\mfrac{k^2}{2\bk^2}\Bigr) \Bigl(1-\mfrac{i \omega}{2|\bk|} \log \mfrac{|\bk|+i \omega}{|\bk|-i \omega}\Bigr) \Bigr], \\
		&\varPi_\|^{(0),\mathrm{reg}}(k, M) = - \frac{k^{2}}{\pi^2} \frac{T^3}{|\bk|^3} \sum_{\{\sigma\}=\pm1} \Bigl[ \sigma_k  \Li^{(1)}_{3}\!\bigl(-e^{\, \sigma_k\frac{|\boldsymbol{k}|}{2T} +\sigma_\omega\frac{i \omega}{2T} + \sigma \frac{\mu}{T}} \bigr) -\mfrac{|\bk|}{2T} \Li^{(1)}_{2}\!\bigl(-e^{\, \sigma_k\frac{|\boldsymbol{k}|}{2T} +\sigma_\omega\frac{i \omega}{2T} + \sigma \frac{\mu}{T}} \bigr) \Bigr] \nonumber,			
		\\[-0.5ex]&\hphantom{\varPi_\|^{(0),\mathrm{reg}}(k, M)} =  \frac{k^{2}}{\pi^{2}} 	\sum_{r=0}^{\infty} \frac{(-)^{r}}{\,(2r+3)!} \Bigl(\frac{ k}{2T}\Bigr)^{\!2r} \bigl( \Li^{(1)}_{-2r}(-e^{\frac{\mu}{T}}) + \Li^{(1)}_{-2r}(-e^{-\frac{\mu}{T}}) \bigr) 
		\, C_{2r}^{(2)}(\tfrac{\omega}{k}).
	\end{align}
	\end{widetext}
	For the regular part we provide the two representations, namely resummed closed form and its high-temperature expansion. Note that though the closed form contains overall coefficient $T^3$, diverging at high-temperature limit, this growth is compensated by the combination of polylogarithm terms. This can be read off the expanded form of regular part. The $M^2$-order of small-mass expansion reduces to the constant
	\begin{equation}
		\varPi_\|^{(1)}(k, M) = -\mfrac{M^{2}}{2\pi^{2}}  ,
	\end{equation}
	whereas the expression for $M^4$-order is more involved
	\begin{widetext}
	\begin{align}
		\varPi_\|^{(2)}(k, M) &= -\frac{1}{4\pi^{2}} \frac{M^{4}}{k^{2}} \Bigl[  \sum_{\{\sigma\}=\pm1} \Li^{(1)}_{0}\!\bigl(-e^{\, \sigma_k\frac{|\boldsymbol{k}|}{2T} +\sigma_\omega\frac{i \omega}{2T} + \sigma \frac{\mu}{T}} \bigr) -4
		\, \bigl( \Li^{(1)}_{0}(-e^{\frac{\mu}{T}}) + \Li^{(1)}_{0}(-e^{-\frac{\mu}{T}}) \bigr)	 \Bigr] \nonumber \\&=
		\frac{M^{2}}{2\pi^2}  \Bigl(\frac{ M}{2T}\Bigr)^{\!2} \,
		\sum_{r=0}^\infty \frac{(-)^{r}}{(2r+1)!} \, \Bigl(\frac{k}{2T}\Bigr)^{\!2r}  \bigl( \Li^{(1)}_{-2(r+1)} (-e^{\frac{\mu}{T}}) +  \Li^{(1)}_{-2(r+1)} (-e^{-\frac{\mu}{T}})  \bigr) \, \tilde{C}^{(0)}_{2r+2}(\tfrac{\omega}{k}).
	\end{align}
	\end{widetext}
	As in the case of regular part of $M^2$-order, we provide both the resummed closed form and its expansion. The polarization operators displayed here have the overall factor $e^2$ suppressed. Their physical values are obtained by multiplying the expressions by $e^2$. 
		
\subsubsection*{Debye masses and susceptibilities}

	We now apply the small-mass expansion of the polarization functions to the static, long-wavelength quantities defined above. The static limit is taken first, by setting the external Matsubara frequency to zero, and is followed by the limit $\bk\to0$. The transverse Debye mass vanishes, $M_{\mathrm{D},\perp}^2=0$, expressing the absence of static magnetic screening at the one-loop order. We therefore give the longitudinal Debye mass and the electric and magnetic susceptibilities $\xi$ and $\chi$. We parameterize the spacetime dimension as $D=2(q+1)$, with $q\ge1$, so that $q=1$ corresponds to $D=4$.
	
	Throughout this subsection, $\mu$ denotes the real physical chemical potential, obtained from the imaginary one as $\mu\mapsto \pi T+ i \mu$. No continuation of the external frequency is needed for these static quantities.
	
	At each order in $M^2$, only finitely many terms of the momentum expansion contribute to the static response coefficients. Our form factor construction thus yields their small-mass expansions to arbitrary order. In particular, the susceptibility formulas extend the mass corrections studied in $D=4$ in~\cite{Ferreira:2023cqw} to arbitrary order in $M^2$ and to general even dimensions. The longitudinal Debye mass reads
	\begin{widetext}
		\begin{multline}
			M^2_{\mathrm{D},\|}=-(-)^{q}2\frac{(2T)^{2q}}{(2\pi)^{q+1}}  \Biggl[ \sum_{m=0}^{q} \frac{(-)^{m+q}}{m!} \Bigl(\frac{M}{2T}\Bigr)^{\!2m} (q-m)! \Bigl(\Li_{2(q-m)}(-e^{\frac{\mu}{T}})+\Li_{2(q-m)}(-e^{-\frac{\mu}{T}})\Bigr) \\  - \sum_{m=1}^{\infty} \frac{2}{(m-1)!(m+q)!} \Bigl(\frac{M}{2T}\Bigr)^{\!2(m+q)} \Bigl(\Li^{(1)}_{-2m}(-e^{\frac{\mu}{T}}) +\Li^{(1)}_{-2m}(-e^{-\frac{\mu}{T}})\Bigr) \Biggr]
		\end{multline}
	whereas the susceptibilities are	
	\begin{align}
			&\xi = \frac{(-)^q}{6}\frac{(2T)^{2(q-1)}}{(2\pi)^{q+1}}\Biggl[ \sum_{m=0}^{q-2}(-)^{m+q} \frac{(q-m-2)!(q-m)}{m!}  \Bigl(\frac{M}{2T}\Bigr)^{\!2m} \bigl( \Li_{2(q-m-1)}(-e^{\frac{\mu}{T}}) + \Li_{2(q-m-1)}(-e^{-\frac{\mu}{T}}) \bigr)\nonumber\\
			&\hspace{1em} - \frac{2}{(q-1)!}\Bigl(\frac{M}{2T}\Bigr)^{\!2(q-1)} \! \bigl(\log \tfrac{\Lambda}{2T} {-} \tfrac12 {+} \gamma  \bigr)
			+\sum_{m=0}^\infty \frac{2(m-1)}{m!(m+q-1)!}  \Bigl(\frac{M}{2T}\Bigr)^{\!2(m+q-1)} \!
			\bigl( \Li^{(1)}_{-2m} (-e^{\frac{\mu}{T}}) + \Li^{(1)}_{-2m} (-e^{-\frac{\mu}{T}})  \bigr) \Biggr]\\
			&\chi = -\frac{(-)^q}{6}\frac{(2T)^{2(q-1)}}{(2\pi)^{q+1}}
			\Biggl[ \sum_{m=0}^{q-2}(-)^{m+q} \frac{(q-m-2)!}{m!}  \Bigl(\frac{M}{2T}\Bigr)^{\!2m}\,  \bigl( \Li_{2(q-m-1)}(-e^{\frac{\mu}{T}}) + \Li_{2(q-m-1)}(-e^{-\frac{\mu}{T}}) \bigr)\nonumber \\ &\hspace{1em}
			-  \frac{2}{(q-1)!}\Bigl(\frac{M}{2T}\Bigr)^{\!2(q-1)} \bigl(\log \tfrac{\Lambda}{2T} + \gamma  \bigr)
			-\sum_{m=0}^\infty \frac{2}{m!(m+q-1)!}  \Bigl(\frac{ M}{2T}\Bigr)^{\!2(m+q-1)} 
			\bigl( \Li^{(1)}_{-2m} (-e^{\frac{\mu}{T}}) +  \Li^{(1)}_{-2m} (-e^{-\frac{\mu}{T}})  \bigr) \Biggr]			
		\end{align}
	\end{widetext}
	Finite sums with an upper limit smaller than their lower limit are understood to vanish. As in the case of the polarization operators the overall factor $e^2$ was suppressed, and the physical values are restored by multiplying the expressions by $e^2$.

	For convenience, we also give representations of these quantities in terms of modified Bessel functions, which retain the full mass dependence without invoking a small-mass expansion. The expression for the Debye mass and susceptibilities the straightforward generalizations of the standard massive fermion result. See e.g.~\cite{Laine:2016hma} for the textbook discussion, where the integral representation is provided for finite $\mu$ along with the explicit in terms of Bessel functions series at $\mu=0$. The Bessel representations of $\xi$ and $\chi$ are direct analogs for the momentum-dependent static response coefficients
	\begin{widetext}
			\begin{align} 
			&M^2_{\mathrm{D},\|} = 
			8 \frac{(2T)^{2q}}{(2\pi)^{q+1}} \Bigl(\frac{M}{2T}\Bigr)^{\!q+1}
			\sum_{n=1}^{\infty}(-)^{n-1}\frac{\cosh(n\frac{\mu}{T})}{n^{q-1}}
			K_{q+1}(\tfrac{M}{T} n),\\	
			&\xi = \frac{(-)^q}6 \frac{(2 T)^{2(q-1)}}{(2\pi)^{q+1}}\biggl[4 (-)^q \Bigl(\frac{M}{2T}\Bigr)^{\!q-1} 
			\sum_{n=1}^{\infty}(-)^n\frac{\cosh(n\frac{\mu}{T})}{n^{q-1}}
			\Bigl(
			K_{q-1}(\tfrac{M}{T}n)
			+\frac{M n}{2T} \, K_q(\tfrac{M}{T}n)
			\Bigr) \nonumber\\&\hspace{20em}
			-\frac{2}{(q-1)!}\Bigl(\frac{M}{2T}\Bigr)^{\!2(q-1)}
			\bigl(\log\tfrac{\Lambda}{M}+\tfrac12 H_{q-1}\bigr) \biggr] \\
			&\chi = - \frac{(-)^q}6 \frac{(2T)^{2(q-1)}}{(2\pi)^{q+1}} 
			\biggl[
			4 (-)^q\Bigl(\frac{M}{2T}\Bigr)^{\!q-1}
			\sum_{n=1}^{\infty}(-)^n\frac{\cosh(n\tfrac{\mu}{T})}{n^{q-1}}
			K_{q-1}(\tfrac{M}{T} n) 
			-\frac{2 }{(q-1)!}
			\Bigl(\frac{M}{2T}\Bigr)^{\!2(q-1)}\,
			\bigl(\log\tfrac{\Lambda}{M}{+}\tfrac12 H_{q-1}\bigr)
			\biggr].
		\end{align}
	\end{widetext}
	Here $K_\nu$ is the modified Bessel function of the second kind. The alternating signs encode fermionic statistics. For $M > 0$, the displayed series converge absolutely for $|\mu| < M$. Their values at $|\mu| > M$ can be obtained by analytic continuation from the convergence domain.
	
	Although the individual susceptibilities depend on the renormalization scale $\Lambda$, their sum $\xi+\chi$ is renormalization group invariant at the one loop order. In the small-mass representation, the scale-dependent logarithms cancel at order $M^{2(q-1)}$. In the Bessel representation, the entire common vacuum contribution proportional to $\log\frac{\Lambda}{M}+\frac12 H_{q-1}$ cancels, making this scale independence manifest.

\begin{widetext}

\end{widetext}

\section{Discussion}
%
	
	We have derived explicit closed form expressions for the nonlocal one-loop effective action form factors and the corresponding polarization operators in scalar and spinor QED at finite temperature and chemical potential. The calculation is performed to quadratic order in the background fields in arbitrary even spacetime dimensions. The massless form factors are obtained in closed form, while the massive results are organized as a small-mass expansion whose coefficients are determined by the same construction. At each mass order, the expressions retain the full dependence on the external momentum and therefore do not require the soft momentum approximation underlying the hard thermal loop expansion. Their static, long-wavelength limits yield the Debye mass and electromagnetic susceptibilities, providing applications of the general results to equilibrium response.
	
	An important consequence of summing the momentum expansion is that the resulting expressions provide access to analytic properties that cannot be resolved by a finite truncation of the expansion in $k / T$. In particular, they allow us to investigate poles, branch points, and cancellations between the different contributions to the polarization operators. Their physical interpretation, however, requires analytic continuation from discrete Euclidean frequencies to real frequencies with the appropriate causal prescription. This continuation must consistently account for the branches of the logarithms and polylogarithms entering the form factors. The analysis of collective excitations then requires locating the poles of the resulting photon propagator, rather than identifying singularities of individual form factors alone. A systematic treatment of this continuation and the associated real-time response remains a subject for further work.
	
	The closed form expressions also provide a direct starting point for other expansions, including the low-temperature expansion of the massless form factors. Instead of returning to the proper-time integrals or Matsubara sums, one can apply the large argument asymptotics of the functions appearing in the resummed answers. In this way, the same representation describes both the high-temperature expansion developed here and the opposite regime of large $k/T$. At finite density, the limiting prescription must be specified, so keeping the physical chemical potential fixed as $T\to0$ differs from keeping the dimensionless twist $\mu/T$ fixed, since it grows in the former case.
	
	The restriction to even dimensions is closely connected with the resummation method. For integer values of the index of the fundamental form factor, the dependence on the ratio of temporal and spatial momenta can be expressed through the shift-like operator $\varOmega$ and a finite number of momentum derivatives. Factoring these operations out of the $\beta k$ expansion reduces the remaining sum to combinations of polylogarithms and their order derivatives. In odd dimensions, the corresponding indices are half-integer, and the finite-order factorization used here is no longer available. Although the underlying integral and series representations remain useful, obtaining comparably explicit resummed expressions requires an alternative to the present $\varOmega$-operator construction.
	
	Another possible extension concerns systems with several commuting charges and their associated chemical potentials. Since these charges can be diagonalized simultaneously, each matter component experiences a chemical-potential shift, determined by its charge eigenvalues. Such a construction arises in QCD for a constant temporal gauge background chosen in the Cartan subalgebra. Specifically, its eigenvalues induce color-dependent shifts of the Matsubara frequencies, equivalent to imaginary color chemical potentials. This suggests extending the present calculation by resolving the matter fields into weight sectors and assembling the corresponding kernels with the appropriate group-theoretical factors. General non-Abelian backgrounds would additionally require accounting for transitions between these sectors and the resulting matrix structure of the response.
	
	Finally, the present construction is closely related to the non-vacuum gravitational effective action studied in \cite{Barvinsky:2026qzr,tbp}, where the same resummation technique is applied at vanishing chemical potential. The results obtained here extend this technique to non-vanishing chemical potential and suggest exploring a corresponding generalization in gravity. In that setting, chemical potentials would be conjugate to gravitational surface charges, with their specification tied to the boundary conditions and asymptotic symmetries of the geometry. It would be interesting to determine whether such thermodynamic data can be incorporated through an analogous modification of the Euclidean identifications, allowing the resummation method to describe nonlocal gravitational response in ensembles with these chemical potentials.
	
\newpage

\section*{Acknowledgments}
	The authors express their especial thanks to A.O.~Barvinsky for the useful discussions. N.K.~is grateful to the organizers of the DIAS Summer School ``Quantum Field Theory and Applications,'' and in particular to I.G.~Pirozhenko, for their warm hospitality during the important stages of this work. The authors thank ``BASIS'' Foundation for the Advancement of Theoretical Physics and Mathematics for support.

\appendix

\renewcommand{\theequation}{\Alph{section}.\arabic{equation}}

\makeatletter
\let\savedaddcontentsline\addcontentsline
\renewcommand{\addcontentsline}[3]{%
	\def\@tempa{#2}%
	\def\@tempb{subsection}%
	\ifx\@tempa\@tempb
	\else
	\savedaddcontentsline{#1}{#2}{#3}%
	\fi
}
\makeatother

\section{Recurrence relations} \label{sec:app_Theta_rec}

	In this Appendix we show that the auxiliary quantities $\Theta^{(l,l')}_{p,p'}(\tau,k)$ introduced in (\ref{eq:theta_Chem}), which we reproduce here for convenience
	\begin{align}\label{eq:app_chem_Theta}
		&\Theta_{p, p'}^{(l,l')}(\tau, k) = \sum_{\sigma=\pm1} \int_0^1 \!\! d\alpha \, (1{-}2\alpha)^p \alpha^{l'}(1{-}\alpha)^{l'} e^{-\tau \alpha(1-\alpha)k^2}  \nonumber\\
		&\hspace{0.25em}\times \frac1{\beta}\sum_{n} \sigma^{p'} (\omega_n - \alpha\omega_m + \sigma \mu)^{l} \, e^{-\tau (\omega_n-\alpha\omega_m + \sigma \mu)^2},
	\end{align}
	can be reduced to two basic quantities $\Theta^{(0,0)}_{0,0}$ and $\Theta^{(1,0)}_{0,1}$.
	For this purpose, we derive the recurrence relations, which allow one to lower indices in (\ref{eq:app_chem_Theta}).
	
	The powers $p$ and $p'$ may be immediately reduced to zero or one by using $\sigma^2=1$ and $(1-2\alpha)^2 = 1 - 4\alpha(1-\alpha)$. A useful parity relation implies that $\Theta_{p, p'}^{(l,l')}$ vanishes for set of parameters, obeying certain parity conditions. Indeed, under the substitution
	\begin{equation}
		\alpha\to1-\alpha, \qquad
		\sigma\to-\sigma, \qquad
		\omega_n\to\omega_m-\omega_n,
	\end{equation}
	where the shift of the Matsubara frequency is merely a relabeling of the summation index, the integrand in (\ref{eq:app_chem_Theta}) acquires the factor	$(-)^{p+p'+l}$. Therefore, $\Theta_{p, p'}^{(l,l')}$ vanishes whenever $p + p' + l$ is odd. This yields the following parity induced reductions
	\begin{equation} \label{eq:parity-reduction}
		\Theta^{(2j,l')}_{1,0}
		=\Theta^{(2j+1,l')}_{0,0}
		=\Theta^{(2j+1,l')}_{1,1}
		=\Theta^{(2j,l')}_{0,1} = 0.
	\end{equation}

	Differentiation with respect to $\tau$ gives the following recurrence relation
	\begin{equation}\label{eq:l-reduction}
		\Theta^{(l+2,l')}_{p,p'} = -\partial_\tau\Theta^{(l,l')}_{p,p'} - k^2\Theta^{(l,l'+1)}_{p,p'}.
	\end{equation}
	Repeated application of this relation reduces the first upper index to zero for even $l$ and to one for odd $l$.
	Thus, taking into account (\ref{eq:l-reduction}) and (\ref{eq:parity-reduction}), only the four families remain
	\begin{equation} \label{eq:app_families}
		\begin{aligned}
			&A_r=\Theta^{(0,r)}_{0,0}, &
			&B_r=\Theta^{(1,r)}_{1,0},\\
			&C_r=\Theta^{(1,r)}_{0,1}, &
			&D_r=\Theta^{(0,r)}_{1,1}.
		\end{aligned}
	\end{equation}

	Starting from the vanishing combinations in (\ref{eq:parity-reduction}) for $j=0$, we insert $d/d\alpha$ into the corresponding integrands and apply the Newton-Leibniz formula. By evaluating the derivative explicitly using (\ref{eq:app_chem_Theta}) on the one hand, and calculating the boundary terms on the other hand, we obtain the following system of recurrences
	\begin{widetext}
	\begin{subequations}\label{eq:recurrences}
		\begin{align}
			&-\tau k^2(A_r-4A_{r+1})+2\tau \omega B_r-2A_r
			+r(A_{r-1}-4A_r) =-2\delta_{r,0}\theta_0, \\
			&-\tau k^2B_r+2\tau\omega(-A_r'-k^2 A_{r+1})
			+rB_{r-1}-\omega A_r =0,	\\
			&-\tau k^2(C_r-4 C_{r+1})+2\tau\omega(-D_r'-k^2 D_{r+1})
			-2C_r+r(C_{r-1}-4 C_r)-\omega D_r
			=-2\delta_{r,0}\theta_1, \\
			&-\tau k^2 D_r+2\tau\omega C_r + r D_{r-1}
			=0,
		\end{align}
	\end{subequations}
	\end{widetext}
	where we omit the subscript of $\omega_m$ and introduce additional auxiliary quantities
	\begin{equation}
		\begin{aligned}
			&\theta_0(\tau)=\sum_{\sigma=\pm1}\frac1\beta\sum_n
			e^{-\tau(\omega_n+\sigma\mu)^2}, \\
			&\theta_1(\tau)=\sum_{\sigma=\pm1}\frac1\beta\sum_n
			\sigma(\omega_n+\sigma\mu)e^{-\tau(\omega_n+\sigma\mu)^2},
		\end{aligned}
	\end{equation}
	arising from the boundary terms at $\alpha=0,1$. Prime denotes differentiation with respect to $\tau$, that arise from applying~(\ref{eq:l-reduction}).
	
	Let us finally describe the order in which the recurrences are applied. We
	begin with a particular quantity $\Theta^{(l,r)}_{p,p'}$ that has to be reduced. We first reduce $p$ and $p'$ to $0$ or $1$ and then check the parity rule (\ref{eq:parity-reduction}). If $p+p'+l$ is odd, the quantity vanishes. Otherwise we apply (\ref{eq:l-reduction}) repeatedly to lower $l$ until $l=0$ or $l=1$. Since (\ref{eq:l-reduction}) preserves $p$,~$p'$ and the parity of $l$, the resulting expression is a linear combination of members of families $A$,~$B$,~$C$, and $D$, defined in (\ref{eq:app_families}).
	
	The four families (\ref{eq:app_families}) naturally split into two sectors, namely $(A,B)$ and $(C,D)$. The recurrences (\ref{eq:recurrences}) do not mix different sectors, so they can be handled separately.
	In the $(A,B)$ sector, the first equality in (\ref{eq:recurrences}) can be solved  for $B_r$ giving
	\begin{align} \label{eq:B-rule}
		&B_r=\frac1{2\tau\omega} \bigl((\tau k^2+2+4r)A_r - r A_{r-1} \nonumber \\
		& \hspace{7em} - 4\tau k^2A_{r+1}-2\delta_{r,0}\,\theta_0\bigr).
	\end{align}
	Thus, any $B_r$ is expressed in terms of the $A$-family.
	Taking the second relation in (\ref{eq:recurrences}) at
	$r-1$ and replacing $B_{r-1}$ and $B_{r-2}$ by using
	(\ref{eq:B-rule}) gives a linear recurrent equation for $A_r$
	\begin{align}
		&A_r=\frac{1}{4\tau^2 k^2(k^2{-}\omega^2)}\Bigl((\tau^2 k^4{+}8r\tau k^2{+}2\tau \omega^2{-}6\tau k^2)A_{r-1} \nonumber\\& \hspace{2em}+4\tau^2 \omega^2A'_{r-1} -2(r-1)(\tau k^2+2r-3)A_{r-2} \nonumber\\&\hspace{2em}+(r-2)(r-1)A_{r-3}+2\theta_0 \delta_{r-2,0}-2\tau k^2\theta_0\delta_{r-1,0}\Bigr)
	\end{align}
	Its r.h.s.\ contains only $A_{r-1}$,~$A_{r-2}$,~$A_{r-3}$ and their $\tau$ derivatives. Iterating this recurrence reduces every $A_r$ to the basic quantity $A_0\equiv\Theta^{(0,0)}_{0,0}$. If the original quantity
	is a $B_r$, one first applies (\ref{eq:B-rule}) and then reduces the resulting $A$-terms in the same way.
	
	The $(C, D)$ sector is handled analogously. The last relation in (\ref{eq:recurrences}) is first solved for~$C_r$
	\begin{equation}\label{eq:C-rule}
		C_r=\frac1{2\tau\omega} \bigl(\tau k^2D_r-rD_{r-1}\bigr).
	\end{equation}
	Substituting (\ref{eq:C-rule}) into the third equality in (\ref{eq:recurrences}) at index $r-1$ yields a linear recurrence for $D_r$
	\begin{align}
		&D_r = \frac{1}{4\tau^2 k^2(k^2{-}\omega^2)}
		\Bigl((\tau^2 k^4 {-} 2\tau((1{-}4r)k^2 {-} \omega^2)) D_{r-1} \nonumber \\ &\hspace{2em}+ 4\tau^2\omega^2 D_{r-1}' - 2(r-1)(\tau k^2 + 2r - 1) D_{r-2} \nonumber\\ & \hspace{2em}+ (r-2)(r-1) D_{r-3}
		- 4\tau\omega \delta_{r-1,0}\theta_1	\Bigr).
	\end{align}
	Repeated application reduces $D_r$ to $D_0$, which is expressed through the second basic quantity by $D_0=\frac{2\omega}{k^2}C_0 =\frac{2\omega}{k^2}\Theta^{(1,0)}_{0,1}$.
	Thus the recurrences are applied directly to the quantity being reduced, and
	the process terminates at $\Theta^{(0,0)}_{0,0}$ or $\Theta^{(1,0)}_{0,1}$. The remaining terms are	their $\tau$-derivatives and the auxiliary functions $\theta_0$ and $\theta_1$.

\section{Alternative form factor parametrization} \label{sec:f_0_eta_par}
	In this Appendix, we derive an alternative representation for the basic form factor $f_0(\tau, k)$, defined in (\ref{eq:f0_def}). To achieve this, we utilize the Dawson function, which is defined as a finite‑range integral and has an alternative representation \cite[Eq.~7.7.3]{NIST:DLMF} via the one‑sided Fourier–Laplace transform of a Gaussian 
	\begin{equation} \label{eq:Dawson}
		F_D(z) = e^{-z^2} \int_0^z du \, e^{u^2} = \frac{1}{2} \int_0^\infty dt \, e^{-\frac{t^2}{4}} \sin(z t).
	\end{equation}
	For convenience, we first reproduce the initial expression for $f_0(\tau, k)$
	\begin{align} \label{eq:f_0_app}
		&f_0(\tau, k) = \frac12 \sum_{\sigma=\pm1} \int_0^1 d\alpha \, e^{-\alpha(1-\alpha)\tau k^2} \nonumber\\&\hspace{5em}\times \frac{\sqrt{4\pi\tau}}{\beta} \sum_{n} e^{-\tau (\omega_n - \alpha\omega_m + \sigma\mu)^2}.
	\end{align}
	We begin by reducing the $\alpha$-integral to match the first definition of the Dawson function. This is achieved through the following change of variables:
	\begin{equation}
		u(\alpha) = \frac{\sqrt{\tau}}{2|\bk|} \Bigl( -(1-2\alpha)k^2 + 2\omega_m(\omega_n - \alpha \omega_m + \sigma \mu)\Bigr).
	\end{equation}
	This substitution transforms the exponentials in (\ref{eq:f_0_app}) into $e^{u^2}$, up to an overall $u$-independent factor. The new integration limits are determined by $u(0)$ and $u(1)$.
	A symmetry relates these boundary values under a simultaneous shift of the summation index $n$ and a sign flip of~$\sigma$
	\begin{align}
		\begin{aligned}
			&u(0) = -\frac{\sqrt{\tau}}{2|\bk|}(k^2 - 2\omega_m(\omega_n + \sigma\mu)), \\
			&u(1)\Bigr|_{\substack{\omega_n \mapsto \omega_m - \omega_n\\\sigma\mapsto-\sigma\hspace{1.05em}}} = - u(0).
		\end{aligned}
	\end{align}
	By leveraging this symmetry and the even parity of the integrand $e^{u^2}$, we can reduce the integration range to $[0, u(1)]$, introducing an overall factor of $2$. This yields:
	\begin{align}
		&f_0(\tau, k) = \frac{1}{\sqrt{\tau}|\bk|}\frac{\sqrt{4\pi\tau}}{\beta}\nonumber\\ &\times \sum_{n} \sum_{\sigma=\pm1} e^{-\frac{\tau k^2}{4\bk^2}(k^2+4(\omega_n+\sigma\mu)(\omega_n-\omega_m+\sigma\mu))}
		\int_0^{u(1)}\!\! du \, e^{u^2}.
	\end{align}
	Next, we convert the finite‑range $u$-integration into an infinite‑range $t$-integration using the second definition of the Dawson function (\ref{eq:Dawson}). We then perform a change of variable $t(\eta) = 2 \frac{|\bk|}{k} \eta$. This leads to
	\begin{equation}\hspace{-5em}
	\begin{aligned}
		&f_0(\tau, k) = \frac{1}{\sqrt{\tau}k}\frac{\sqrt{4\pi\tau}}{\beta} \sum_{n} \sum_{\sigma=\pm1}  e^{-\tau(\omega_n +\sigma\mu)^2}  \\ &
		\hspace{1em}\times\int_0^\infty d\eta\, e^{-(1-\frac{\omega_m^2}{k^2})\eta^2} \sin\bigl[ \sqrt{\tau}(k - 2\tfrac{\omega_m}{k}(\omega_n + \sigma\mu))\eta\bigr].
	\end{aligned} \hspace{-5em}
	\end{equation}
	Using the symmetry of $e^{-\tau(\omega_n +\sigma\mu)^2}$ under the transformations $\omega_n \mapsto -\omega_n$ and $\sigma \mapsto -\sigma$, we can effectively replace the sine term with
	\begin{align}
		&\sin\bigl[ \sqrt{\tau}(k - 2\tfrac{\omega_m}{k}(\omega_n + \sigma\mu))\eta\bigr] \nonumber \\ & \hspace{3em} \simeq \sin (\sqrt{\tau}k\eta) \exp\bigl(-2i\tfrac{\omega_m}{k}(\omega_n+\sigma\mu)\sqrt{\tau}\eta\bigr).
	\end{align}
	Finally, after completing the square in the exponentials, we arrive at the desired alternative form of the basic form factor
	\begin{align}
		&f_0(\tau, k) =
		\sum_{\sigma=\pm1}\int_0^\infty d\eta\, e^{-\eta^2} \, \frac{\sin(\sqrt{\tau} k\eta)}{\sqrt{\tau}k} \nonumber\\ &\hspace{4em}\times \frac{\sqrt{4\pi\tau}}{\beta} \sum_{n} e^{-\tau(\omega_n + \sigma\mu + i \frac{\eta}{\sqrt{\tau}} \frac{\omega_m}{k})^2},
	\end{align}
	that will be used in derivation of the explicit form of the effective action form factors.

	\onecolumngrid 
\section{Form factor derivation} 
	\subsection{Calculation of $\tilde{f}_{0}(s, k)$} \label{sec:app_f0_Mellin}

	Thus, our aim is to calculate the Mellin transform $\tilde{f}_0(s, k)$ of the basic effective action form factor $f_0(\tau, k)$. Substituting its explicit form (\ref{eq:f_0_eta}) to (\ref{eq:ff_0_Mellin}) and performing the change of variables $y = \beta^2/(4 \pi^2 \tau)$, $u = 4\pi \eta \sqrt{\tau}/\beta$, we get
	\begin{align}
		\tilde{f}_0(s, k) =
		\frac{\sqrt{\pi}}{z} \Bigl( \frac{\beta}{2\pi}\Bigr)^{\!2s} \sum_{n}\int_0^\infty du \, \sin(\tfrac{z u}{2\pi}) \int_0^\infty dy \, y^{-s-\frac12} \, e^{-y (\frac{u \cos\vartheta}{2})^2 - y^{-1} (n+\frac{\theta}{2\pi})^2} \cos(|n + \tfrac{\theta}{2\pi}| u \sin{\vartheta})
	\end{align}
	where we introduce the variables $z = \beta k / 2$ and $\vartheta = \arcsin \frac{\omega}{k}$. The $y$-integral yields modified Bessel function of the second kind \cite[Eq.~8.432.7]{gradshteyn2014table}
	\begin{equation}
		\tilde{f}_0(s, k) =  2^{-s+\frac32}\frac{\sqrt{\pi}}{z}  \Bigl( \frac{\beta}{2\pi}\Bigr)^{\!2s} (\cos \vartheta)^{s-\frac12} \sum_{n} |n + \tfrac{\theta}{2\pi}|^{-s+\frac12}
		\int_0^\infty du \, \sin(\tfrac{z u}{2\pi}) \cos(|n + \tfrac{\theta}{2\pi}| u \sin{\vartheta}) \, u^{s-\frac12} K_{s-\frac12}(|n + \tfrac{\theta}{2\pi}| u \cos\vartheta).
	\end{equation}
	Expanding the product of sine and cosine into sum of sines, and changing the integration variable to $v = u \, |n +\frac{\theta}{2\pi}|\cos \vartheta$, we get
	\begin{equation}
		\tilde{f}_0(s, k) =  2^{-s+\frac12}\frac{\sqrt{\pi}}{z\cos\vartheta}  \Bigl( \frac{\beta}{2\pi}\Bigr)^{\!2s} \sum_{n} |n + \tfrac{\theta}{2\pi}|^{-2s} \;
		\sum_{\pm}\int_0^\infty dv \,\sin\bigl( \tfrac{\frac{z}{2\pi} \pm  |n + \frac{\theta}{2\pi}| \sin{\vartheta}}{|n + \frac{\theta}{2\pi}| \cos\vartheta}v\bigr) \, v^{s-\frac12} K_{s-\frac12}(v)
	\end{equation}
	The remaining $v$-integral has the form of the tabulated one \cite[Eq.~6.699.3]{gradshteyn2014table},
	yielding
	\begin{equation}
		\tilde{f}_0(s, k) =  \frac{\sqrt{\pi}}{z\cos^2\vartheta}  \Bigl( \frac{\beta}{2\pi}\Bigr)^{\!2s}   \Gamma(s+\tfrac12) 
		\sum_{\pm}\sum_{n} \frac1{|n + \tfrac{\theta}{2\pi}|^{2s+1}} \bigl(\tfrac{z}{2\pi} \pm  |n + \tfrac{\theta}{2\pi}| \sin{\vartheta} \bigr) {}_2F_1\bigl(s+\tfrac12, 1, \tfrac32, -\bigl(\tfrac{\frac{z}{2\pi} \pm  |n + \frac{\theta}{2\pi}| \sin{\vartheta}}{|n + \frac{\theta}{2\pi}| \cos\vartheta}\bigr)^{\!2\,}\bigr),
	\end{equation}
	Now, we expand the hypergeometric function into Gauss series
	\begin{equation}
		\tilde{f}_0(s, k) =  \frac{\pi}{2z}  \Bigl( \frac{\beta}{2\pi}\Bigr)^{\!2s} 
		\sum_{l=0}^\infty  \frac{\Gamma(s+l+\frac12)}{\Gamma(l+\frac32)} \frac{(-)^l}{(\cos\vartheta)^{2l+2}} \sum_{\pm}\sum_{n}  \frac{(\frac{z}{2\pi} \pm  |n + \frac{\theta}{2\pi}| \sin{\vartheta})^{2l+1}}{|n + \frac{\theta}{2\pi}|^{2(s+l)+1}}.
	\end{equation}
	To calculate the sum over $n$ explicitly, we expand the numerator in the rightmost fraction using the binomial theorem
	\begin{equation}
		\tilde{f}_0(s, k) = \frac12 \Bigl( \frac{\beta}{2\pi}\Bigr)^{\!2s} 
		\sum_{l=0}^\infty  \frac{\Gamma(s+l+\frac12)}{\Gamma(l+\frac32)} \frac{(-)^l}{(\cos\vartheta)^{2l+2}}
		\sum_{r=0}^{l}  \frac{(2l+1)!}{(2r+1)!(2l-2r)!} \Bigl(\frac{z}{2\pi}\Bigr)^{\!2r} (\sin{\vartheta})^{2l-2r} \sum_{n}\frac1{|n + \frac{\theta}{2\pi}|^{2(s+r)+1}}.
	\end{equation}
	where the sum over signs eliminates the even powers of $z$, and the overall $1/z$ prefactor reduces the surviving odd powers to the even ones. Exchanging the order of the summation over $r$ and $l$ and taking the sum over $n$ explicitly in terms of the Hurwitz zeta function $\zeta(s, a) = \sum_{n=0}^\infty (n+a)^{-s}$, we get
	\begin{multline}
		\tilde{f}_0(s, k) = \frac{\sqrt{\pi}}2\Bigl( \frac{\beta}{2\pi}\Bigr)^{\!2s} 
		\sum_{r=0}^\infty (-)^r \bigl[\zeta(2(s+r)+1, \tfrac{\theta}{2\pi})+\zeta(2(s+r)+1, 1-\tfrac{\theta}{2\pi})\bigr] \frac{1}{(\cos\vartheta)^{2r+2}} \Bigl(\frac{z}{2\pi}\Bigr)^{\!2r}   \\\times 
		\frac{1}{\Gamma(r+1) \Gamma(r+\frac32)}\sum_{l=0}^\infty  (-)^{l} \frac{\Gamma(r+s+l+\frac12)\Gamma(l+r+1)}{\Gamma(l+\frac12)\Gamma(l+1)}  (\tan{\vartheta})^{2l},
	\end{multline}
	where we transform the gamma functions using the Legendre duplication formula. Recognizing the hypergeometric series in the sum over $l$, we obtain
	\begin{multline}
		\tilde{f}_0(s, k) =  \frac12\Bigl( \frac{\beta}{2\pi}\Bigr)^{\!2s} 
		\sum_{r=0}^\infty (-)^r   \bigl[\zeta(2(s+r)+1, \tfrac{\theta}{2\pi})+\zeta(2(s+r)+1, 1-\tfrac{\theta}{2\pi})\bigr]  \frac1{(\cos\vartheta)^{2r+2}} \Bigl(\frac{z}{2\pi}\Bigr)^{\!2r} \\\times 
		\frac{\Gamma(r+s+\frac12)}{\Gamma(r+\frac32)} {}_2F_1(s+r+\tfrac12, r+1, \tfrac12, -\tan^2\vartheta).
	\end{multline}
	Applying the Pfaff transformation to the hypergeometric function, we find
	\begin{equation} 
		\tilde{f}_0(s, k) =  \frac12\Bigl( \frac{\beta}{2\pi}\Bigr)^{\!2s} 
		\sum_{r=0}^\infty (-)^r   \bigl[\zeta(2(s+r)+1, \tfrac{\theta}{2\pi})+\zeta(2(s+r)+1, 1-\tfrac{\theta}{2\pi})\bigr]  \Bigl(\frac{z}{2\pi}\Bigr)^{\!2r} 
		\frac{\Gamma(r+s+\frac12)}{\Gamma(r+\frac32)}  {}_2F_1(-s-r, r+1, \tfrac12, \sin^2\vartheta).
	\end{equation}
	Finally, using the consequence of Hurwitz's formula, relating the Hurwitz zeta function and the polylogarithm
	\begin{equation}
		\zeta(s, \tfrac{\theta}{2\pi}) + \zeta(s, 1-\tfrac{\theta}{2\pi}) = \frac{2\Gamma(1-s)}{(2\pi)^{1-s}} \sin \bigl(\mfrac{\pi s}{2}\bigr) \Bigl[ \Li_{1-s} (e^{i\theta}) +  \Li_{1-s} (e^{-i \theta})  \Bigr].
	\end{equation}	
	we get
	\begin{equation} \label{eq:app_ff_Mellin_Li}
		\tilde{f}_0(s, k) =   \Bigl(\frac{\beta}{2}\Bigr)^{\!2s} 
		\sum_{r=0}^\infty (-)^r \frac{r!}{(2r+1)!}  \, z^{2r}  \Bigl[ \Li_{-2(s+r)} (e^{i\theta}) +  \Li_{-2(s+r)} (e^{-i \theta})  \Bigr]  
		\Gamma(-r-s) \, {}_2F_1(-s-r, r+1, \tfrac12, \sin^2\vartheta).
	\end{equation}
	
\subsection{Calculation of $\varphi^n_{0}(k)$, $n\ge0$} \label{sec:app_vaphi_massless}
	
	In this section we derive the closed form expression (\ref{eq:vphi_n_ans}) for the pure thermal part of the massless form factors $\varphi^n_{0\beta}(k)$ with $n\ge 0$.
	
	Substituting (\ref{eq:app_ff_Mellin_Li}) and (\ref{eq:ff_Mellin_vac}) into the definition (\ref{eq:vphi_beta_def}) of $\varphi^n_{0\beta}$, we observe that the only singularity in $\tilde{f}_0$, that is canceled by the vacuum part occurs in the $r=n$ term in the answer (\ref{eq:app_ff_Mellin_Li}) for $\tilde{f}_0(s, k)$. Thus, it is natural to split $\varphi_{0\beta}^{q}(k)$ into three parts
	\begin{equation} \label{eq:app_vphi_three}
		\varphi_{0\beta}^{n}(k) = \varphi_{0\beta}^{<n}(k) + \varphi_{0\beta}^{=n}(k) + \varphi_{0\beta}^{>n}(k),
	\end{equation}
	corresponding to $r < n$, $r = n$ and $r > n$ in (\ref{eq:app_ff_Mellin_Li}), respectively. 
	
	Let us handle these three terms in order. Setting $s=-n$ in the $r < n$ terms in (\ref{eq:app_ff_Mellin_Li}) we immediately obtain the expression for $\varphi^{<n}_{0\beta}(k)$
	\begin{equation}
		\varphi^{<n}_{0\beta}(k) = \Bigl(\frac{\beta}{2}\Bigr)^{\!\!-2n} 
		\sum_{r=0}^{n-1} (-)^{r}  \frac{r!}{(2r+1)!} \, z^{2r}   \bigl( \Li_{2n-2r}(e^{i\theta}) + \Li_{2n-2r}(e^{-i\theta}) \bigr) \, (n-r-1)! \, {}_2F_1(n-r, r+1, \tfrac12, \sin^2\vartheta). \label{eq:app_vphi_ls}
	\end{equation}
	The calculation of $\varphi^{>n}_{0\beta}(k)$ is slightly more subtle. Indeed, due to the presence of $\Gamma(-r-s)$ in the summand, which is singular in the limit $s\to -n$ for integers $r>n$. Nevertheless, this singularity is canceled by the factor $\Li_{-2(s+r)} (e^{i\theta}) +  \Li_{-2(s+r)} (e^{-i \theta})$, which vanishes in the limit $s \to -n$ for integers $r>n$. Hence, the product gives a finite quantity, namely
	\begin{equation}
		\lim_{s\to -n} \Gamma(-r-s) \,\bigl(\Li_{-2(s+r)} (e^{i\theta}) +  \Li_{-2(s+r)} (e^{-i \theta})\bigr) =  (-)^{r-n}\frac{2}{(r-n)!} \bigl(\Li^{(1)}_{2n-2r} (e^{i\theta}) +  \Li^{(1)}_{2n-2r} (e^{-i \theta})\bigr).
	\end{equation}
	The remaining factors in the summand in (\ref{eq:app_ff_Mellin_Li}) are finite in this limit, yielding
	\begin{equation} \label{eq:app_vphi_gr_sum}
		\varphi^{>n}_{0\beta}(k) =  2 (-)^n\Bigl( \frac{\beta}{2}\Bigr)^{\!\!-2n} 
		\sum_{r=n+1}^{\infty} \frac{r!}{(2r+1)!} \, z^{2r} \bigl( \Li^{(1)}_{2n-2r}(e^{i\theta}) + \Li^{(1)}_{2n-2r}(e^{-i\theta}) \bigr)  \,
		\frac{1}{(r-n)!}  {}_2F_1(n-r, r+1, \tfrac12, \sin^2\vartheta). 
	\end{equation}	
	Finally, according to (\ref{eq:vphi_beta_def}), the contribution $\varphi^{=n}_{0\beta}(k)$ is given by the following expression
	\begin{equation} \label{eq:app_vphi_eq_lim}
		\varphi^{=n}_{0\beta}(k) = \lim_{s\to 0} \Bigl[ \tilde{f}_{0}^{=n}(s-n, k) - \tilde{f}_{0v}(s-n, k) \Bigr],
	\end{equation}
	where $\tilde{f}_{0}^{=n}(s-n, k)$ denotes the $n$-th term of the series representation (\ref{eq:app_ff_Mellin_Li}) of $\tilde{f}_0(s, k)$. We need to verify the cancellation of the pole terms between $\tilde{f}_0^{=n}(s, k)$ and its vacuum part $\tilde{f}_{0v}(s, k)$, and then calculate the finite part.
	For the first term of the difference we have the following Laurent expansion near the pole at $s=0$
	\begin{align}
		\tilde{f}_{0}^{=n}(s-n, k) &=  (-)^n\Bigl(\frac{\beta}{2}\Bigr)^{\!2(s-n)} 
		 \frac{n!}{(2n+1)!}  \Gamma(-s) \Bigl[ \Li_{-2s} (e^{i\theta}) +  \Li_{-2s} (e^{-i \theta})  \Bigr]  
		z^{2n} \, {}_2F_1(-s, n+1, \tfrac12, \sin^2\vartheta) \nonumber \\&=
		(-)^n\Bigl(\frac{\beta}{2}\Bigr)^{\!\!-2n} 
	 	\frac{n!}{(2n+1)!} z^{2n} \, \frac1{s} \Bigl[1 + s \bigl(2\log \tfrac{\beta}{2} + \gamma + 2\Li_{0}^{(1)} (e^{i\theta}) +  2\Li_{0}^{(1)} (e^{-i \theta})   \nonumber\\& \hspace{13.875em}-2(n+1) \, \sin^2\vartheta \, {}_3F_2(1,1,n+2;2,\tfrac32; \sin^2\vartheta)\bigr)   + \mathcal{O}(s^2)\Bigr]. \label{eq:f0_Mellin_pole}
	\end{align}
	Here the generalized hypergeometric function ${}_3F_2$ comes from the expansion of the Gauss one ${}_2F_1$ in its first parameter (cf.~\ref{eq:app_2F1_par_exp})
	\begin{align}
		{}_2F_1(-s, n+1, \tfrac12, x) &= 
		1 - 2(n+1) \, s \, x \, {}_3F_2(1,1,n+2;2,\tfrac32; x) + \mathcal{O}(s^2).
	\end{align}
	The second term of the difference in (\ref{eq:app_vphi_eq_lim}) expands as
	\begin{equation}
		\tilde{f}_{0v}(s-n,k) = (-)^n k^{2n} \frac{n!}{(2n+1)!} \, \frac1{s} \Bigl[1 + s \, (-2 \log k -\gamma - H_n +2 H_{2n+1}) + \mathcal{O}(s^2)\Bigr].
	\end{equation}
	Thus, the pole terms cancel as expected, so isolating the finite part, we get the answer
	\begin{align} 
		&\varphi^{=n}_{0\beta}(k) =
		2 (-)^n \Bigl(\frac{\beta}{2}\Bigr)^{\!\!-2n} \!\! \frac{n!}{(2n+1)!} \, z^{2n} \Bigl(\log z + \gamma  -
		(n+1) \sin^2\vartheta \, {}_3F_2(1,1,n+2;2,\tfrac32;\sin^2\vartheta) \nonumber \\ & \hspace{20.5em}  + \tfrac12 H_n - H_{2n+1} +
		\Li_{0}^{(1)}(e^{i\theta})+\Li_{0}^{(1)}(e^{-i\theta}) \Bigr). \label{eq:app_vphi_eq}
	\end{align}
	This finalizes the calculation of the components of three-term decomposition (\ref{eq:app_vphi_three}) of $\varphi^n_{0\beta}$.
	
	The sum of the vacuum and pure thermal parts of the massless fundamental form factor $\varphi_0^n(k) = \varphi_{0v}^n(k) + \varphi^n_{0\beta}(k)$ suggests another useful three-term splitting $\varphi_0^n(k) = \varphi_0^{n,\mathrm{sing}}(k) + \varphi_0^{n,\mathrm{log}}(k) + \varphi_0^{n,\mathrm{reg}}(k)$, according to the behavior of the terms in the $\beta \to 0$ limit (recall that $z = \beta k / 2$). The singular part, which is power-divergent, comes entirely from $ \varphi_{0\beta}^{<n}(k)$. The logarithmic part comes from the sum $\varphi_{0\beta}^{=n}(k) + \varphi_{0v}^{n}(k)$, in which harmonic number terms cancel. It is also useful to extract the polylogarithmic part from this combination and combine it with $\varphi_{0\beta}^{>n}(k)$, since it has exactly the same form as the $r=n$ term, missing in (\ref{eq:app_vphi_gr_sum}). The explicit form after the rearrangement reads
	\begin{subequations}
		\begin{align}
			&\varphi^{n,\mathrm{sing}}_{0}(k) = \, (-)^n k^{2n}
			\sum_{r=1}^{n} (-)^{r}  \frac{(r-1)!(n-r)!}{(2n-2r+1)!} \, z^{-2r}   \bigl( \Li_{2r}(e^{i\theta}) + \Li_{2r}(e^{-i\theta}) \bigr) \, {}_2F_1(r, n-r+1, \tfrac12, \sin^2\vartheta), \\
			&\varphi^{n,\mathrm{log}}_{0}(k) =
			2 (-)^n k^{2n} \frac{n!}{(2n+1)!} \, \Bigl(\log \tfrac{\beta \Lambda}{2} + \gamma -
			(n+1) \sin^2\vartheta \,\, {}_3F_2(1,1,n+2;2,\tfrac32;\sin^2\vartheta) \Bigr),  \\
			&\varphi^{n,\mathrm{reg}}_{0}(k) =  2 (-)^n k^{2n} 
			\sum_{r=0}^{\infty} \frac{(r+n)!}{r!\,(2r+2n+1)!} \, z^{2r} \bigl( \Li^{(1)}_{-2r}(e^{i\theta}) + \Li^{(1)}_{-2r}(e^{-i\theta}) \bigr)\, {}_2F_1(-r, r+n+1, \tfrac12, \sin^2\vartheta).
		\end{align}
	\end{subequations}	
	where we also shift summation indices in singular and regular parts.

	Now, our aim is to employ the operator $\varOmega$ to absorb all $\vartheta$-dependence and bring the components of three-term decomposition (\ref{eq:app_vphi_three}) of $\varphi^n_{0\beta}$ into a simpler form. In particular, this will allow us to explicitly take the infinite sum in the series representation (\ref{eq:app_vphi_gr}) of its third part $\varphi^{>n}_{0\beta}$. For the first term $\varphi^{<n}_{0\beta}$ we start with its representation (\ref{eq:app_vphi_ls}) and absorb the $\vartheta$-dependence into the operator $\varOmega$, using the identity (\ref{eq:app_Omega_log_less})
	\begin{equation}
		\partial_{z^2}^q\bigl(\varOmega z^{2r} \log z\bigr)	= \tfrac12 (-)^{q-r-1} r!(q-r-1)! z^{2r-2q} {}_2 F_1(-r+q,r+1;\tfrac12;\sin^2\vartheta), \qquad  r < q
	\end{equation}
	Hence, $\varphi^{<n}_{0\beta}$ takes the following form
	\begin{equation}  \label{eq:vphi_ls_ans}
		\varphi^{<n}_{0\beta}(k) =   (-)^{n}k^{2n} \, 
		\partial_{z^2}^n \biggl[\varOmega \biggl( -  \sum_{r=0}^{n-1}   \frac{2}{(2r+1)!}  \bigl( \Li_{2n-2r}(e^{i\theta}) + \Li_{2n-2r}(e^{-i\theta}) \bigr) z^{2r} \log z\biggr) \biggr]
	\end{equation}
	To obtain the $\varOmega$-form of $\varphi^{=n}_{0\beta}$, we apply the identities 
	\begin{equation}
		\begin{aligned}
			&\partial_{z^2}^n\bigl(\varOmega z^{2n} \log z\bigr)	= n! (\log z + \tfrac12 H_n)  - (n+1)! \sin^2\vartheta \, {}_3 F_2(1,1,n+2;\tfrac32,2;\sin^2\vartheta), \\
			&\partial_{z^2}^n\bigl(\varOmega z^{2n}\bigr) = n!
		\end{aligned}
	\end{equation}
	both derived in Appendix~\ref{sec:app_Omega}, to (\ref{eq:app_vphi_eq}) and obtain
	\begin{equation} \label{eq:app_vphi_eq_ans}
		\varphi^{n,=}_{0\beta}(k) = (-)^n
		k^{2n} \, \partial_{z^2}^n\biggl(  \varOmega \frac{2}{(2n+1)!} \Bigl[
		- \bigl(\Li_{0}(e^{i\theta})+\Li_{0}(e^{-i\theta})\bigr) \bigl(\log z + \gamma -  H_{2n+1} \bigr) + +\Li_{0}^{(1)}(e^{i\theta})+\Li_{0}^{(1)}(e^{-i\theta})  \Bigr] z^{2n} \biggr),
	\end{equation}
	where we inserted $\Li_{0}(e^{i\theta})+\Li_{0}(e^{-i\theta}) \equiv -1$ to mimic the general pattern of the answer (cf.~(\ref{eq:vphi_n_ans})).
	To handle the third part $\varphi^{>n}_{0\beta}$ of $\varphi^{n}_{0\beta}$, we first absorb $\vartheta$ into $\varOmega$ in its series representation (\ref{eq:app_vphi_gr}) using the identity that follows directly from (\ref{eq:app_Omega_z_d})
	\begin{equation}
		\partial_{z^2}^n\bigl(\varOmega z^{2r}\bigr)	= \frac{r!}{(r-n)!} z^{2r-2n} {}_2F_1(-r+n,r+1;\tfrac12;\sin^2\vartheta)
	\end{equation}
	so it takes the form
	\begin{equation} 
		\varphi^{>n}_{0\beta}(k) =(-)^n k^{2n} \, \partial_{z^2}^{n} \biggl[ \varOmega \biggl( 2 
		\sum_{r=n+1}^{\infty} \frac{1}{(2r+1)!} \bigl( \Li^{(1)}_{2n-2r}(e^{i\theta}) + \Li^{(1)}_{2n-2r}(e^{-i\theta}) \bigr)  z^{2r} \biggr) \biggr]
	\end{equation}
	Employing the Taylor series for polylogarithm (\ref{eq:Li_Taylor}), we resum the series as 
	\begin{multline} \label{eq:app_vphi_gr}
		\varphi^{>n}_{0\beta}(k) = (-)^n k^{2n} \, \partial_{z^2}^{n} \biggl[ \varOmega \biggl( \frac{1}{z}\bigl( \Li^{(1)}_{2n+1}(e^{z+i\theta}) +\Li^{(1)}_{2n+1}(e^{z-i\theta}) -\Li^{(1)}_{2n+1}(e^{-z+i\theta}) -\Li^{(1)}_{2n+1}(e^{-z-i\theta}) \bigr) \\  -2\sum_{r=0}^{n} \frac{1}{(2r+1)!} \bigl( \Li^{(1)}_{2n-2r}(e^{i\theta}) +\Li^{(1)}_{2n-2r}(e^{-i\theta}) \bigr)z^{2r} \biggr) \biggr].
	\end{multline}
	To combine the three contributions in (\ref{eq:app_vphi_three}), we first simplify (\ref{eq:app_vphi_gr}). In the second line of this equation, the terms with $r=0,\ldots n-1$ may be omitted, since $\partial_{z^2}^n (\varOmega z^{2r}) = 0$ for $r<n$. The surviving term with $r=n$ is then canceled by the corresponding term in the second line of (\ref{eq:app_vphi_eq_ans}). Next, in (\ref{eq:vphi_ls_ans}) we replace $\log z$ by $\log z + \gamma - H_{2r+1}$. This shift does not change the result, because the additional terms are annihilated by $\partial_{z^2}^n$. After this shift, the term in the second line of (\ref{eq:app_vphi_eq_ans}) supplies the missing $r=n$ term in the sum in (\ref{eq:vphi_ls_ans}). Combining these simplifications result yields (\ref{eq:vphi_n_ans}).
\subsection{Series representation of fundamental form factor}
	\label{sec:app_alt_vphi}
	
	In this appendix we derive an alternative form of the form factor $\varphi_0^{q}(k, M)$, which does not rely on smallness of the mass $M$. We begin with the zeta-regularized form (\ref{eq:vphi_0_spectral}) of the form factor. It is expressed in terms of the spectral function $\tilde\varphi(s, k, M)$, having the contour integral representation (\ref{eq:spectral_mellin}). We reproduce both here for the convenience
	\begin{equation}
		\varphi_0^{q}(k, M) = \lim_{s\to0}\frac{\partial}{\partial s} \frac{\Lambda^{2s}}{\Gamma(s)} \tilde{\varphi}(s-q, k, M), \qquad
		\tilde{\varphi}(s, k, M) = \frac1{2\pi i} \int_{c-i\infty}^{c+i\infty} dw \, M^{-2w} \, \Gamma(w) \, \tilde{f}_0(s-w, k).
	\end{equation}
	The contour integral representation for the spectral function is valid for $s > 0$ and the contour defined by the constant $c$ subject to $0<c<s$. To define $\tilde\varphi(s-q, k, M)$ near $s = 0$, we analytically continue in $s$.
	The function $\tilde{f}_0(s, k)$ has poles at $s=-r$, for all nonnegative integers $r$. Thus, the integrand in the integral representation (\ref{eq:spectral_mellin}) of spectral function $\tilde\varphi(s, k, M)$, containing $\tilde{f}_0(s-w, k)$, has poles at $w = s+r$. Thus, analytically continuing $\tilde{\varphi}$ from $s$ to $s-q$ we pick up poles of $\tilde{f}_0(s-q-w, k)$ at $w = s-q+r$ for $r = 0, \dots q$. Their residues yield the pole tail. Thus, the spectral function $\tilde{\varphi}(s-q, k, M)$ splits into an integral part and a pole tail as $\tilde{\varphi}(s-q, k, M) = \tilde{\varphi}^{\mathrm{int}}(s-q, k, M) + \tilde{\varphi}^{\mathrm{tail}}(s-q, k, M)$, where 
	\begin{align}
		&\tilde{\varphi}^{\mathrm{int}}(s-q, k, M) = \frac1{2\pi i} \int_{c-i\infty}^{c+i\infty} dw \, M^{-2w} \, \Gamma(w) \, \tilde{f}_0(s-q-w, k), \\
		&\tilde{\varphi}^{\mathrm{tail}}(s-q, k, M) =
		- \sum_{r=0}^{q} \operatorname{res}_{w=s-q+r} M^{-2w} \Gamma(w) \tilde{f}_0(s-q-w, k) \nonumber\\ &\hphantom{\tilde{\varphi}^{\mathrm{tail}}(s-q, k, M)}=
		M^{2(q-s)} \sum_{r=0}^{q} (-)^r \, \frac{r!}{(2r+1)!}  \,\Bigl(\frac{k}{M}\Bigr)^{\!2r}   \,  \Gamma(r+s-q) 
	\end{align}
	The constant $c$ in the integral part satisfies $0 < c < 1$.
	Going from the first to the second line for $\tilde{\varphi}^{\mathrm{tail}}$ we calculate the residues explicitly, using the known behavior of $\tilde{f}_{0}(s-r, k)$ at the limit $s\to 0$, following from (\ref{eq:f0_Mellin_pole}).
	
	Now, we substitute $\tilde{\varphi}(s-q, k, M)$ into the expression for $\varphi_0^q(k, M)$. The splitting of $\tilde{\varphi}$ into sum of $\tilde{\varphi}^{\mathrm{int}}$ and $\tilde{\varphi}^{\mathrm{tail}}$ induces the corresponding splitting of the fundamental form factor $\varphi_0^q(k, M) = \varphi_0^{q,\mathrm{int}}(k, M) + \varphi_0^{q,\mathrm{tail}}(k, M)$.
	For the integral part $\tilde{\varphi}^{\mathrm{int}}(s-q, k M)$, the zeta-regularization is lifted, since it is regular in the $s\to 0$ limit. The tail part is singular in this limit, hence, zeta regularization leads to a nontrivial dependence on the scale parameter
	\begin{equation}
		\varphi^{q,\mathrm{tail}}_0 (k, M) = (-)^{q}  M^{2q} \sum_{r=0}^q\frac{1}{(q-r)!} \frac{r!}{(2r+1)!} \Bigl(\frac{k}{M}\Bigr)^{\!2r}  \bigl(2 \log\tfrac{\Lambda}{M} + H_{q-r}\bigr)
	\end{equation}
	Now, let us calculate the integral part
	\begin{equation}
		\varphi_0^{q,\mathrm{int}}(k, M) = \frac1{2\pi i} \int_{c-i\infty}^{c+i\infty} dw \, M^{-2w} \, \Gamma(w) \, \tilde{f}_0(-q-w, k) 
	\end{equation}
	Substitution of the explicit expression (\ref{eq:ff_Mellin_Li}) for $\tilde{f}_0(s, k)$	gives
	\begin{multline}
		\varphi_0^{q,\mathrm{int}}(k, M) = \Bigl(\frac{\beta}{2}\Bigr)^{\!\!-2q} \sum_{r=0}^\infty (-)^r \frac{r!}{(2r+1)!} \, 	z^{2r}  \frac1{2\pi i} \int_{c-i\infty}^{c+i\infty} dw \,\Bigl(\frac{\beta M}{2}\Bigr)^{\!\!-2w} \, \Gamma(w) \,  \Bigl[ \Li_{2(w+q-r)} (e^{i\theta}) +  \Li_{2(w+q-r)} (e^{-i \theta})  \Bigr]  \\\times 
		\Gamma(w+q-r)\, {}_2F_1(w+q-r, r+1, \tfrac12, \sin^2\vartheta)
	\end{multline}
	Now, for each term in the sum over $r$, we shift the contour to the right of the pole at $w = r - q$, denoting the new contour parameter $c_r$. For $r = 0, \ldots q$ no shift is needed, so we take $c_r = c$. The poles with $r \ge q + 1$ lie to the right of the original contour, defined by $c$. We shift the contour to the right of this pole, denoting the new contour parameter by $c_r$. Shifting the contour past them, we pick up their residues, yield the additional terms
	\begin{multline}
		\varphi_0^{q,\mathrm{int}}(k, M) = \Bigl(\frac{\beta}{2}\Bigr)^{\!\!-2q} \sum_{r=0}^\infty (-)^r \frac{r!}{(2r+1)!} \, 	z^{2r}  \frac1{2\pi i} \int_{c_r-i\infty}^{c_r+i\infty} dw \,\Bigl(\frac{\beta M}{2}\Bigr)^{\!\!-2w} \, \Gamma(w) \,  \Bigl[ \Li_{2(w+q-r)} (e^{i\theta}) +  \Li_{2(w+q-r)} (e^{-i \theta})  \Bigr]  \\\times 
		\Gamma(w+q-r)\, {}_2F_1(w+q-r, r+1, \tfrac12, \sin^2\vartheta)\\
		+ M^{2q} \sum_{r=q+1}^\infty (-)^r \frac{r!}{(2r+1)!} \, (r-q-1)!  \, \Bigl(\frac{k}{M}\Bigr)^{\!2r} 
	\end{multline}

	Expanding polylogarithm and hypergeometric function in the standard series we get
	\begin{multline}
		\varphi_0^{q,\mathrm{int}}(k, M) = 2\Bigl(\frac{\beta}{2}\Bigr)^{\!\!-2q} \sum_{r=0}^\infty (-)^r \frac{r!}{(2r+1)!} \, 	z^{2r} \sum_{n=1}^{\infty} \frac{\cos(n\theta)}{n^{2(q-r)}}\sum_{\ell=0}^\infty \frac{(r+1)_\ell}{(\frac12)_\ell} \frac{(\sin\vartheta)^{2\ell}}{\ell!}  \\ \times \frac1{2\pi i} \int_{c_r-i\infty}^{c_r+i\infty} dw \,\Bigl(\frac{\beta M n}{2}\Bigr)^{\!\!-2w} \, \Gamma(w) 
		\Gamma(\ell+w+q-r) \\
		+ M^{2q} \sum_{r=q+1}^\infty (-)^r \frac{r!}{(2r+1)!} \, (r-q-1)!  \, \Bigl(\frac{k}{M}\Bigr)^{\!2r} 
	\end{multline}
	where the $n$-sum and $\cos(n\theta)$ arise from the series expansion of polylogarithm, whereas the $\ell$-sum comes from the Gauss series for the hypergeometric function.
	To evaluate the remaining $w$-integral, we employ the Mellin-Barnes representation \cite[10.32.13]{NIST:DLMF} of the modified Bessel function of the second kind
	\begin{equation}
		K_{\nu}(z) = \frac{(\frac12 z)^{\nu}}{4\pi i} \int_{c-i\infty}^{c+i\infty} dw \, (\tfrac12 z)^{-2w} \, \Gamma(w) \Gamma(w-\nu) 
	\end{equation}
	which is valid for $c > \max\{0, \nu\}$. This condition is the reason for the contour shifting made above. Thus, we calculate the contour integral explicitly and obtain
	\begin{multline}
		\varphi_0^{q,\mathrm{int}}(k, M) = 4\Bigl(\frac{\beta}{2}\Bigr)^{\!\!-2q} \sum_{r=0}^\infty (-)^r \frac{r!}{(2r+1)!} \, 	z^{2r} \sum_{\ell=0}^\infty \frac{(r+1)_\ell}{(\frac12)_\ell} \frac{(\sin\vartheta)^{2\ell}}{\ell!} \sum_{n=1}^{\infty} \frac{\cos(n\theta)}{n^{q-r-\ell}}  \Bigl(\frac{\beta M }{2}\Bigr)^{\!\ell+q-r}  K_{\ell+q-r}(\beta M n)
		\\
		+ M^{2q} \sum_{r=q+1}^\infty (-)^r \frac{r!}{(2r+1)!} \, (r-q-1)!  \, \Bigl(\frac{k}{M}\Bigr)^{\!2r} 
	\end{multline}
	collecting all terms together, we obtain
	\begin{multline}
		\varphi_0^{q}(k, M) = 4\Bigl(\frac{\beta}{2}\Bigr)^{\!\!-2q} \sum_{r=0}^\infty (-)^r \frac{r!}{(2r+1)!} \, 	z^{2r} \sum_{\ell=0}^\infty \frac{(r+1)_\ell}{(\frac12)_\ell} \frac{(\sin\vartheta)^{2\ell}}{\ell!} \sum_{n=1}^{\infty} \frac{\cos(n\theta)}{n^{q-r-\ell}}  \Bigl(\frac{\beta M }{2}\Bigr)^{\!\ell+q-r}  K_{\ell+q-r}(\beta M n)
		\\ + (-)^{q}  M^{2q} \sum_{r=0}^q\frac{1}{(q-r)!} \frac{r!}{(2r+1)!} \Bigl(\frac{k}{M}\Bigr)^{\!2r}  \bigl(2 \log\tfrac{\Lambda}{M} + H_{q-r}\bigr)
		+ M^{2q} \sum_{r=q+1}^\infty (-)^r \frac{r!}{(2r+1)!} \, (r-q-1)!  \, \Bigl(\frac{k}{M}\Bigr)^{\!2r}.
	\end{multline}
	
%

\section{Properties of $\varOmega$ operator} \label{sec:app_Omega}
	The explicit form (\ref{eq:vphi_n_ans}),~(\ref{eq:vphi_neg_ans}) of the massless effective action fundamental form factor $\varphi_{0}^q(k)$ involves the shift-like operator $\varOmega$, defined by its action on a test function of $|k|$
	\begin{equation}
		\varOmega \, \varPhi(k) = \frac{e^{i\vartheta}\varPhi(k e^{i\vartheta}) + e^{-i\vartheta} \varPhi(k e^{-i\vartheta})}{e^{i\vartheta} + e^{-i\vartheta}},
	\end{equation}
	where $\vartheta = \arcsin (\omega / k)$.
	We need the explicit form of its action on the power function
	\begin{equation} \label{eq:app_Omega_z}
		\varOmega z^{\lambda} = z^{\lambda} \,  \frac{\cos(\lambda+1) \vartheta}{\cos\vartheta} = z^{\lambda} {}_2F_1(-\tfrac{\lambda}2,\tfrac{\lambda}2+1,\tfrac12,\sin^2\vartheta),
	\end{equation}
	where $z = \beta k /2$ and in the second equality we used the known property of the hypergeometric function \cite[Eq.~9.121.21]{gradshteyn2014table}.
	
	We also require the explicit form of the derivatives $\partial_{z^2}^p(\varOmega z^\lambda)$. When taking the derivative w.r.t.\ $z^2$, we treat $\omega$, entering the expressions via $\vartheta$, as constant. Then $\sin \vartheta = y / z$, where $y = \beta \omega / 2$ is fixed during the differentiation. 
	Now, let us apply the derivative w.r.t.\ $z^2$ to both sides of (\ref{eq:app_Omega_z}). For this purpose, we use the identity
	\begin{equation}
		\partial_t^p = (-)^p t^{-p-1} \, \partial_{t^{-1}}^p \, t^{-p+1}
	\end{equation}
	for $t = y^2 / z^2$, with $t = t(z)$, and the identity \cite[Eq.~15.5.3]{NIST:DLMF} on the derivative of the hypergeometric function. This yields
	\begin{subequations}
		\begin{align}
			\partial_{z^2}^p\bigl(\varOmega z^{\lambda}\bigr) &= \frac{\Gamma(\frac{\lambda}2+1)}{\Gamma(\frac{\lambda}2-p+1)} z^{\lambda-2p} {}_2F_1(-\tfrac{\lambda}2+p,\tfrac{\lambda}2+1;\tfrac12;\sin^2\vartheta) \label{eq:app_Omega_z_d}\\
			&=\frac{\Gamma(\frac{\lambda}2+1)}{\Gamma(\frac{\lambda}2-p+1)} z^{\lambda-2p} (\cos \vartheta)^{-2p-1} {}_2F_1( \tfrac{\lambda}2-p + \tfrac12,-\tfrac{\lambda}2-\tfrac12;\tfrac12;\sin^2\vartheta),  \label{eq:app_Omega_z_d_Euler}
		\end{align}
	\end{subequations}
	where, in passing from the first line to the second one, we applied Euler's transformation to the hypergeometric function. In particular, substituting $\lambda = 2p+2r+1$ and applying argument inversion and duplication identities to gamma function, we get (\ref{eq:Omega_d_z}).
	
	Note that for $\frac{\lambda}2 - p + 1$ is equal to a nonpositive integer $-m$ with $m = 0, \ldots (p-1)$, the r.h.s.\ in (\ref{eq:app_Omega_z_d}) vanishes due to $\Gamma(\frac{\lambda}2 - p + 1)$ in the denominator.
	In particular, this means that $\partial_{z^2}^p\bigl(\varOmega z^{2n}\bigr) = 0$ for integer $n = 0, \ldots (p-1)$ as if there were no operator $\varOmega$. However a nontrivial consequence arises if one expands both sides of (\ref{eq:app_Omega_z_d}) around this vanishing case. Indeed, setting $\frac{\lambda}2=n+s$ and taking the derivative w.r.t.\ $s$ at $s=0$, we get
	\begin{equation} \label{eq:app_Omega_log_less}
		\partial_{z^2}^p\bigl(\varOmega z^{2n} \log z\bigr)	= \tfrac12 (-)^{p-n-1} n!(p-n-1)! z^{2n-2p} {}_2 F_1(-n+p,n+1;\tfrac12;\sin^2\vartheta).
	\end{equation}
	Similarly, for $\frac{\lambda}2 = p$ we have a trivial identity $\partial_{z^2}^p\bigl(\varOmega z^{2p}\bigr) = p!$. However, setting $\frac{\lambda}2 = p + s$ and taking the $s$-derivative of both sides of (\ref{eq:app_Omega_z_d}) at $s=0$, we get
	\begin{equation}
		\partial_{z^2}^p\bigl(\varOmega z^{2p}\log z\bigr) =  p! (\log z + \tfrac12 H_p)  - (p+1)! \sin^2\vartheta \, {}_3 F_2(1,1,p+2;\tfrac32,2;\sin^2\vartheta),
	\end{equation}
	where $H_p = \sum_{j=1}^p 1/j$ is the harmonic number, and we have used the Taylor expansion of the hypergeometric function in its first parameter
	\begin{equation} \label{eq:app_2F1_par_exp}
		{}_2F_1(-s, p+1, \tfrac12, \sin^2\vartheta) =
		1 - 2(p+1) \, s \, \sin^2\vartheta \, {}_3F_2(1,1,p+2;2,\tfrac32; \sin^2\vartheta) + \mathcal{O}(s^2).
	\end{equation}

\section{Explicit expressions for $\varPi_\perp$ in $D=4$} \label{sec:app_Pi_explicit}

		\begin{align}
		&\varPi_\perp^{(0),\mathrm{sing}}(k,M) = - \Bigl(\mfrac{T^2}{6}+\mfrac{\mu^2}{2\pi^2}\Bigr) \Bigl[ \mfrac{k^2}{\bk^2} \Bigl(1-\mfrac{k_0}{2|\bk|}\log \mfrac{|\bk|+k_0}{|\bk|-k_0}\Bigr) - 1\Bigr],\\
		&\varPi_\perp^{(0),\mathrm{log}}(k,M) = \frac{k^2}{6\pi^{2}} \Bigl[\log \mfrac{\Lambda}{2T} + \gamma + \mfrac{3}{4}  - \Bigl(1 -\mfrac{k^2}{4\bk^2}\Bigr) \Bigl(1-\mfrac{k_0}{2|\bk|} \log \mfrac{|\bk|+k_0}{|\bk|-k_0}\Bigr) \Bigr] , \\
		&\varPi_\perp^{(0),\mathrm{reg}}(k,M) = \frac{k^{2}}{2\pi^{2}}  \frac{T^3}{|\bk|^3} \sum_{\{\sigma\}=\pm1} \Bigl[ \sigma_k  \Li^{(1)}_{3}\!\bigl(-e^{\, \sigma_k\frac{|\boldsymbol{k}|}{2T} +\sigma_\omega\frac{k_0}{2T} + \sigma \frac{\mu}{T}}\bigr) -\mfrac{|\bk|}{2T} \Li^{(1)}_{2}\!\bigl(-e^{\, \sigma_k\frac{|\boldsymbol{k}|}{2T} +\sigma_\omega\frac{k_0}{2T} + \sigma \frac{\mu}{T}}\bigr)  \nonumber \\[-0.5ex] &\hspace{27em}+ 
		\mfrac{|\bk|^2}{4T^2} \sigma_k \Li^{(1)}_{1}\!\bigl(-e^{\, \sigma_k\frac{|\boldsymbol{k}|}{2T} +\sigma_\omega\frac{k_0}{2T} + \sigma \frac{\mu}{T}}\bigr)\Bigr] \nonumber\\ &
		\hphantom{\varPi_\perp^{(0),\mathrm{reg}}(k,M)} = \frac{k^2}{(2\pi)^{2}} \sum_{r=0}^{\infty} \frac{(-)^{r}}{\,(2r+3)!} \Bigl(\frac{k}{2 T}\Bigr)^{\!2r} \bigl( \Li^{(1)}_{-2r}(-e^{\frac{\mu}{T}}) + \Li^{(1)}_{-2r}(-e^{-\frac{\mu}{T}}) \bigr) 
		\Bigl((2r+2)(2r+3)C_{2r}^{(1)}(\tfrac{ik_0}{k})-2 C_{2r}^{(2)}(\tfrac{ik_0}{k})\Bigr)
	\end{align}
	\subsubsection*{Order $M^2$ }

	\begin{align}
		&\varPi_\perp^{(1),\mathrm{log}}(k,M) = -\mfrac{M^{2}}{2\pi^{2}} \,
		\mfrac{k_0}{2|\bk|}\log \mfrac{|\bk|+k_0}{|\bk|-k_0},\\
		&\varPi_\perp^{(1),\mathrm{reg}}(k,M) = -\frac{1}{(2\pi)^{2}} M^{2}  
		\mfrac{T}{ |\boldsymbol{k}|} \sum_{\{\sigma\}=\pm1} \Bigl[ \sigma_k  \Li^{(1)}_{1}\!\bigl(-e^{\, \sigma_k\frac{|\boldsymbol{k}|}{2T} +\sigma_\omega\frac{k_0}{2T} + \sigma \frac{\mu}{T}} \bigr) - \mfrac{ |\boldsymbol{k}|}{2T} \Li^{(1)}_{0} ( -e^{\, \sigma_k\frac{|\boldsymbol{k}|}{2T} +\sigma_\omega\frac{k_0}{2T} + \sigma \frac{\mu}{T}}) \Bigr] \\ &=
		-\frac{M^{2}}{(2\pi)^{2}}
		\sum_{r=0}^\infty \frac{(-)^{r}}{(2r+3)!} \, \Bigl(\frac{k}{2T}\Bigr)^{2(r+1)}  \bigl( \Li^{(1)}_{-2(r+1)} (-e^{\frac{\mu}{T}}) +  \Li^{(1)}_{-2(r+1)} (-e^{-\frac{\mu}{T}})  \bigr) \, \Bigl((2r+2)(2r+3)\tilde{C}^{(0)}_{2r+2}(\tfrac{\omega}{k}) - 2 C_{2r+2}^{(1)}(\tfrac{\omega}{k})\Bigr) 
	\end{align}
	
	\subsubsection*{Order $M^4$}

	\begin{align}
		&\varPi_\perp^{(2)}(k,M) =  -\frac{1}{8\pi^2} \frac{M^{4}}{k^{2}}  \biggl(\sum_{\{\sigma\}=\pm1}  \Bigl[ 2 \Li^{(1)}_{0} ( -e^{\, \sigma_k\frac{|\boldsymbol{k}|}{2T} +\sigma_\omega\frac{k_0}{2T} + \sigma \frac{\mu}{T}}) - \sigma_k \mfrac{ |\bk|}{2T}  \Li^{(1)}_{-1} ( -e^{\, \sigma_k\frac{|\boldsymbol{k}|}{2T} +\sigma_\omega\frac{k_0}{2T} + \sigma \frac{\mu}{T}} )  \Bigr] \nonumber \\[-0.75ex]& \hspace{32em}
		-8 \bigl( \Li^{(1)}_{0}(e^{\frac{\mu}{T}}) + \Li^{(1)}_{0}(e^{-\frac{\mu}{T}}) \bigr)
		\biggr)	\nonumber \\ &= \frac{M^{2}}{(2\pi)^{2}}  \Bigl(\frac{M\beta}{2}\Bigr)^{\!2}  \biggl(
		\bigl( \Li^{(1)}_{-2} (e^{i\frac{\mu}{T}}) +  \Li^{(1)}_{-2} (e^{-\frac{\mu}{T}})  \bigr) \, (\tilde{C}^{(0)}_{2}(\tfrac{\omega}{k})+1) \nonumber\\&\hspace{1em}+
		\frac12 \,
		\sum_{r=0}^\infty \frac{(-)^{r}}{(2r+3)!} \, \Bigl(\frac{ k}{2T}\Bigr)^{\!2(r+1)}  \bigl( \Li^{(1)}_{-2(r+2)} (e^{\frac{\mu}{T}}) +  \Li^{(1)}_{-2(r+2)} (e^{-\frac{\mu}{T}})  \bigr) \, \Bigl((2r+2)(2r+3)\tilde{C}^{(-1)}_{2(r+2)}(\tfrac{\omega}{k}) - 2 \tilde{C}^{(0)}_{2(r+2)}(\tfrac{\omega}{k}) \Bigr)
		\biggr)
	\end{align}

\twocolumngrid

\bibliography{main}

\end{document}